\documentclass[twocolumn]{aastex701}

\definecolor{holliscolor}{HTML}{0077BB}

\usepackage{xspace}
\usepackage{hyperref}
\begin{document}

% --------------------------------------------------------------
% -------------------------- COMMANDS --------------------------
% --------------------------------------------------------------
% This is a list of common commands that will be used, including
% lines, measurements, math, units, etc.

% Additionally, color definitions and commands color-coding
% text are also listed in here, now.  Co-authors, feel free to 
% add your own color command for your comments (if that's how
% you want to leave comments).
% --------------------------------------------------------------

\newcommand{\sref}{\S\ref} % to reference a section -- not for figures or tables
\newcommand{\etal}{+}

% colors
% \definecolor{notes}{HTML}{C70039} % red color
%\definecolor{notes}{HTML}{7A3803} % orange color
%\definecolor{emphasis_box}{HTML}{D1E1EC}
%\definecolor{aggiemaroon}{HTML}{500000} % added by CJP

% coloring text
\newcommand{\editone}[1]{\textcolor{red}{\bf #1}}
\newcommand{\note}[1]{\textcolor{red}{#1}}

%coloring links
% \hypersetup{
%     colorlinks=true,
%     linkcolor=blue,
%     filecolor=magenta,      
%     urlcolor=arxiv,
% }
%\urlstyle{same}

% box for emphasizing things
\newcommand{\colorme}[1]{\begin{tcolorbox}[width=\textwidth,
                            colback={emphasis_box}]
                            #1    
                         \end{tcolorbox}}

% LINES
\newcommand{\be}{$\beta$}
\newcommand{\lam}{$\lambda$}
\newcommand{\hi}{\hbox{\sc H\,i}}           % neutral HI gas
\newcommand{\lya}{\hbox{Ly$\alpha$}}        % Lya
\newcommand{\nv}{\hbox{\sc N\,v}}           % NV 1239,1243
\newcommand{\niv}{\hbox{\sc N\,iv]}}      % CIII] 1907,1909
\newcommand{\ariv}{\hbox{\sc Ar\,iv]}}      % CIII] 1907,1909
\newcommand{\civ}{\hbox{\sc C\,iv}}         % CIV 1548,1551
\newcommand{\heii}{\hbox{He\,{\sc ii}}}     % HeII 1640
\newcommand{\cii}{\hbox{\sc C\,ii]}}      % CIII] 1907,1909
\newcommand{\hei}{\hbox{He\,{\sc i}}}     % HeI 1640
\newcommand{\oiiisemi}{\hbox{\sc O\,iii]}}  % OIII] 1661,1666
\newcommand{\ciii}{\hbox{\sc C\,iii]}}      % CIII] 1907,1909
\newcommand{\siiii}{\hbox{Si\,{\sc iii]}}}  % SiIII] 1883,1892
\newcommand{\mgii}{\hbox{Mg\,{\sc ii}}}     % MgII 2796,2803
\newcommand{\oi}{\hbox{O\,{\sc i}}}     % MgII 2796,2803
\newcommand{\oii}{\hbox{\sc [O\,ii]}}     % [OII]
\newcommand{\hb}{\hbox{\sc H$\beta$}}       % Hbeta 4863
\newcommand{\oiii}{\hbox{\sc [O\,iii]}}     % [OIII] 4959,5007
\newcommand{\ha}{\hbox{\sc H$\alpha$}}      % Ha 6563
\newcommand{\hd}{\hbox{\sc H$\delta$}}      % Hd 4102
\newcommand{\hg}{\hbox{\sc H$\gamma$}}      % Hg 4341
\newcommand{\he}{\hbox{\sc H$\epsilon$}}      % Hg 4341
\newcommand{\het}{\hbox{\sc H$\eta$}}      % Hg 4341
\newcommand{\sii}{\hbox{[S\,{\sc ii}]}}     % [SII] 6717,6713
\newcommand{\siii}{\hbox{[S\,{\sc iii}]}}   % [SIII] 9069,9531
\newcommand{\paa}{\hbox{Pa$\alpha$}}      % PaB 1.282micron
\newcommand{\pab}{\hbox{Pa$\beta$}}      % PaB 1.282micron
\newcommand{\pag}{\hbox{Pa$\gamma$}}
\newcommand{\neiii}{\hbox{[Ne\,{\sc iii]}}}  % [NeIII] 15.6micron
\newcommand{\siiiIR}{\hbox{[S\,{\sc iii]}}}  % [SIII] 18.7micron
\newcommand{\feii}{\hbox{[Fe\,{\sc ii]}}}  % [SIII] 18.7micron
\newcommand{\nev}{\hbox{[Ne\,{\sc v]}}}  % [SIII] 18.7micron
\newcommand{\oiv}{\hbox{O\,{\sc iv]}}}  % [SIII] 18.7micron
\newcommand{\siv}{\hbox{[S\,{\sc iv]}}}  % [SIV] 10.5micron

\DeclareRobustCommand{\ion}[2]{%
\relax\ifmmode
\ifx\testbx\f@series
{\mathbf{#1\,\mathsc{#2}}}\else
{\mathrm{#1\,\mathsc{#2}}}\fi
\else\textup{#1\,{\mdseries\textsc{#2}}}%
\fi}

% Measurements from lines
\newcommand{\vlya}{\hbox{$\Delta$v$_{Ly\alpha}$}}
\newcommand{\vciv}{\hbox{$\Delta$v$_{CIV}$}}
\newcommand{\wciii}{\hbox{W$_{\text{\sc C\,iii}]}$}~}
\newcommand{\woiii}{\hbox{W$_{[\text{\sc O\,iii}]}$}~}
\newcommand{\wciv}{\hbox{W$_{\text{\sc C\,iv}}$}~}
\newcommand{\wlya}{\hbox{W$_{\text{Ly$\alpha$}}$}~}
\newcommand{\fesc}{\hbox{$f_{esc}$}}

% Telescopes
\newcommand{\hst}{{\it HST}}
\newcommand{\jwst}{{\it JWST}}
\newcommand{\spitzer}{{\it Spitzer}}
\newcommand{\Spitzer}{{\it Spitzer}}

% objects/targets
\newcommand{\gndone}{z7\_GND\_42912}
\newcommand{\gndtwo}{z7\_GND\_16863}
\newcommand{\gndthree}{z7\_GND\_22483}

% code & models
\newcommand{\cloudy}{{\sc Cloudy}}
\newcommand{\bpass}{BPASS}

% Units  (just a very big list)
\newcommand{\unit}[1]{\ensuremath{\mathrm{\,#1}}\xspace}
\newcommand{\yr}{\unit{yr}}
\newcommand{\Gyr}{\unit{Gyr}}
\newcommand{\Myr}{\unit{Myr}}
\newcommand{\eV}{\unit{eV}}
\newcommand{\keV}{\unit{keV}}
\newcommand{\MeV}{\unit{MeV}}
\newcommand{\GeV}{\unit{GeV}}
\newcommand{\TeV}{\unit{TeV}}
\newcommand{\MB}{\unit{MB}}
\newcommand{\GB}{\unit{GB}}
\newcommand{\TB}{\unit{TB}}
\newcommand{\degree}{\ensuremath{{}^{\circ}}\xspace}
\newcommand{\degrees}{\degree}
\newcommand{\mas}{\unit{mas}}
\newcommand{\amin}{\unit{arcmin}}
\newcommand{\asec}{\unit{arcsec}}
\newcommand{\arcsecond}{$^{\prime\prime}$}
\newcommand{\angstrom}{\unit{\AA}}
\newcommand{\um}{\unit{$\mu$m}}
\newcommand{\cm}{\unit{cm}}
\newcommand{\km}{\unit{km}}
\newcommand{\pc}{\unit{pc}}
\newcommand{\kpc}{\unit{kpc}}
\newcommand{\second}{\unit{s}}
\newcommand{\us}{\unit{$\mu$s}}
\newcommand{\sr}{\unit{sr}}
\newcommand{\Msolar}{M$_\odot$}
\newcommand{\Msun}{M$_\odot$}
\newcommand{\Mstar}{\unit{M_{*}}}
\newcommand{\Zsolar}{$Z_\odot$}
\newcommand{\Lsolar}{\unit{L_\odot}}
\newcommand{\Lsun}{\unit{L_\odot}}
\newcommand{\Lstar}{\unit{L_{*}}}
\newcommand{\Lum}{\ensuremath{ L }\xspace}
\newcommand{\Dsun}{\unit{D_\odot}}
\newcommand{\Dgc}{\ensuremath{D_{GC}}\xspace}
\newcommand{\Rgc}{\ensuremath{R_{GC}}\xspace}
\newcommand{\cmcubes}{\ensuremath{\cm^{3}\second^{-1}}\xspace}
\newcommand{\magn}{\unit{mag}}
\newcommand{\mmag}{\unit{mmag}}
\newcommand{\e}{\unit{e^{-}}}
\newcommand{\rms}{\unit{rms}}
\newcommand{\pix}{\unit{pix}}
\newcommand{\rmspix}{\unit{rms/pix}}
\newcommand{\ermspix}{\e \rmspix}
\newcommand{\sol}{$_{\odot}$}

% Math
\mathchardef\mhyphen="2D
\newcommand{\vect}[1]{\boldsymbol{#1}}
\newcommand{\roughly}{\ensuremath{ {\sim}\,} }
\newcommand{\gtr}{\ensuremath{ {>}\,} }
\newcommand{\less}{\ensuremath{ {<}\,} }
\newlength{\dhatheight}
\newcommand{\doublehat}[1]{%
    \settoheight{\dhatheight}{\ensuremath{\hat{#1}}}%
    \addtolength{\dhatheight}{-0.35ex}%
    \hat{\vphantom{\rule{1pt}{\dhatheight}}%
    \smash{\hat{#1}}}}
\newcommand{\code}[1]{\texttt{#1}\xspace}
\newcommand{\dd}{\ensuremath{\rm d}}

\title{SPAM: Survey Overview and JWST/NIRCam Multi-Band Photometry Catalog}

\author[0000-0003-2366-8858]{Rebecca L. Larson}
\altaffiliation{Giacconi Postdoctoral Fellow}
\affiliation{Authors contributed equally to this work.}
\affiliation{Space Telescope Science Institute, 3700 San Martin Drive, Baltimore, MD 21218, USA}
\email[show]{rlarson@stsci.edu}  

\author[0000-0001-8047-8351]{Kelcey Davis}
\altaffiliation{NSF Graduate Research Fellow}
\affiliation{Authors contributed equally to this work.}
\affiliation{Department of Physics, 196A Auditorium Road, Unit 3046, University of Connecticut, Storrs, CT 06269, USA}
\affiliation{Los Alamos National Laboratory, Los Alamos, NM 87545, USA}
\email[show]{kelcey.davis@uconn.edu}

\author[0000-0003-3596-8794]{Hollis B. Akins}
\email{hollis.akins@gmail.com}
\altaffiliation{NSF Graduate Research Fellow}
\affiliation{Department of Astronomy, The University of Texas at Austin, Austin, TX 78712}
\affiliation{Cosmic Frontier Center, The University of Texas at Austin, Austin, TX 78712}

\author[0000-0001-8519-1130]{Steven L.\ Finkelstein}
\affiliation{Department of Astronomy, The University of Texas at Austin, Austin, TX 78712}
\affiliation{Cosmic Frontier Center, The University of Texas at Austin, Austin, TX 78712}
\email{stevenf@astro.as.utexas.edu}

\author[0000-0003-4922-0613]{Katherine Chworowsky}
\affiliation{Department of Astronomy, The University of Texas at Austin, Austin, TX 78712}
\affiliation{Cosmic Frontier Center, The University of Texas at Austin, Austin, TX 78712}
\email{k.chworowsky@utexas.edu}

\author[0000-0001-9298-3523]{Kartheik G. Iyer}
\affil{Institute for Astronomy, University of Hawai'i, 2680 Woodlawn Drive, Honolulu, HI 96822, USA}
\affil{Center for Computational Astrophysics, Flatiron Institute, 162 5th Ave, New York, NY 10010, USA}
\email{kiyer@hawaii.edu}

\author[0000-0002-1410-0470]{Jonathan R. Trump}
\affil{Department of Physics, 196A Auditorium Road, Unit 3046, University of Connecticut, Storrs, CT 06269, USA}
\email{jonathan.trump@uconn.edu}

\author[0000-0002-0243-6575]{Jacqueline Antwi-Danso}
\altaffiliation{Dunlap Postdoctoral Fellow}
\affiliation{University of Toronto, 50 St George St,Toronto, ON, M5S 3H4, Canada}
\email{j.antwidanso@utoronto.ca}

\author[0000-0002-7959-8783]{Pablo Arrabal Haro}
\affiliation{Center for Space Sciences and Technology, UMBC, 5523 Research Park Dr, Baltimore, MD 21228 USA }
\affiliation{Astrophysics Science Division, NASA Goddard Space Flight Center, 8800 Greenbelt Rd, Greenbelt, MD 20771, USA}
\email{pablo.arrabalharo@nasa.gov}

\author[0000-0001-8534-7502]{Bren E. Backhaus}
\affil{Department of Physics and Astronomy, University of Kansas, Lawrence, KS 66045, USA}
\email{bren.backhaus@ku.edu}

\author[0000-0002-9921-9218]{Micaela B. Bagley}
\affiliation{Southeastern Universities Research Association, Washington, DC 20005, USA}
\affiliation{Astrophysics Science Division, NASA Goddard Space Flight Center, Greenbelt, MD 20771, USA}
\affiliation{Department of Astronomy, The University of Texas at Austin, Austin, TX 78712}
\affiliation{Cosmic Frontier Center, The University of Texas at Austin, Austin, TX 78712}
\email{micaela.b.bagley@nasa.gov}

\author[0000-0001-6813-875X]{Guillermo Barro}
\affiliation{University of the Pacific, Stockton, CA 90340 USA}
\email{gbarro@pacific.edu}

\author[0000-0003-0492-4924]{Laura Bisigello}
\affiliation{INAF-Osservatorio Astronomico di Padova, Via dell'Osservatorio 5, 35122 Padova, Italy}
\email{laura.bisigello@inaf.it}

\author[0000-0001-5384-3616]{Madisyn Brooks}
\altaffiliation{NSF Graduate Research Fellow}
\affil{Department of Physics, 196A Auditorium Road, Unit 3046, University of Connecticut, Storrs, CT 06269, USA}
\email{madp.brooks@gmail.com}

\author[0000-0002-6184-9097]{Jaclyn B. Champagne}
\affiliation{Space Telescope Science Institute, 3700 San Martin Drive, Baltimore, MD 21218, USA}
\email{jchampagne@stsci.edu}

\author[0000-0001-8551-071X]{Yingjie Cheng}
\affiliation{Department of Astronomy, University of Washington, Seattle, WA 98195, USA}
\email{yingjiec@uw.edu}

\author[0000-0001-7151-009X]{Nikko J. Cleri}
\affiliation{Department of Astronomy and Astrophysics, The Pennsylvania State University, University Park, PA 16802, USA}
\affiliation{Institute for Computational and Data Sciences, The Pennsylvania State University, University Park, PA 16802, USA}
\affiliation{Institute for Gravitation and the Cosmos, The Pennsylvania State University, University Park, PA 16802, USA}
\email{cleri@psu.edu}

\author[0000-0001-7410-7669]{Dan Coe}
\affiliation{Space Telescope Science Institute, 3700 San Martin Drive, Baltimore, MD 21218, USA}
\affiliation{Center for Astrophysical Sciences, Department of Physics and Astronomy, The Johns Hopkins University, 3400 N Charles St. Baltimore, MD 21218, USA}
\affiliation{Association of Universities for Research in Astronomy (AURA), Inc.~for the European Space Agency (ESA)}
\email{dcoe@stsci.edu}

\author[0000-0001-6820-0015]{Luca Costantin}
\affiliation{Centro de Astrobiolog\'{\i}a (CAB), CSIC-INTA, Ctra. de Ajalvir km 4, Torrej\'on de Ardoz, E-28850, Madrid, Spain}
\email{lcostantin@cab.inta-csic.es}

\author[0000-0002-6219-5558]{Alexander de la Vega}
\affiliation{Department of Physics and Astronomy, University of California, 900 University Avenue, Riverside, CA 92521, USA}
\email{alexandd@ucr.edu}

\author[0000-0001-5414-5131]{Mark Dickinson}
\affiliation{NSF's National Optical-Infrared Astronomy Research Laboratory, 950 N. Cherry Ave., Tucson, AZ 85719, USA}
\email{mark.dickinson@noirlab.edu}

\author[0000-0002-7622-0208]{Callum T. Donnan}
\affiliation{NSF's National Optical-Infrared Astronomy Research Laboratory, 950 N. Cherry Ave., Tucson, AZ 85719, USA}
\email{callum.donnan@noirlab.edu}

\author[0000-0001-7113-2738]{Henry C. Ferguson}
\affiliation{Space Telescope Science Institute, 3700 San Martin Drive, Baltimore, MD 21218, USA}
\email{ferguson@stsci.edu}

\author[0000-0003-1530-8713]{Eric Gawiser}
\affiliation{Department of Physics and Astronomy, Rutgers, the State University of New Jersey, Piscataway, NJ 08854, USA}
\email{gawiser@physics.rutgers.edu}

\author[0009-0006-1252-206X]{Ananya Ganapathy}
\affiliation{Center for Astrophysical Sciences, Department of Physics and Astronomy, The Johns Hopkins University, 3400 N Charles St. Baltimore, MD 21218, USA}
\email{aganapa1@jh.edu}

\author[0000-0003-3248-5666]{Giovanni Gandolfi}
\affiliation{INAF – Osservatorio Astronomico di Roma, via Frascati 33, 00078, Monteporzio Catone, Italy}
\email{giovanni.gandolfi@inaf.it}

\author[0000-0002-7831-8751]{Mauro Giavalisco}
\affiliation{University of Massachusetts Amherst, 710 North Pleasant Street, Amherst, MA 01003-9305, USA}

\email{mauro@umass.edu}

\author[0000-0001-9440-8872]{Norman A. Grogin}
\affiliation{Space Telescope Science Institute, 3700 San Martin Drive, Baltimore, MD 21218, USA}
\email{nagrogin@stsci.edu}

\author[0000-0002-3301-3321]{Michaela Hirschmann}
\affiliation{Institute of Physics, Lab for galaxy evolution and spectral modelling, Observatory of Sauverny, Chemin Pegasi 51, 1290 Versoix, Switzerland}
\email{michaela.hirschmann@epfl.ch}

\author[0000-0003-3424-3230]{Weida Hu}
\email{weidahu@shao.ac.cn}
\affiliation{Key Laboratory for Research in Galaxies and Cosmology, Shanghai Astronomical Observatory, Chinese Academy of Sciences, 80 Nandan Road, Shanghai 200030, People's Republic of China}

\author[0000-0001-6251-4988]{Taylor A. Hutchison}
\altaffiliation{NASA Postdoctoral Fellow}
\affiliation{Astrophysics Science Division, NASA Goddard Space Flight Center, 8800 Greenbelt Rd, Greenbelt, MD 20771, USA}
\affiliation{Department of Astronomy, University of Maryland, Baltimore County, MD 21250, USA}
\affiliation{Center for Research and Exploration in Space Science and Technology, NASA/GSFC, Greenbelt, MD 20771 USA}
\email{astro.hutchison@gmail.com}

\author[0000-0003-1187-4240]{Intae Jung}
\affiliation{Department of Astronomy and Space Science, Chungbuk National University, Cheongju, 28644, Republic of Korea}
\email{ijung@cbnu.ac.kr}

\author[0000-0001-9187-3605]{Jeyhan S. Kartaltepe}
\affiliation{Laboratory for Multiwavelength Astrophysics, School of Physics and Astronomy, Rochester Institute of Technology, 84 Lomb Memorial Drive, Rochester, NY 14623, USA}
\email{jeyhan@astro.rit.edu}

\author[0000-0002-8360-3880]{Dale D. Kocevski}
\affiliation{Department of Physics and Astronomy, Colby College, Waterville, ME 04901, USA}
\email{}

\author[0000-0002-9393-6507]{Gene C. K. Leung}
\affiliation{MIT Kavli Institute for Astrophysics and Space Research, 77 Massachusetts Ave., Cambridge, MA 02139, USA}
\email{gckleung@mit.edu}

\author[0000-0003-1354-4296]{Mario Llerena}
\affiliation{INAF Osservatorio Astronomico di Roma, Via Frascati 33, 00078 Monte Porzio Catone, Rome, Italy}
\email{mario.llerenaona@inaf.it}

\author[0000-0003-1581-7825]{Ray A. Lucas}
\affiliation{Space Telescope Science Institute, 3700 San Martin Drive, Baltimore, MD 21218, USA}
\email{lucas@stsci.edu}

\author[0009-0008-4976-3216]{Faezeh Manesh}
\affiliation{Department of Physics and Astronomy, University of California, 900 University Avenue, Riverside, CA 92521, USA}
\email{fakhl001@ucr.edu}

\author[0000-0001-8688-2443]{Elizabeth J. McGrath}
\affiliation{Department of Physics and Astronomy, Colby College, Waterville, ME 04901, USA}
\email{test@gmail.com}

\author[0009-0006-9587-9311]{Jessica A. Mitlehner} 
\affiliation{Department of Physics and Astronomy, University of Iowa, Iowa City, IA, USA}
\email{jessica-mitlehner@uiowa.edu}

\author[0000-0001-7503-8482]{Casey Papovich}
\affiliation{Department of Physics and Astronomy, Texas A\&M University, College
Station, TX, 77843-4242 USA}
\affiliation{George P.\ and Cynthia Woods Mitchell Institute for
 Fundamental Physics and Astronomy, Texas A\&M University, College
 Station, TX, 77843-4242 USA}
 \email{papovich@tamu.edu}

\author[0000-0003-4528-5639]{Pablo G. P\'erez-Gonz\'alez}
\affiliation{Centro de Astrobiolog\'{\i}a (CAB), CSIC-INTA, Ctra. de Ajalvir km 4, Torrej\'on de Ardoz, E-28850, Madrid, Spain}
\email{pgperez@cab.inta.csic.es}

\author[0000-0002-9415-2296]{Giulia Rodighiero}
\affiliation{Dipartimento di Fisica e Astronomia, Università degli Studi di Padova, Vicolo dell'Osservatorio 3, I-35122 Padova, Italy}
\email{giulia.rodighiero@unipd.it}

\author[0000-0002-6386-7299]{Raymond C.\ Simons}
\affiliation{Department of Engineering and Physics, Providence College, 1 Cunningham Sq, Providence, RI 02918 USA}
\email{rcsimons@providence.edu}

\author[0000-0001-9495-7759]{Lu Shen}
\affiliation{Center for Astronomy and Astrophysics and Department of Physics, Fudan University, Shanghai 200438, People's Republic of China}
\email{lushen@fudan.edu.cn}

\author[0000-0002-6748-6821]{Rachel S.~Somerville}
\affiliation{Center for Computational Astrophysics, Flatiron Institute, 162 5th Avenue, New York, NY 10010, USA}
\email{rsomerville@flatironinstitute.org}

\author[0000-0002-8163-0172]{Brittany N. Vanderhoof}
\affiliation{Space Telescope Science Institute, 3700 San Martin Drive, Baltimore, MD 21218, USA}
\email{bvanderhoof@stsci.edu}

\author[0000-0003-3466-035X]{{L. Y. Aaron} {Yung}}
\altaffiliation{Giacconi Postdoctoral Fellow}
\affiliation{Space Telescope Science Institute, 3700 San Martin Drive, Baltimore, MD 21218, USA}
\affiliation{Institute of Astronomy and Department of Physics, National Tsing Hua University, Hsinchu 30013, Taiwan}
\email{l.y.aaronyung@gmail.com}

\author[0000-0002-7051-1100]{Jorge A. Zavala}
\affiliation{University of Massachusetts Amherst, 710 North Pleasant Street, Amherst, MA 01003-9305, USA}
\email{jzavala@umass.edu}

\collaboration{all}{The SPAM collaboration}

\shorttitle{SPAM Survey Overview and Data Release}
\shortauthors{Larson \& Davis et al.}

\begin{abstract}

%mention medium bands

We present the Star-formation from Photometry through the Addition of Medium-bands (SPAM) survey%(GO\#8559; PIs: Davis \& Larson)
, a 62.9 hr JWST/NIRCam imaging program to add ten new filters, nine medium-band filters, and one broad-band filter to the CEERS %Early-Release Science (ERS \#1345; PI: Finkelstein; \citealt{finkelstein25}) 
legacy dataset. We describe the observation strategy, present the data, and detail the data-reduction process. 
We include a complete reduction of imaging in the legacy CEERS footprint, inclusive of all publicly available photometric parallel and GO observations within the legacy CEERS footprint.  We present a multi-band photometric and photometric redshift catalog which includes all medium- and broad-band filters available on NIRCam.  We find that adding medium bands from the SPAM and MINERVA programs improves photometric redshifts. We discuss how these data advance the SPAM survey's key science goals: better-constrained photometric redshifts, identification of high-redshift overdensity members, improved recovery of emission-line strengths, and more accurate star formation histories. %, which we showcase alongside the improved photometry.

%We also provide the complete photometric catalog, inclusive of legacy CEERS photometry %, additional F090W coverage from GO\#2234%(PI: Ba\~nados)
%and the %Medium-band Imaging with NIRCam to Explore ReVolutionary
%Astrophysics 
%MINERVA program. %, GO\#7814, PIs: Muzzin, Marchesini, \& Suess; \citealt{Muzzin2025}) 

\end{abstract}

\keywords{\uat{Galaxies}{573}}


\section{Introduction} \label{sec:intro}

\setcounter{footnote}{0} 

Early Universe ($z>4$) discovery with \jwst\ has been driven by the legacy extragalactic deep fields, such as the Cosmic Evolution Early Release Science Survey (CEERS; \citealt{finkelstein25}), the Public Release IMaging for Extragalactic Research (PRIMER), the Ultradeep NIRSpec and NIRCam ObserVations before the Epoch of Reionization (UNCOVER; \citealt{bezanson24}), and the JWST Advanced Deep Extragalactic Survey (JADES; \citealt{Eisenstein26}), among others. Deep near-IR photometry from NIRCam \citep{Rieke23} in these fields, combined with extensive \hst/ACS and WFC3 photometry from large surveys such as the Great Observatories Origins Deep Survey (GOODS; \citealt{Giavalisco04}), the Cosmic Assembly Near-infrared Deep Extragalactic Legacy Survey (CANDELS; \citealt{grogin11,koekemoer11}), and the All-Wavelength Extended Groth Strip International Survey (AEGIS; \citealt{davis07}), provides photometric coverage from the rest-UV through the rest-optical at $4<z<9$. This coverage enables, among other discovery spaces, searches for early luminous galaxies, probes of star-formation-history timescales, and identification of the earliest forming active galactic nuclei (AGN).

%and others, %has shed new light on the early Universe.

However, many extragalactic fields largely use NIRCam broad-band (BB) filters and often lack the photometric coverage necessary to de-blend emission lines and resolve the shape of the rest-UV to rest-optical continuum at $2<z<9$.  
 %Identification of new and exciting sources is limited in scope by the available photometric coverage.
% Even after four years of \jwst\ spectroscopy programs targeting CEERS, photometrically observed targets outnumber those covered spectroscopically by at least 2 orders of magnitude. 
One possible avenue to mitigate this is to break up the spectral energy distribution (SED) by adding NIRCam medium-band (MB) filters, which provide higher wavelength resolution when combined with the BB filters. This technique has a long history in imaging predating \jwst, including ground-based surveys such as the NEWFIRM Medium-Band Survey (NMBS; \citealt{vanDokkum2009}), which showed that better-sampled SEDs yield improved photometric redshifts \citep{Whitaker2011}, and the Survey for High-z Absorption Red and Dead Sources (SHARDS; \citealt{Perez-Gonzales2013}) which used 24 MBs to provide an intermediate spectral resolution (R$\sim$50) over the GOODS-N field. The \texttt{FourStar} Galaxy Evolution Survey (ZFOURGE; \citealt{Straatman2016}) similarly provided MB coverage across a wide ($\sim400 \, \rm{arcmin}^2$) field of view with ground-based IR imaging and demonstrated improvement for SED sampling at $1<z<3$.

Such uses of MBs have proved crucial in the JADES field \citep{Eisenstein26}, where the \jwst\ Extragalactic Medium-band Survey (JEMS; \citealt{williams23}) covers $\sim40\,\rm{arcmin}^2$ across 5 MBs. In the lensed Canadian NIRISS Unbiased Cluster Survey (CANUCS; \citealt{Sarrouh2026}), 12 MBs cover $\sim 30  \,\rm{arcmin}^2$, where the addition of MBs has been used to rigorously identify high-z galaxies \citep{Asada2026}. MBs have also been utilized in UNCOVER \citep{Suess2024}, where 11 MBs cover $\sim 30  \,\rm{arcmin}^2$. Such surveys have also shown improvements in SED sampling from other telescopes, such as Spitzer (i.e., the F2 Extragalactic NearIR K-Split Survey (FENIKS) \citealt{Zaidi2024}). However, these surveys consistently probe small ($\leq 40  \,\rm{arcmin}^2$) areas of sky and/or rely on coverage from lower-resolution telescopes.

% \citep{Zaidi2024} mbs with spitzer

% ground based mbs:
% NEWFIRMMedium-Band Survey (NMBS), NOAO survey program on the Kitt Peak
% 4m telescope, \citep{vanDokkum2009} improved phot zs by adding medium bands \citep{Whitaker2011}

% this paragraph should end with limited area and limited # of MBs

%  e.g., NMBS,  ZFOURGE, FENIKS. 

The CEERS survey covers $\sim100 \,\rm{arcmin}^2$ with JWST Cycle 1 coverage including one MB and six BB \jwst/NIRCam photometric filters: F115W, F150W, F200W, F277W, F356W, F410M, and F444W \citep{finkelstein25}. This limited-resolution coverage introduces degeneracies, especially when emission lines are photometrically blended, which can produce spuriously high-redshift solutions in galaxies with emission lines in the BB filters \citep{Donnan2023, Arrabal2023a}.
NIRSpec observations effectively break degeneracies between source types, %(e.g., $z=16$ galaxies and $z=5$ dusty galaxies \citealt{Arrabal2023b, Donnan2023}), 
but they require careful targeting of individual sources, are time-expensive, and are incomplete over large survey areas. Such spectroscopic observations can constrain a galaxy's properties, such as redshift or star formation history (SFH), but we can recover them only for a small subset of imaged galaxies. Improved SED sampling, enabled by the increased resolution of MBs, enables this science on a larger scale within a single imaging footprint.

Adding MBs is therefore essential to expand this discovery space in the early universe within legacy fields. In this work, we present the Star Formation from the Addition of Medium Bands (SPAM, GO\#8559, PIs: Davis \& Larson) survey, which adds ten new filters, nine MB filters and one BB filter, to the CEERS legacy footprint at depths comparable to the legacy CEERS coverage. The SPAM program is presented here alongside photometry from a complementary program in the same \jwst\ cycle over the CEERS footprint: Medium-band Imaging with NIRCam to Explore ReVolutionary
Astrophysics (MINERVA, GO\#7814, PIs: Muzzin, Marchesini, \& Suess; \citealt{Muzzin2025}). Together, the programs add all remaining BB and MB filters from NIRCam over a total of $\sim 100 \, \rm{arcmin}^2$. %within the legacy CEERS footprint. 
The CEERS footprint now represents the legacy extragalactic field with the most complete SED sampling over the largest sky area, enabling a wide variety of science.

% The Cycle 4 \jwst\ observing cycle has brought unprecedented coverage of these MBs to the legacy extragalactic fields through the 
% Medium-band Imaging with NIRCam to Explore ReVolutionary
% Astrophysics (MINERVA, GO\#7814, PIs: Muzzin, Marchesini, \& Suess; \citealt{Muzzin2025}) and Star-formation from Photometry through the Addition of Medium-bands (SPAM, GO\#8559, PIs: Davis \& Larson) MB programs. Together, the programs cover the legacy CEERS footprint with all remaining 
% MB and broad-band NIRCam photometric filters over a total of $\sim 100 \, \rm{arcmin}^2$. This unique dataset will serve as an unmatched testbed for early-universe discovery. 

% SPAM is an efficient program that will have a large impact as it can, in combination with legacy CEERS data, simultaneously 1) measure the extent and occupation of two overdensities that may be powering the earliest-known reionized bubbles at $z= $ 7.7 and 8.7, 2) provide necessary filters to break photometric redshift degeneracies and reliably map galaxy evolution at $z>4$, 3) identify and characterize emission line galaxies, and 4) constrain star-formation histories of massive galaxies in the early Universe.

In this paper, we present the SPAM MB survey, provide reduced imaging mosaics for all MB and BB filters in the CEERS field, which includes a complete photometric catalog, alongside photometric redshifts for every source, and describe the program's key science cases. In Section \ref{sec:data}, we describe the survey, present the imaging data-reduction process, and produce a photometric catalog inclusive of the SPAM + MINERVA dataset, which we make publicly available here to enable community science. In Section \ref{sec:photoz}, we present the photometric redshift catalog which accompanies the photometric catalog and compare it to $\sim4500$ sources in CEERS with spectroscopic redshifts to assess improvement. In Section \ref{sec:sciencecases}, we show early results to demonstrate the improvement the SPAM + MINERVA dataset brings to (1) quantifying overdensities, (2) identifying emission-line galaxies, and (3) improving star-formation histories (SFHs). We summarize our conclusions in \ref{sec:conclusions}.

%\section{Observation Strategy} \label{sec:observations}

% \noindent{\bf Expected Yield:}
% The UV LF predicts 18 $z\sim 8.7$ galaxies brighter than the CEERS limiting magnitude of 28.6 \citep{Finkelstein2022c}, and with a 4$\times$ overdense structure \citep{Whitler2024}, there are at least 70 EoR sources we will confirm with these observations.
% %The vast majority of these are at CEERS continuum levels, leaving their photometric redshifts currently unconstrained without these MBs. 
% Observations will be obtained for the entire massive galaxy sample presented in \citep{Labbe2023} % and 
% %Chworowsky+23 
% ($\sim$ 150 galaxies), including candidates with log$(M_\star/M_\odot) > 10$ at $z>6$, constraining their SFHs beyond the capabilities that only BB photometry provides. As opposed to targeted spectroscopic observations, SPAM offers an affordable method of studying populations of galaxies across cosmic time, providing further constraints on how galaxies evolve.

\begin{figure*}[ht!]
    \centering
    \includegraphics[width=0.99\linewidth]{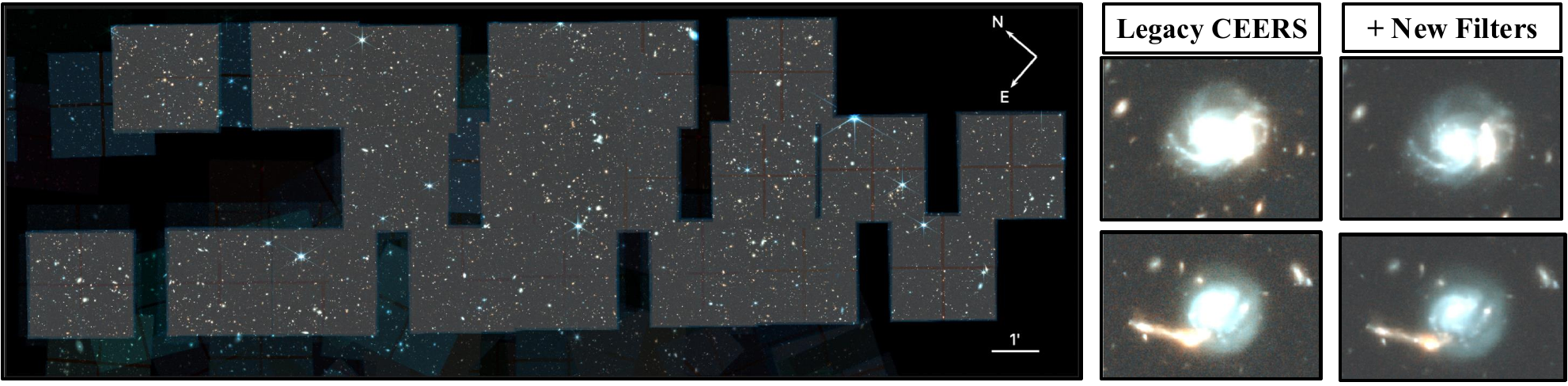}
    \caption{\textbf{Left:} Full 20-filter NIRCam color image of the CEERS legacy field observations using a rainbow color stack where blue colors start at the F070W image and end with red colors in the F480M image. This incorporates data from CEERS, GO \#2344, SPAM, MINERVA, and all other public NIRCam imaging in the vicinity (As described in \S\ref{sec:datareduction}). \textbf{Right:} Two zoom-ins on galaxies with just the legacy CEERS+GO\#2344 8-filter imaging (left), and from the full 20-color image (right). The increased wavelength-space sampling from the 12 additional filters included by SPAM + MINERVA highlights smaller details in these galaxies, enabling more robust scientific measurements.}
    \label{fig:surveyimage}
\end{figure*}

\section{Data} \label{sec:data}

The SPAM photometric dataset follows the CEERS (ERS-\#1345; PI: Finkelstein; \citealt{finkelstein25}) survey's legacy, adding nine new MB filters and one BB photometric filter: F070W, F140M, F162M, F182M, F210M, F300M, F335M, F360M, F430M, and F480M. This unprecedented coverage, combined with the two additional MB filters from MINERVA \citep{Muzzin2025}: F250M and F460M, gives the CEERS legacy footprint coverage across all broad-band and medium-band photometric filters available on NIRCam. MINERVA also adds additional depth to four blue MB filters: F140M, F162M, F182M, and F210M.

We obtained these data with \jwst/NIRCam between 24-26 June 2026 at a position angle of PA=131\degree\ to align with the NIRCam pointings of the CEERS program observed in 2022. 
We choose to use the INTRAMODULEBOX 4 dither pattern to cover all detector gaps and provide the most uniform depth across our survey field, ensuring sensitivity to even the faintest emission lines. We use the FULL subarray with the MEDIUM8 readout pattern (more groups per fixed exposure time, less susceptible to cosmic rays), with 8 groups, 1 integration, and 4 exposures per specification for the F480M/F430M/F360M + F070W/F140M/F162M filters. We reduced our groups to 7 for the F335M + F210M filters and to 6 for the F300M + F182M filters, but kept 1 integration and 4 exposures to take advantage of \jwst's increased sensitivity at these wavelengths and use telescope time more efficiently. All data from the SPAM survey are available in MAST at \dataset[10.17909/get7-sa13]{http://dx.doi.org/10.17909/get7-sa13}.

Our exposure times are those needed to detect a significant MB excess, with a formal signal-to-noise of ${\rm SNR}>3.0$ for the MB/BB flux ratio. The red filter(s) drive the depth, while the blue filters are matched to each red filter based on the sensitivity curves in order to achieve the most uniform depth.  We reach this in 0.9 hours (3349.9s) of integration per pointing in F480M/F430M/F360M (+F070W/F140M/F162M), 0.8 hours (2920.4s) in F335M (+F210M), and 0.7 hours (2490.9s) in F300M (+F182M) due to the increased sensitivity of \jwst\ at those wavelengths. At this depth, we are sensitive to galaxies with observed-frame EW$_{\oiii}$ down to $\sim 400$\AA\ in the red MB filters. %For galaxies with high EW$_{\oiii + \hb}$ similar to those measured at $z=7.7$ and 8.7 already in CEERS ($\sim 1000$\AA), our MB significance would be $>8$ (such galaxies would also be detectable with an excess significance of $\sim3$ with continua a full magnitude fainter). 

\subsection{Data Reduction} \label{sec:datareduction}

The SPAM photometric dataset is processed alongside complementary medium-band data from the MINERVA survey \citep{Muzzin2025}, legacy CEERS broad-band coverage \citep{finkelstein25}, an additional two filters from GO\#2344 (PI: E.\ Ba\~nados), and all other public NIRCam imaging falling within the CEERS footprint as of July 17, 2026 (See Appendix \ref{sec:depths}). We ensure a uniform reduction across the imaging filters presented. Table \ref{tbl:filters} lists all BB and MB filters that cover the CEERS legacy footprint and were reduced for this survey. Figure \ref{fig:surveyimage} shows a full 20-filter NIRCam color image illustrating the CEERS footprint we use for our mosaics and catalogs. This is inclusive of all public NIRCam imaging in these filters as of the date of publication.

% working draft! 

We reduce all available NIRCam imaging data through a custom pipeline that largely follows \citet{Bagley2023} and \citet{franco2026cosmos}. Briefly, we processed the raw NIRCam images through the JWST Calibration Pipeline version 1.20.2 \citep{bushouse_2025_16280965}, with several custom modifications, including subtracting $1/f$ noise and sky background.  We use the Calibration Reference Data System (CRDS)\footnote{\url{jwst-crds.stsci.edu}} pmap 1481, which corresponds to NIRCam instrument mapping imap 0342. We subtract $1/f$ noise through iterative source masking and amp-row median subtraction, as described in \citet{Bagley2023}. 

We remove wisps --- scattered light features present in several NIRCam short wavelength (SW) detectors --- by rescaling and subtracting the templates provided by STScI.\footnote{\url{https://stsci.app.box.com/s/1bymvf1lkrqbdn9rnkluzqk30e8o2bne}}
After stage2 of the pipeline, we visually inspect all exposures for residual unsubtracted wisps or other imaging artifacts (e.g., ``claws,'' ``ginko leaves,'' ``dragon's breath''); we then manually mask and remove these artifacts from the combined mosaics. 

Astrometric calibration is conducted via the \jwst/\hst\ alignment tool \citep[JHAT;][]{rest2023jwsthst} using as an absolute reference a photometric catalog generated from the CEERS DR1 NIRCam imaging \citep{Bagley2023, Finkelstein2024}, itself aligned to the DESI Legacy Imaging Surveys \citep{Dey2019} DR9 catalog as described in \citep{Bagley2023}. 

We create the final mosaics in Stage 3 of the pipeline with a pixel size of $0\farcs03$/pixel on the same pixel grid as the CEERS DR1 imaging. We perform background subtraction on the final mosaics following the iterative procedure described in \citet{Bagley2023}, which yields a highly uniform background without subtracting significant flux in the wings of bright sources.

\begin{figure}[]
    \centering
    \includegraphics[width=0.99\linewidth]{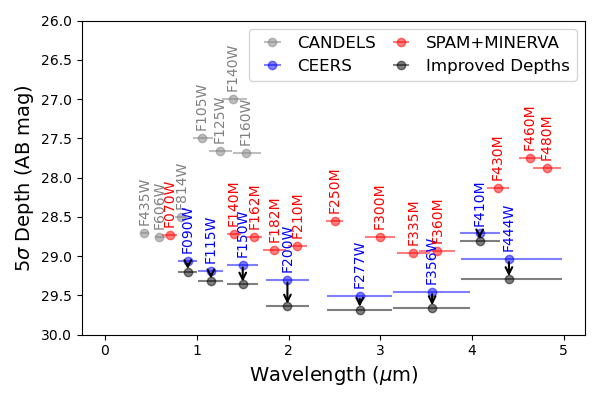}
    \caption{5$\sigma$ depths measured in 0\farcs2 apertures for each \jwst/NIRCam filter in the SPAM catalog (red). Depths of the original public CEERS v1.0 imaging \citep{Bagley2023} and Cycle 1 GO\#2234 data are in blue and are shallower compared to the new depths achieved by the SPAM reductions for those same filters (black; see Appendix \ref{sec:depths}). Legacy CANDELS HST Depths are shown in gray \citep{grogin11, koekemoer11}.}
    \label{fig:depths}
\end{figure}

We show the SPAM data depths in red in Figure \ref{fig:depths}, with the legacy CANDELS \hst\ depths \citep{grogin11, koekemoer11} in gray and the legacy CEERS v1.0 depths in blue \citep{finkelstein25}. Our catalog has improved depths for the CEERS NIRCam filters compared to the public CEERS imaging \citep{Bagley2023}, as shown in black in Figure \ref{fig:depths}. We discuss how the SPAM catalog improves depths over the initial public CEERS reductions in Appendix \ref{sec:depths}. We list the depths of all the \jwst/NIRCam filters in our SPAM catalog in Table \ref{tbl:filters}. Fully reduced image mosaics can be found on the SPAM website\footnote{\href{https://kelceydavis33.github.io/SPAM}{https://kelceydavis33.github.io/SPAM}}.

\begin{table}[h!]
   \begin{centering}
        
    \begin{tabular}{c|c|c|c}
        \hline
         Filter & Wavelength [$\mu$m] & 5$\sigma$ Depth & Survey Coverage \\ [0.5ex] 
         \hline\hline
         
         F070W & 0.705 & 28.73 & \textbf{SPAM} \\ 
         \hline
         F090W & 0.902 & 29.20 & GO\#2234\\ 
         \hline
         F115W & 1.154 & 29.32 & CEERS \\ 
         \hline
         F140M & 1.405 & 28.72 & \textbf{SPAM} + MINERVA \\
         \hline
         F150W & 1.501 & 29.36 & CEERS \\
         \hline
         F162M & 1.627 & 28.76 & \textbf{SPAM} + MINERVA \\
         \hline
         F182M & 1.845 & 28.92 & \textbf{SPAM} + MINERVA \\
         \hline
         F200W & 1.990 & 29.64 & CEERS \\
         \hline
         F210M & 2.096 & 28.87 & \textbf{SPAM} + MINERVA\\
         \hline
         F250M & 2.503 & 28.55 & MINERVA \\ 
         \hline
         F277W & 2.786 & 29.68 & CEERS \\
         \hline
         F300M & 2.996 & 28.75 & \textbf{SPAM} \\
         \hline
         F335M & 3.362 & 28.96 & \textbf{SPAM} \\
         \hline
         F356W & 3.553 & 29.66 & CEERS \\
         \hline
         F360M & 3.623 & 28.93 & \textbf{SPAM} \\
         \hline
         F410M & 4.092 & 28.81 & CEERS \\
         \hline
         F430M & 4.281 & 28.13 & \textbf{SPAM} \\
         \hline
         F444W & 4.421 & 29.29 & CEERS \\
         \hline
         F460M & 4.630 & 27.75 & MINERVA \\
         \hline
         F470N & 4.707 & 26.37 & GO\#2234 \\
         \hline
         F480M & 4.817 & 27.88 & \textbf{SPAM} \\ %[1ex] 
         \hline
         \hline
    \end{tabular}

    \end{centering}
    \caption{Full \jwst/NIRCam filter coverage in the photometry catalog includes the new observations from SPAM + MINERVA and the legacy CEERS and GO\#2234 data. 5$\sigma$ point source depths are in AB Mag \citep[]{Oke1983}, and are measured in 0\farcs2 on the newest reductions for all filters as described in \S\ref{sec:datareduction}.}
    \label{tbl:filters}
\end{table}

\subsection{Photometry Catalog}\label{sec:catalog}

The full photometric catalog presented here can be accessed from the SPAM website\footnote{\href{https://astrosteven.github.io/spam/}{https://astrosteven.github.io/spam/}}. This catalog contains 174,465 sources across the legacy CEERS footprint, plus additional coverage from other public NIRCam data in the vicinity (See Appendix \ref{sec:depths}). This catalog provides photometry across 28 filters: 7 \hst/ACS and WFC3 filters from the CANDELS data \citep{grogin11, koekemoer11}, and other public archival \hst\ data, reprocessed to the latest calibrations,  plus 20 \jwst/NIRCam filters: seven from the CEERS survey \citep{finkelstein25}, inclusive of additional public data in the field, two from GO\#2234, two from MINERVA \citep{Muzzin2025}, and 10 from SPAM, all in nJy. Table \ref{tbl:filters} lists all \jwst/NIRCam filters and depths included in the photometry catalog.

The photometric catalog was created as part of the Uniform Near-Infrared CatalOgs from Robust imaging (UNICORN) project. UNICORN aims to measure highly accurate photometric colors, total fluxes, and photometric redshifts. This full process will be described in  S.\ Finkelstein et al. (in prep), though it is based predominantly on the process described in \citet{Finkelstein2024}. The README associated with this release includes all details, but we provide a summary here.

We match the PSF of all ACS bands and all NIRCam bands bluer than F277W to that of F277W. For bands with larger PSFs (the redder NIRCam bands, and all WFC3 bands), we calculate correction factors by convolving the F277W image to the larger PSF and measuring the flux ratio in the native image to that in the convolved image for each source. By applying this correction factor to fluxes measured in images with larger PSFs, we correct for the missing flux (assuming the morphology does not change significantly).  

We performed photometry with Source Extractor (v2.25.0, \citealt{1996A&AS..117..393B}) in two-image mode, using a weighted sum of the native-resolution F277W and F356W as the detection image. For the detection image, we used a 1/sqrt(WHT) map with map\_rms weighting, so Poisson noise (which is in the ERR maps) did not affect the detections. For the measurement image, we used the relevant ERR map. 

We ran both hot- and cold-mode catalogs. The hot-mode catalog has DETECT\_THRESH=1.75 and DETECT\_MINAREA=5, derived iteratively via visual inspection to minimize spurious sources while maximizing completeness; this has been validated to work reasonably well across a wide range of \jwst\ imaging depths. We ran a cold-mode catalog with DETECT\_THRESH=3.5 and DETECT\_MINAREA=40. For both catalogs, we used a top-hat convolution kernel with a width of 4 pixels and a kernel size of 5x5 pixels, and set DEBLEND\_NTHRESH=32. For the hot catalog, we used DEBLEND\_MINCONT 0.0001, while for the cold catalog we used DEBLEND\_MINCONT 0.005.  

Empirical flux uncertainties were derived by placing random non-overlapping apertures with diameters from 0.1-1.5" in empty regions of the image (avoiding the segmentation map), fitting a four-parameter function to the Normalized Median Absolute Deviation (NMAD) of the fluxes measured in each aperture as a function of the aperture area. We calculate the flux error for each object from this fit for the area in a given object's aperture.

We measure our fiducial fluxes in small Kron apertures and apply two aperture corrections. The first is based on the ratio of the default larger MAG\_AUTO Kron aperture to our small aperture. We then calculate an additional aperture correction to measure flux missed on larger scales. In the NIRCam bands, we ran source injection simulations, injecting small Sersic profiles \citep{Sersic63} with Galfit into the F277W and F356W images. We then create a detection image from F277W+F356W, and measure photometry in F277W. Comparing input and output fluxes, we find the input fluxes are $\sim$5-20\% higher than the output, with this discrepancy slightly increasing for fainter magnitudes. We fit a linear function to this offset and apply it to our fluxes. This is typically a factor of 1.02 at $m=24$, rising to 1.10 at $m=28$ (we cap this value at 1.20, since we cannot reliably measure recovered fluxes near the imaging depth in a given field).

We did the same for the hot- and cold-mode detection parameters. To combine the catalogs, we first included all cold-mode objects. We then included hot-mode objects whose central pixel did not lie in a cold-mode segmentation map, after dilating the cold map with a 5x5 kernel. Finally, we created a separate catalog using only F444W as the detection image in hot mode, and then added any sources from this catalog that were not already in the hot$+$cold catalog. This helps ensure we still include faint, red sources in our catalog.

Our released photometric catalog includes fiducial fluxes and flux errors in all filters, as well as fluxes and errors measured in a number of circular-diameter apertures. Further information is available on the team website.
%The photometric catalog is provided in the \texttt{Photometry} file and includes columns for ID, RA, DEC, and the traditional Source Extractor outputs. Our fiducial fluxes are in columns labeled \texttt{FLUX\_[FILTER]}, with corresponding errors in the \texttt{FLUXERR\_[FILTER]} for all 27 included filters. The catalog also includes aperture photometry for 12 different apertures in arcseconds: [0.05, 0.1, 0.15, 0.2, 0.3, 0.4, 0.5, 0.6, 0.7, 0.8, 0.9, 1.0]. These fluxes are in an array under the column \texttt{FLUX\_APER\_[FILTER]}, and corresponding errors are in an array in the \texttt{FLUXERR\_APER\_[FILTER]} column. We also provide a half-light radius (\texttt{RH\_[FILTER]}) and full-width half-max (\texttt{FWHM\_[FILTER]} for each filter. 

\section{Photometric Redshift Measurements} \label{sec:photoz}

Photometric redshifts are the backbone of many science goals enabled by the SPAM data (e.g., see Section \ref{sec:sciencecases}). We estimated photometric redshifts ($z_{phot}$) with \texttt{LAZY}\footnote{\href{https://hollisakins.com/Lazy.jl/stable/}{https://hollisakins.com/Lazy.jl/stable/}} a Julia adaptation of the \texttt{EAZY} software \citep{Brammer2008}. The methodology is similar to that in CEERS Key Paper I \citep{finkelstein23}, including new templates from \citep{larson22b}, and is described further in S.\ Finkelstein et al.\ (in prep). Our fiducial photometric redshifts use 23 filters in the catalog (as described in \S\ref{sec:catalog}), from both \hst\ (ACS) and \jwst, and were fit over $z=0-35$. We provide a photometric redshift for all sources in the SPAM photometric catalog under the \texttt{ZA} column (the redshift where the goodness-of-fit is maximized), with upper and lower 68th-percentile errors in the \texttt{L68} and \texttt{U68} columns. 95th percentile errors on the $z_{phot}$ are in the \texttt{L95} and \texttt{U95} columns.  We also provide full $z_{phot}$ probability distribution functions and the best-fitting templates.  We also provide full $z_{phot}$ catalogs with the fiducial filter set, including (separately) an extreme emission line galaxy (EELG) and little red dot (LRD) template, as well as our fiducial template set with only broad bands, plus WFC3 filters. The results of these fits are available on the team website. 

%The distribution of sources in the SPAM catalog as a function of photometric redshift can be found in red in the left panel of Figure \ref{fig:distributions}. This is compared to the distribution of sources in the CEERS only catalog (blue). The SPAM catalog includes coverage of additional public NIRCam pointings in the EGS field and thus covers more area than the CEERS only catalog, incorporating more sources overall. The CEERS only catalog contains 123,184 sources compared to the 174,456 sources in the SPAM catalog.

\subsection{Spectroscopic Redshift Catalog}\label{sec:specz}

The spectroscopic redshift ($z_{spec}$) measurements come from several input catalogs. The first is from the CEERS spectroscopic redshift catalog (Arrabal Haro et al., in prep), which uses CEERS NIRSpec data. We also use public \jwst\ spectroscopic redshifts from the Dawn \jwst\ Archive\footnote{\href{https://dawn-cph.github.io/dja/spectroscopy/nirspec/}{https://dawn-cph.github.io/dja/spectroscopy/nirspec/}} (DJA; \citealt{degraaff24, heintz24}) for recent spectroscopic redshifts from public NIRSpec data in EGS. We also measure spectroscopic redshifts ourselves on the public NIRSpec data in EGS and add these measurements to the list. Additionally, we include products from CANDELS \citep{grogin11,koekemoer11}, specifically the public CANDELS \hst+\spitzer\ catalog \citep{stefanon17}. Finally, we include the ground-based redshifts from Keck/MOSFIRE: from the MOSDEF survey \citealt{mosdef}, other MOSFIRE programs \citep{larson22, jung18}, and the Keck/DEIMOS observations in EGS \citep{urbanostawinski24}. We have spectroscopic redshifts for 7,004 sources in the SPAM survey. These values are in the \texttt{SPAM\_Specz\_Catalog} file under the \texttt{ZSPEC} column found on the SPAM website\footnote{\href{https://kelceydavis33.github.io/SPAM}{https://kelceydavis33.github.io/SPAM}}.

% \begin{figure}[h!]
%     \centering
%     \includegraphics[width=0.99\linewidth]{zspec_distributions.png}
%     \caption{%\textbf{Left: } The distribution of sources in the full SPAM (red) and CEERS (blue) catalogs as a function of photometric redshift ($z_{phot}$) in bins of $\Delta z = 0.1$. \textbf{Right: } 
%     The distribution of sources in the SPAM catalog with good spectroscopic redshifts (\texttt{zspec\_vetted = 2}) as a function of $z_{spec}$ in bins of $\Delta z = 0.1$. }
%     \label{fig:distributions}
% \end{figure}

\begin{figure*}[ht!]
    \centering
    \includegraphics[width=1\linewidth]{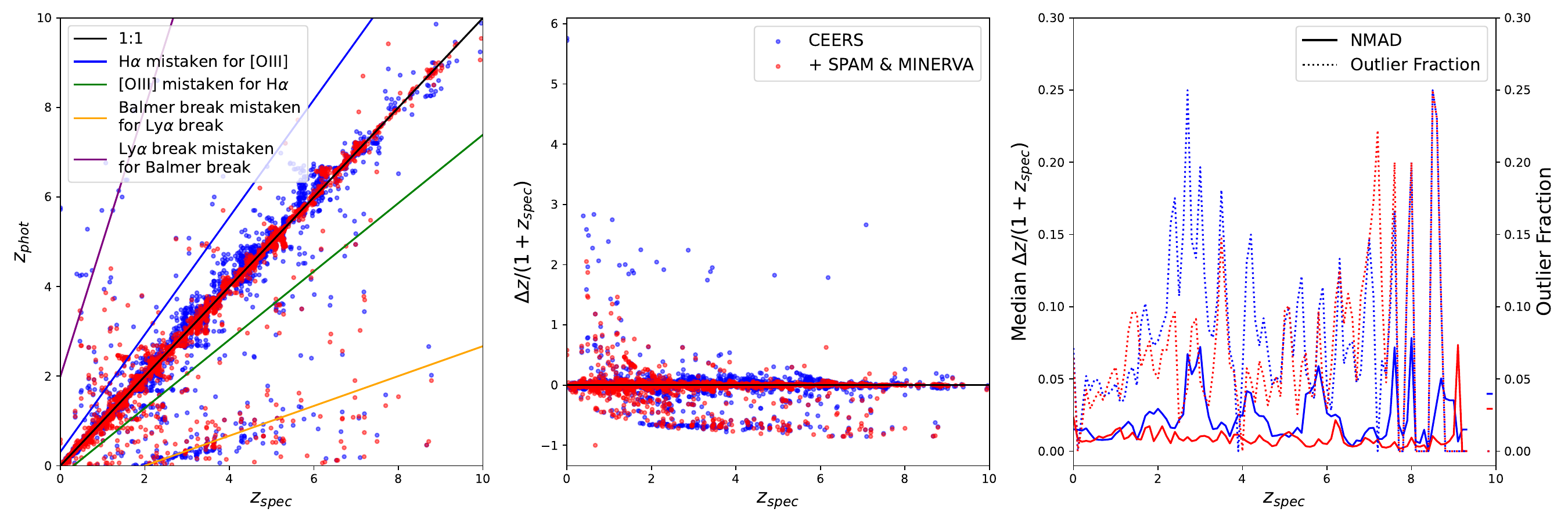}
    \caption{\textbf{Left:} Confident spectroscopic redshifts for $\sim4500$ targets in EGS compared to photometric redshifts before (blue) and after (red) inclusion of the SPAM and MINERVA MBs. Lines in each plot trace potential sources of confusion in SED fitting. The blue line traces \ha\ mistaken for \oiii\ and the green line \oiii\ mistaken for \ha. The orange line traces the Balmer break mistaken for the \lya\ break, and the purple line traces the \lya\ break mistaken for the Balmer break. \textbf{Middle: } $\Delta z$ / ($1 + z$)  vs $z_{spec}$ for the sources in both the CEERS catalog (blue) and the SPAM catalog (red). \textbf{Right:} NMAD (solid) and outlier fraction (dotted) as a function of redshift in $\Delta z$ bins of 0.1, before (blue) and after (red) adding the MBs. At almost all redshifts, adding the MBs to the photometric redshift fitting improves accuracy.}
    \label{fig:zcomp}
\end{figure*}

Each input catalog includes a grade for the reliability of its redshift measurement, and we use those grades to determine a vetting value for the SPAM catalog. We vet spectroscopic redshifts by assigning a value of 2 if the $z_{spec}$ is believed, 1 if it is likely but uncertain, 0 if it is not believed, and $-1$ if it is not vetted. For example, those sources in the DJA that have a grade of 3 (flagged by DJA as robust) are assigned a $z_{spec}$ vetted value of 2, but those that have a DJA grade of 0 (flagged by DJA as a data quality issue) or 1 (flagged by DJA as having no features) are assigned a vetted value of 0. We have confident (grade $= 2$) spectroscopic redshifts for 4,557 sources in the SPAM catalog, and these are what we use to test our photometric redshifts (Figure \ref{fig:zcomp}). Our grades are in the \texttt{SPAM\_Specz\_Catalog} file under the \texttt{ZSPEC\_VETTED} column.

%The distribution of sources in the SPAM catalog as a function of spectroscopic redshift can be found in red in Figure \ref{fig:distributions}. The SPAM catalog includes good (\texttt{zspec\_vetted} =2) redshifts for 4,557 sources in EGS. The SPAM catalog includes additional public NIRCam pointings in the EGS field, covering more depth than the CEERS-only catalog and incorporating more sources overall.

\subsection{Improvements to Photo-z's with Medium Bands}

%We vet photometric redshifts by comparing them with available spectroscopic redshifts in the EGS field. 

We compare photometric redshifts to our catalog of confident (\texttt{zspec\_vetted = 2}) spectroscopic redshifts in Figure \ref{fig:zcomp}. The left panel of Figure \ref{fig:zcomp} compares the photometric redshifts before (blue) and after (red) the addition of SPAM and MINERVA MBs. We include several lines that trace possible sources of degeneracy in SED template fitting. The figure shows that adding medium bands improves photometric redshift accuracy across a broad redshift range.
The blue line in Figure \ref{fig:zcomp} traces \ha\ mistaken for \oiii, and the green line traces \oiii\ mistaken for \ha. We find \ha\ mistaken for \oiii\ to be a rare source of confusion. However, several sources, especially at $4<z<7$, have \oiii\ mistaken for \ha\ (along the green line). This is often tied to AGN, especially LRDs, where \ha\ line strengths are much greater than \oiii, as \citet{Davis2026} noted.

The yellow line in Figure \ref{fig:zcomp} traces cases where the Balmer break has been mistaken for the \lya\ break, and the purple line traces cases where the Lyman break has been mistaken for the Balmer break. We find that the Balmer break is mistaken for the \lya\ break in rare cases, and that several interlopers at $z<2$ are mistaken for higher-redshift sources due to this confusion. The opposite case, where the \lya\ break is mistaken for the Balmer break, is our most significant source of confusion for SED fitting before adding the medium bands, as the largest outlier fraction occurs at these redshifts. 

The middle panel of Figure \ref{fig:zcomp} shows the $\Delta z$ / ($1+z_{spec}$) for both the CEERS (blue) and SPAM (red) catalogs. Here $\Delta z$ = $z_{phot} - z_{spec}$. This highlights the improvements in photometric redshift fitting after including the MBs (red). 
To understand this more statistically, we calculate
the NMAD for our standard deviation calculations, as it is more outlier-resistant:
$$ \mathrm{NMAD = \frac{median(|data-median(data)|)}{X}}$$

Here, the normalization value is $X = 0.5 \times \sqrt{2}$ or 0.67449, which makes NMAD equivalent to a standard deviation for a Gaussian distribution. This is shown as the solid line in the right panel of Figure \ref{fig:zcomp} for our photometric redshift fits both with (blue) and without (red) the addition of the MBs. 

We also calculate the outlier fraction, defined as the fraction of points where $|\Delta z| > 0.2 \times (1 + z_{spec})$, often considered catastrophic failures of our photometric redshift fitting. These fractions are shown as dotted lines in the right panel of Figure \ref{fig:zcomp} both with (red) and without (blue) the new MBs. The outlier fractions are much higher around $z=3$ before the addition of the MBs (blue), which are often the points along the orange line in the left panel of Figure \ref{fig:zcomp} where the Balmer break has been mistaken for the \lya\ break, or along the purple line where the \lya\ break has been confused for the Balmer break. Outlier fractions increase at higher redshift due to low-number statistics.

%This issue is twofold: Extreme Emission Line Galaxies (EELGs; \citealt{vanderWel2011}) suffer from photometrically blended lines when an MB does not probe the continuum between the lines, and other galaxies exhibit atypically strong Balmer breaks that SED templates do not model well. EELGs at $4<z<5$ do not have photometric coverage to de-blend their \ha\ and \hb + \oiii\ emission lines, leading to a ``bump'' in the SED. This is compounded by fitting codes that model a $2\mu m$ stellar bump at a redshift that coincides with confusion between a low-z Balmer break and a high-z \lya\ break. This represents a systemic source of confusion for EELGs with photometric coverage that lacks medium bands \citep{Davis2024, Davis2026}. 

% A substantial population of sources has Balmer and Lyman breaks confused at $2<z<4$, where LRDs drop off in number density. This confusion has been noted in LRD photometric redshifts \citep{Taylor2024, Davis2026} and may point to an understudied population. A complete investigation of the impact of this confusion on LRD populations will be the subject of future work. 

\subsection{Improved Selection for Distant Galaxies}

SPAM adds the bluest NIRCam BB filter to the CEERS legacy dataset: F070W. This filter probes \lya\ break dropouts from $5<z<12$ as it lies between the \hst\ F606W and F814W filters. SPAM achieves fainter depths than CANDELS \hst\ F814W imaging and comparable depths to the F606W. This further constrains high-z candidates by increasing the confidence in \lya\ break dropout galaxies. This coverage, in combination with the addition of the MBs, gives the most complete look to date at the distant ($z>5$) Universe in CEERS.

\begin{figure}
    \centering
    \includegraphics[width=1\linewidth]{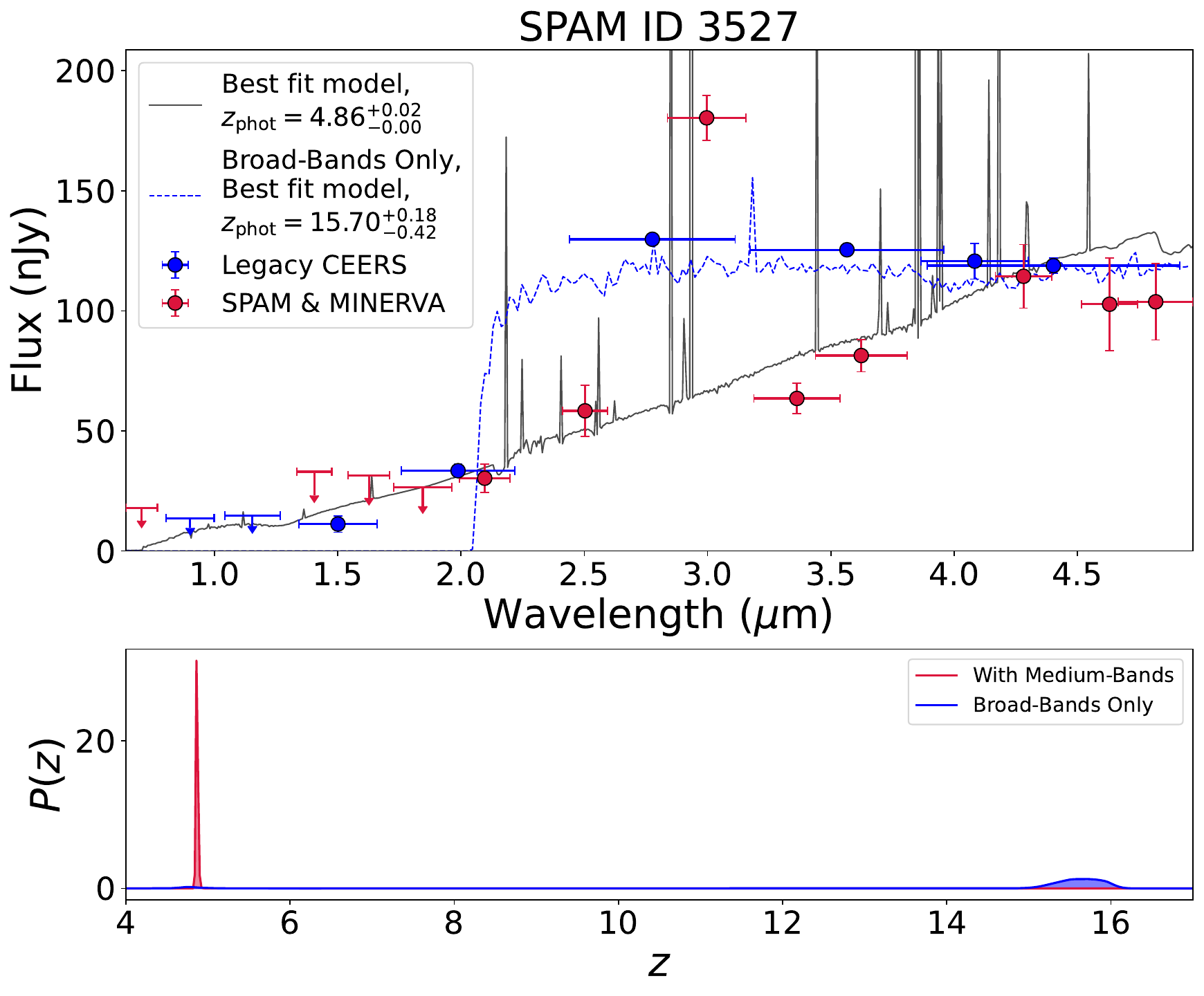}
    \caption{\textbf{Left:} An example of successful emission-line deblending using the SPAM + MINERVA MBs. This source was originally expected to be at $ z\sim 16$ based on the BB-only photometry (blue) from \citet{Donnan2023}. The additional wavelength resolution from the SPAM + MINERVA MBs (red) correctly disentangles the strong emission-line features from the continuum, yielding a photometric redshift of $z=4.86$ (black) that matches the spectroscopic redshift of $z=4.9$ from \citet{Arrabal2023b}. \textbf{Right:} $\rm{P}(z)$ for the BB-only (blue) and SPAM + MINERVA photometry (red) showing the SPAM filters recover the correct $z$ solution.
    }
    \label{fig:highz}
\end{figure}

Early CEERS work included searching for the furthest galaxies \citep{Finkelstein2022a}, with notable early candidates at $z\sim12$ \citep{Finkelstein2022c} and $z\sim16$ \citep{Donnan2023}. However, follow-up spectroscopy confirmed the former at $z=11.4$, but revealed that the $z\sim16$ candidate was a $z\sim4.9$ reddened extreme emission-line galaxy \citep{Arrabal2023b}. We present the SEDs of the candidate $z=16$ source in Figure \ref{fig:highz} using only the BBs (blue) and after adding the SPAM + MINERVA coverage (red), where we recover the correct photometric redshift at $z=4.86$ (black). The WB only fits presented here are fit through a methodology identical to \ref{sec:photoz} with the MBs removed. Using only BBs produced an SED for this source with photometrically blended lines that mimicked those of a high-z source with a \lya\ break at $z\sim16$ (blue). However, the MBs localize the photometric impact of the strong emission lines at certain wavelengths, enabling identification of the continuum between the strong emission-line features and accurate photometric redshift measurement (black).

\section{Science Cases} \label{sec:sciencecases}

Rare sources are typically identified via photometric surveys and then followed up via expensive spectroscopic observations. Compared to these expensive targeted observations, SPAM offers an efficient way to enable science over a large sample of sources in only a few (10) NIRCam pointings with $\sim2-3$ orders of magnitude more sources than available from all NIRSpec surveys in CEERS combined. SPAM is also a non-proprietary program. This ensures the community can use the added SPAM filters, enabling a wide range of discoveries when combined with the already prolific CEERS dataset and the similarly non-proprietary MINERVA program. Here, we demonstrate how adding MBs improves the three key science cases of the SPAM Survey: detecting overdensities in the CEERS field, identifying emission-line galaxies, and constraining star formation histories.

\subsection{Quantifying Overdensities Powering Reionization}

The history of cosmic reionization is poorly constrained, both in which objects drove it and how fast the process occurred. Current theory favors ``late reionization" \citep[e.g.,][]{Robertson2015}, where UV emission from galaxies accounts for the ionizing photons needed to match the constraints. This model requires an ionizing photon escape fraction of 20\% for galaxies across all luminosities; if this holds, then the brightest galaxies ($>$0.01L$^{\star}$) naturally dominate reionization. Other simulations predict that the ionizing photon escape fraction depends on galaxy halo mass, with a higher escape fraction from galaxies in the lowest-mass halos (log M$_h$/M$_{\odot}<$9). In this scenario, the faintest galaxies ($<$0.01L$^{\star}$) instead dominate reionization \citep{Rosdahl2022}, and the high abundance of faint galaxies at very early times \citep{Mason2015, Finkelstein2022c} drives an earlier start to reionization and a smoother time evolution \citep{Finkelstein2019}. In either scenario, over-dense regions of the universe ionize first due to the cumulative contribution of galaxies' ionizing radiation output \citep{Robertson2015}. If large, early reionized patches can be robustly identified and characterized at $z > 8$, they would support an earlier reionization start. 

\begin{figure}[h!]
    \centering
    \includegraphics[width=0.99\linewidth]{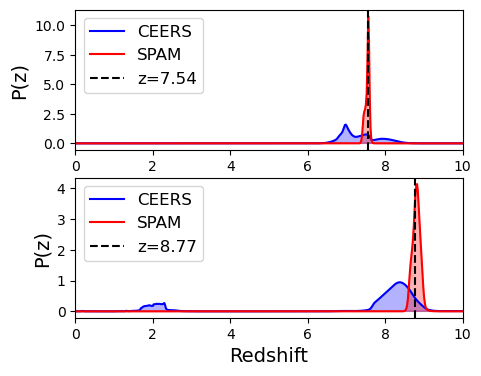}
    \caption{%\textbf{Left:} Distribution of sources in the catalog as a function of photometric redshift vs F200W magnitude. Robust overdensity candidates are marked in pink at $z=7.8$ and purple at $z=8.7$. Spectroscopically confirmed sources at these redshifts are marked as stars. The redshift range where \hb\ + \oiii\ falls into the SPAM MB filters, producing an increased flux over the CEERS BB filter, is shaded in the respective colors. \textbf{Right:} 
    The redshift probability distribution, P(z), for a representative source at each overdensity redshift, both before (blue) and after (red) the addition of the medium bands. Spectroscopic redshifts of these two sources are drawn as vertical lines. These better-constrained photometric redshifts allow more robust overdensity member identification and a full mapping of these structures' extent.}
    \label{fig:Overdensity}
   % \vspace{-22pt}
\end{figure}

While the confirmed galaxies in these epochs are bright, they are unlikely to dominate the local ionizing budget alone. This is not surprising, however, as overdensities should be traced not only by these bright galaxies but also by an overabundance of faint galaxies.  Detecting an abundance in the fainter galaxy population would not only confirm these overdensities, but also provide high-value targets for \jwst\ spectroscopy to probe faint \lya\ from these systems and map the full extent of these ionized regions.

SPAM enables an efficient method to detect and quantify these additional faint sources, as galaxies at this redshift are known to exhibit very strong \hb\ + \oiii\ emission (combined rest-frame EW $\geq$ 800\AA; \citep{Davis2024, Llerena2024, Tang2023}).
SPAM covers the \hb\ + \oiii\ emission line complex at $z=7.3-8.0$ and $z=8.3-8.9$, respectively, through the F430M and F480M filters.
This encompasses two known overdensities at $z=7.8$ \citep{Tang2023} and $z=8.7$ \citep{larson22, Whitler2024, Whitler2026}. SPAM selects members by detecting excess emission in the MB imaging relative to the CEERS BB image, as the emission lines pass through the MB filters. This tightens constraints on the redshift probability, $P(z)$, as shown in Figure \ref{fig:Overdensity}, where the $P(z)$ before (blue) is much broader than after adding the medium bands (red). This lets us identify robust overdensity candidates, as those with tightly constrained $P(z)$ distributions around the overdensity redshifts. Preliminary tests on the SPAM catalog identify 25 candidate galaxies at $z=7.8$  and 13 candidate galaxies at $z=8.7$. More detailed analysis on this can be found in Larson et al. (in prep).

\begin{figure*}[ht!]
    \centering
    \includegraphics[width=1\linewidth]{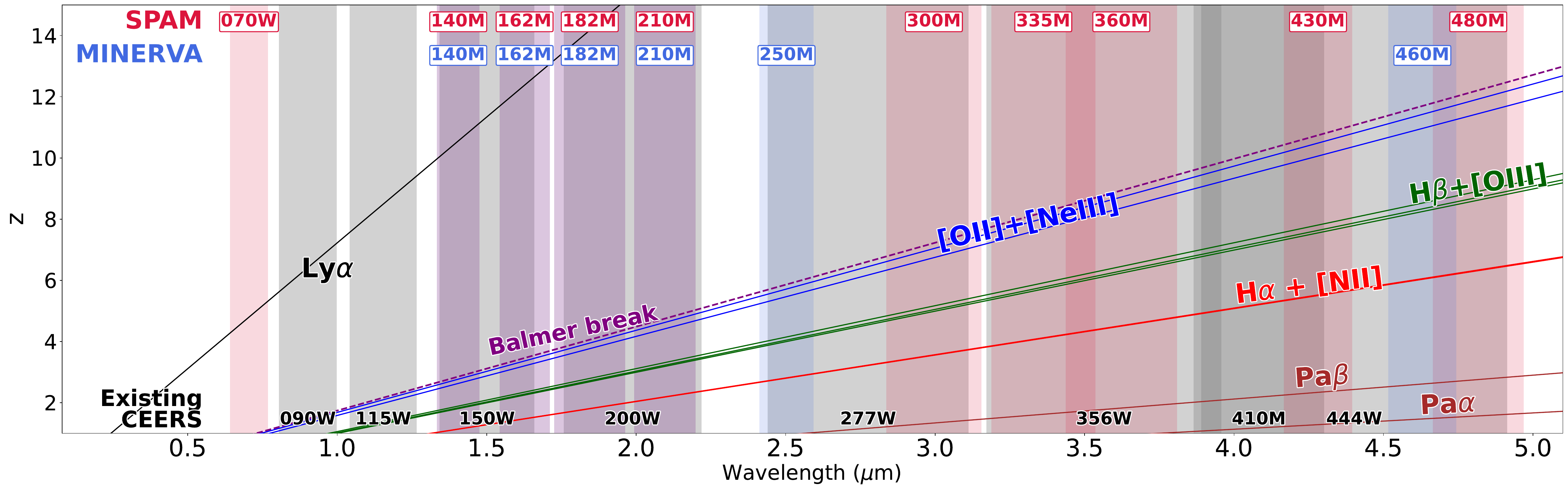}
    \caption{The new and legacy photometric coverage in the CEERS field. Photometric coverage prior to JWST Cycle 4 is shown in gray. New SPAM and MINERVA filters are indicated in red and blue, respectively. Solid lines trace the location of emission lines at $1<z<15$ recoverable from the photometry. A command-line interactive version of this plot is available  \href{https://github.com/jannemit/redpy}{here}. }
    \label{fig:emissionline}
\end{figure*}

\subsection{Emission-Line Galaxies at $2 < z < 9$} \label{sec:ELGs}

\begin{figure}[hb!]
    \centering
    \includegraphics[width=1\linewidth]{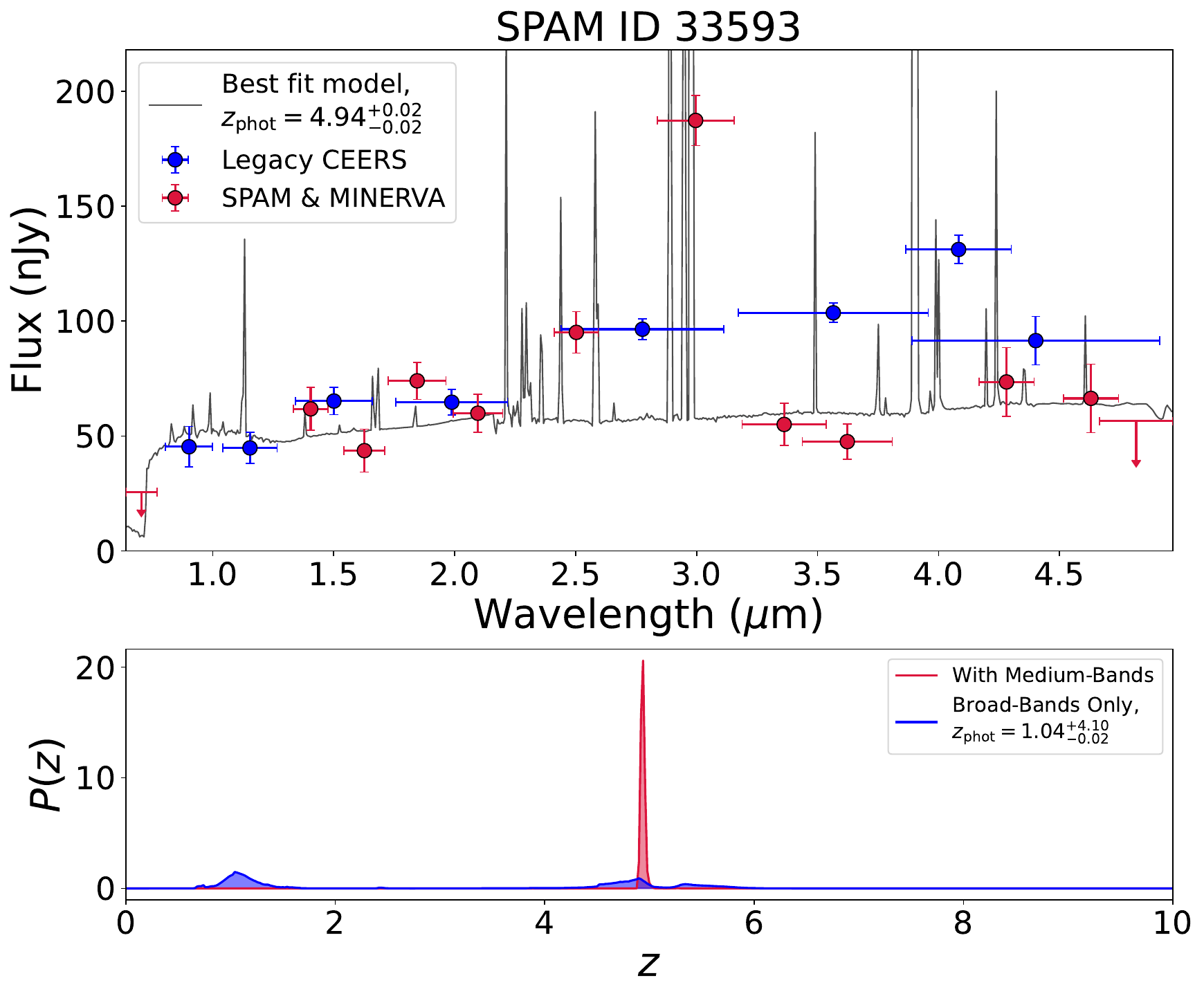}
    \includegraphics[width=1\linewidth]{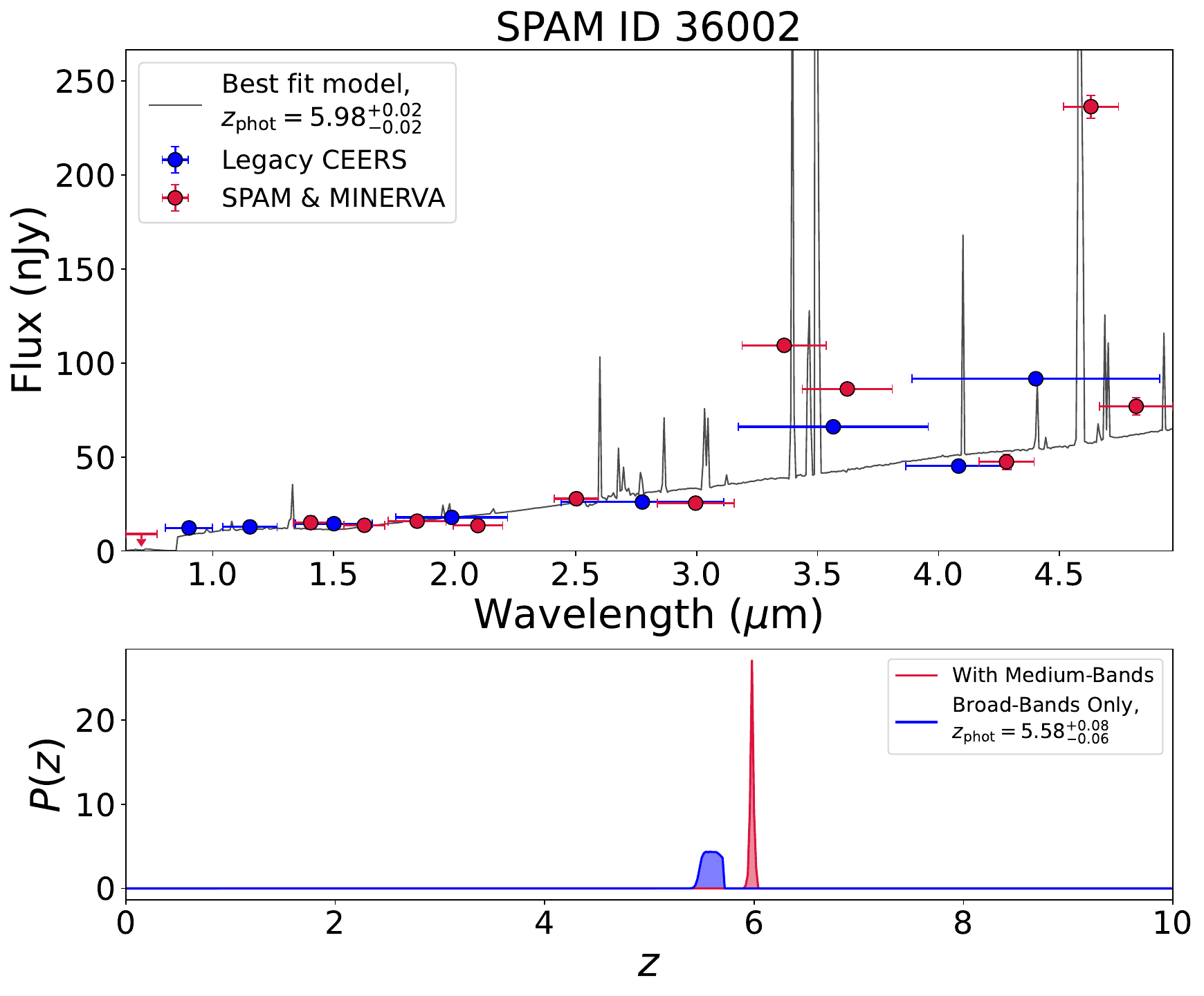}
    \caption{Two examples of sources with SEDs improved by the SPAM + MINERVA data set with the SEDs plotted alongside their best-fit models (top) and $\rm{P}(z)$ distributions for the SPAM + MINERVA and BB-only fits (bottom). Blue points indicate legacy CEERS photometry, red points indicate SPAM + MINERVA bands. The best-fit template is plotted in black. \textbf{Top:} An EELG from \citet{Davis2024} which has not been targeted by follow-up spectroscopy. The \ha\ emission line is in the F410M filter from the legacy CEERS coverage, but it contaminates both nearby BB filters while \oiii\ boosts the F277W filter. This creates a degenerate SED that the SED fitting code mistakes for a low-z galaxy, while the SPAM MBs successfully deblend both strong emission lines from the photometry. \textbf{Bottom:} An LRD photometrically identified in \citet{Kocevski2024}, confirmed with broad \ha\ and $\rm{z_{spec}} = 5.95$ by the Skyfire survey \citep{Kocevski2026}. Both the \ha\ and \oiii\ emission are successfully deblended from the reddened rest-optical continuum. }
    \label{fig:sed_ex}
\end{figure}

NIRSpec detections have confirmed strong \hb\ + \oiii\ emission in $4<z<9$ galaxies \citep{Tang2023, Harikane2024, Jung2024, Llerena2025, Davis2026}, first identified in NIRCam photometry. In some cases, equivalent widths (EWs) $>2000$ \AA\ have been reported \citep{Davis2024, Endsley2024, Llerena2024}, dwarfing low-z emission line strengths. These high EWs indicate high ionization parameters and consequently high ionizing production rates ($\xi_{ion}$, \citep{Tang2023, Fujimoto2023, Whitler2024, Rinaldi2023, Llerena2025}). The strength of these lines, combined with the sensitivity of \jwst's NIRCam, has enabled studies of \ha\ at $4<z<7$ and \hb\ + \oiii\ at $4<z<9$ from broadband photometry alone \citep{Davis2024, Llerena2024, Withers2023}.

However, the confidence with which these emission lines can be disentangled from the broadband photometry depends directly on the availability of MB photometric filters \citep{Davis2024, Guo2024, Wold2024}. Emission lines are most reliably detected when they are captured by a MB filter, while \ha\ and \hb\ + \oiii\ are most confidently de-blended from the photometry when a medium-band filter probes the continuum between the lines \citep{Davis2024}. However, the legacy CEERS observing design included only one MB filter, F410M. %Further, sparse MB coverage can introduce degeneracies where LRD-like continuum shapes can be mistaken for strong rest-optical emission lines \citep{Davis2026}. 

Consequently, high-confidence photometric \ha\ and \hb\ + \oiii\ studies in CEERS at $z>4$ are limited to a narrow redshift range. Figure \ref{fig:emissionline} shows how we can now deblend these emission lines, with legacy CEERS coverage in gray, SPAM in red, and MINERVA in blue. Solid lines indicate which emission lines the new filters probe at $1<z<15$. SPAM provides coverage of F300M, F335M, and F360M, which, in combination with existing CEERS F410M coverage, samples the continuum between \ha\ and \hb\ + \oiii\ at $4<z<6.7$ with no gap in coverage. The 430M and 480M filters further break up the 444W band to isonalte \oiii\ emission at $6.7<z<9$, which enables emission-line deblending and improved $\rm{z_{phot}}$ measurements at the highest redshift available from NIRCam photometry alone.  Anticipating that probing the continuum between emission lines would also improve detection of emission-line galaxies at lower redshifts, SPAM includes two bluer filters, F210M to probe the wavelength range between the F200W and F277W broadband filters and deblend \ha\ and \hb\ + \oiii\ at $2<z<3.5$. Similarly, SPAM provides F182M to probe the region between the F150W and F200W filters, deblending these emission lines at $1.5<z<2$. We also include the 140M and 162M filters to similarly break apart the wavelengths covered by the F150W filter. 
%At lower redshifts, $4<z<5$, these emission lines are difficult to disentangle from broadband photometry. 
%This excludes a key epoch in cosmic time from photometric studies and biases any derived population. 
%Spectroscopic observations are inefficient for population studies of these emission lines because $\sim 2-3$ orders of magnitude fewer sources are observed spectroscopically than photometrically and the continuum depths probed by both high- and medium-resolution NIRSpec spectroscopy are often shallower. MB photometry enables emission line strength measurements from photometry directly.

%The addition of the dually-observed bands from SPAM + MINERVA: F140M, F162M, F182M, and F210M probes fainter continuum emission than is accessible by any available spectroscopy from large surveys in EGS at $3<z<9$.

% We show all new MBs (and one wide band) added to the CEERS legacy photometry in Figure \ref{fig:emissionline}. Legacy CEERS coverage is indicated by gray shaded regions, new SPAM coverage is in red and new MINERVA coverage is in blue. Figure lines indicate which emission lines are probed by the new filters at $1<z<15$.

\begin{figure*}[ht!]
    \centering
   % \vspace{-8pt}
    \includegraphics[width=0.9\textwidth]{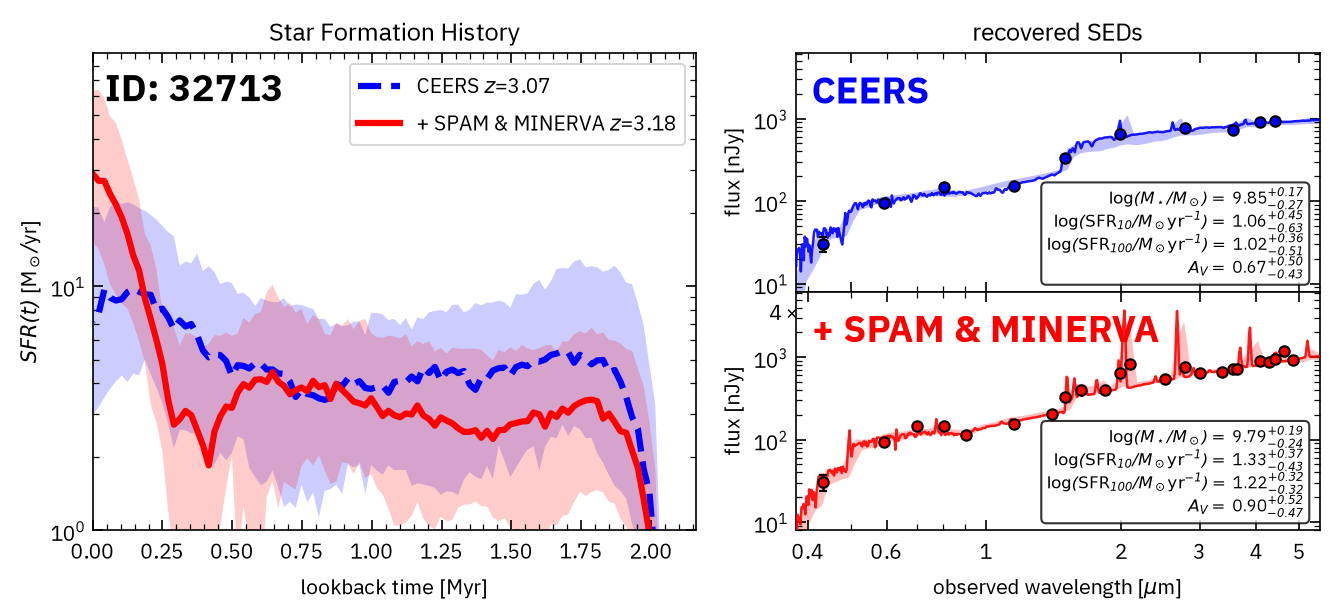}
    %\vspace{-10pt}
    \caption{Example star-formation history (SFH) recovery of a galaxy with CEERS BB photometry alone (blue) and with the addition of the SPAM+MINERVA MB observations (red). The MB information significantly alters the measured SFH by providing exceptional coverage of the Balmer break and identifying strong emission lines. 
    %identifies younger stellar populations.
    Additional medium-band coverage further breaks the rest-frame optical degeneracy between strong nebular emission lines and stellar mass, revealing more accurate SFHs.}
  %  \vspace{-22pt}
    \label{fig:SFH}
\end{figure*}

%Figure \ref{fig:sed_ex} shows two examples of galaxies in CEERS with SEDs improved by adding MBs. 
The two sources in Figure \ref{fig:sed_ex} show the emission-line isolation now possible photometrically with SPAM + MINERVA, enabling new emission-line science across expanded populations derived photometrically.
The source in the top panel, SPAM 33593, is an EELG from \citep{Davis2023} whose \ha\ and \hb\ + \oiii\ lines were previously blended in the BB filters. In the legacy CEERs photometry (blue), the source was placed at $\rm{z}_{phot} = 1.04^{+4.10}_{-0.02}$, with wide error reflecting the degenerate nature of the SED. The \ha\ line is captured by the single MB from the legacy CEERS coverage, F410M, but falls at a redshift where the line boosts the neighboring BB filters (F356W and F444W) while the \oiii\ emission boosts F277W. This creates a false continuum shape and the SED fitting finds a $\rm{z}\sim4$ EELG and a $\rm{z}\sim1$ galaxy with a stellar bump boosting the BBs degenerate.
The improved MB coverage (red) successfully isolates the continuum emission between the two lines, and our new SED places the galaxy at $\rm{z}_{phot} = 4.94^{+.02}_{-0.02}$, increasing the accuracy of the photometric redshift through isolation of the \ha\ and \hb\ + \oiii\ emission lines without a need for follow-up spectroscopy. 

The bottom panel of Figure \ref{fig:sed_ex} shows a Little Red Dot (LRD)  identified in \citet{Kocevski2024}. The WB-only fit places the LRD at  $\rm{z}_{phot} = 5.58^{+0.08}_{-0.06}$, and our fit from the SPAM+MINERVA coverage places the source at $\rm{z}_{phot} = 5.98^{+0.02}_{-0.02}$, in agreement with $\rm{z_{spec}} = 5.95$ from the Skyfire survey \citep{Kocevski2026}. While the z recovery from the BB only fit performs well, the strength of the \ha\ and \oiii\ emission lines are difficult to recover, blended with the  reddened rest-optical continuum. Without the emission-line isolation enabled by MBs, the rest-optical continuum is under-constrained. The complete MB photometry enables recovery of the strengths of these lines and continuum in a system with sub-extreme emission lines, demonstrating the expanded line recovery enabled by MBs.
% The $\rm{P}(z)$ is highly degenerate in the WB only fit. The code indicates two scenarios are likely and favors the slightly lower-z solution ($\sim7.20$) with weak \oiii\ falling in F410M and the rest-optical SED dominated by reddened continuum emission. The alternative scenario places the LRD at a much higher z ($\sim9$) where the rest-optical continuum is highly reddened and no strong emission lines shape the SED. In fact, neither of these solutions were correct and the F430M SPAM filter isolates a strong \hb\ + \oiii\ emission line, resolving both the emission line and the continuum and recovering the $\rm{z}_{phot} = 7.68^{+0.04}_{-0.04}$ solution. 

 \subsection{SFHs of Massive Galaxies at $z > 4$} \label{sec:SFHs}

% \begin{figure*}[ht]
%     \centering
%     \includegraphics[width=0.95\linewidth]{Template.png}
%     \caption{Model spectrum of a high-redshift galaxy made using BPASS binary models \citep{Eldridge2017} and the Cloudy software \citep{Ferland2017} redshifted and normalized to $z\sim7.7$ (maroon), and an observed spectrum from NIRSpec $z\sim8.7$ (purple, \citep{Larson2023}). Bright \hb\ + \oiii\ emission falls into our proposed F430M \& F480M filters at high redshifts and in the F300M \& F335M with \ha\ in the F360M \& F430M at $z\sim4.5$ \& $z\sim5.5$ (blue \& green).}
%     \label{fig:filters}
%    % \vspace{-6mm}
% \end{figure*}

Early \jwst\ results have revealed an apparent overabundance of high-redshift, massive galaxies at $z>4$ \citep{Valentino23, Weibel2024, Akins2025, Chworowsky2024, Gentile2024},
challenging galaxy formation models to create adequate amounts of stellar mass in a short enough timescale to reproduce the observed sources \citep{Xiao2023, Lagos24, Baker25}, motivating theories requiring highly efficient star formation in the early Universe \citep{Dekel2023, Li2024, Mason2023}. Furthermore, observations have confirmed galaxies that have ceased their star formation entirely, hosting an older-than-expected stellar population \citep{Carnall2022, Glazebrook2023, deGraaff2025Nature, Edward26, Hamadouche26} at early times. Together, this implies a Universe in which galaxies cannot only form extreme levels of stellar mass in a short time but also rapidly quench star formation on a galactic scale. To properly constrain galaxy formation histories in the early Universe, medium bands are necessary to isolate line emission from the underlying continuum flux. SPAM provides coverage of F300M, F335M, and F360M to deblend \ha\ and \hb\ + \oiii\ from the photometry at $4<z<5.5$ and F140M, F162M, F182M, and F210M to probe the UV continuum at $2<z<12$.

Both \ha\ and UV fluxes not only constrain photometric redshift-derived distances, but also directly probe star formation histories (SFHs). \ha\ emission from galaxies requires massive, short-lived O- and B-type stars, directly probing SF on recent ($\sim 10$ Myr) timescales \citep{Kennicutt2012}. The medium bands isolate this line emission from the broadband continuum and enable 2D emission-line studies, building SEDs from pixels as in \citet{Perez-Gonzlez2023}, and directly constraining SF from \ha. 

UV continuum emission traces sustained SF populations over $\sim 100$ Myr timescales. %Comparing the UV implied SFRs with \Ha\ SFRs directly probes the ``burstiness'' of star formation in the early Universe traced by scatter. 
Studies of SFHs in \jwst\ have begun to uncover evidence for burstiness in early stellar populations \citep{Sun2023, Cole2025}, and MB coverage from SPAM + MINERVA further quantifies population-level statistics for bursty star formation by isolating \ha\ and UV fluxes.

%The addition of SPAM MBs tackles two additional science goals: (i) measuring the burstiness of star-forming galaxies in the time leading up to cosmic noon and (ii) quantifying the quenching behavior of massive, high-redshift galaxies. Both of these questions carry implications for our current understanding of galaxy evolution and will help answer questions including (i) do galaxies  high redshift (and lower stellar masses) harbor younger stellar populations and does this affect their inferred properties? (ii) when do the first galaxies quench? and (iii) what is the quenched fraction of galaxies at high redshift, and how does that correlate with AGN or other modes of feedback thought to be responsible for quenching? 

In the more evolved Universe, the four blue MB filters cover the Balmer break region at $3<z<5$, one of the most information-rich regions of the earliest-known massive, quiescent galaxies. The shape of the Balmer break can be effectively determined using multiple MB filters \citep{AntwiDanso25}, adding necessary constraints on the age and dust content of these galaxies and breaking the degeneracy between Balmer breaks and dust reddening. The SPAM F070W photometric filter extends this coverage to the Balmer break at $0.5<z<1.25$, which, combined with existing F090W coverage, completes Balmer-break detection over cosmic history in CEERS. 

In Figure \ref{fig:SFH}, we show an example star-formation history (SFH) of a galaxy using only CEERS BB photometry (blue) versus that recovered with the addition of the SPAM+MINERVA MBs (red). The MB information significantly improves the galaxy's SFH by capturing emission lines and capturing the exact location (improving redshifts) and strength of the Balmer break, leading to much stronger constraints on galaxy stellar populations.

\section{Summary}\label{sec:conclusions}

We present a photometric catalog from the SPAM and MINERVA surveys which, combined, leave the CEERS legacy footprint with coverage from all BB and MB NIRCam photometric filters. The only other fields with such complete filter coverage are CANUCS and JADES, with coverage over  $\sim30 \,\rm{arcmin^2}$ and $\sim10 \,\rm{arcmin^2}$, respectively, compared to  $\sim 100 \,\rm{arcmin^2}$ presented here, a greater area than all surveys predating Cycle 5 combined. This photometric catalog includes both the legacy CEERS coverage and the SPAM + MINERVA coverage in a uniform reduction, along with photometric redshifts. The catalog is publicly available and will be maintained in the most up-to-date version on the SPAM website\footnote{\href{https://kelceydavis33.github.io/SPAM/}{https://kelceydavis33.github.io/SPAM/}}.

% SPAM adds 9 MBs and 1 BB and, combined with the additional 2 MBs from the MINERVA program, brings the photometric coverage in the CEERS legacy footprint from 8 photometric filters to 20, now including all MB and BB photometric filters on NIRCam. 

The SPAM coverage improves photometric redshifts across all redshifts, and we include photometric redshifts with the photometry catalog. We compare these photometric redshifts to $\sim4500$ spectroscopic redshifts available in the EGS field. The SPAM-inclusive photometric redshifts improve on the legacy CEERS BB coverage, especially in cases of chronic degeneracy tied to unresolved emission lines and confused SED breaks. 

This coverage not only makes the CEERS field an excellent test bed for photometric science, but also provides an equivalent to very-low-resolution spectroscopy that can recover the continuum shape and photometrically isolate \ha\ and \hb\ + \oiii. This will enable a wide range of science goals. Here we highlight three key improvements:

\begin{itemize}
    \item \textbf{Overdensities in CEERS: } Improved photometric redshifts enable identification of resolved overdensities in CEERS, partly due to photometric deblending of strong rest-optical emission lines. 
    \item \textbf{Emission Line Galaxy Identification:} Emission lines, especially strong rest-optical lines, can now be isolated photometrically for all sources in the CEERS legacy footprint. This improves identification of these strong lines and will help characterize many source types, from EELGs to LRDs, in future work.
    \item \textbf{Star Formation Histories:} Better Balmer-break fitting and emission-line detection with the additional MB data enable more well-constrained SFHs, as they break degeneracies between continuum and emission-line contributions. 
\end{itemize}

%% Please use the acknowledgment and contribution environments. This will 
%% be anonomyized when the "anonymous" style option is used. 
\begin{acknowledgments}

We thank the MINERVA team for maintaining open communication, which enabled this unprecedented dataset. 
The work is based on observations made with the NASA/ESA/CSA \jwst. The data were obtained from the Mikulski Archive for Space Telescopes at the Space Telescope Science Institute, which is operated by the Association of Universities for Research in Astronomy, Inc., under NASA contract NAS 5-03127 for \jwst. These observations are associated with \jwst\ Cycle 4 GO programs 8559 and 7814. Support for the program JWST-GO-8559 was provided by NASA through a grant from the Space Telescope Science Institute (STScI), which is operated by the Association of Universities for Research in Astronomy (AURA), Incorporated, under NASA contract NAS5-26555.
The material is based upon work supported by NASA under award number 80GSFC24M0006.

RLL and LYAY appreciate support from a Giacconi Fellowship at the Space Telescope Science Institute, which is operated by the Association of Universities for Research in Astronomy, Inc., under NASA contracts NAS 5-26555 and NAS5-03127. KD and RLL acknowledge support from NASA JWST-GO-8559. KD acknowledges support from an NSF Graduate Research Fellowship award number 2040433. KD and JRT acknowledge support from NASA JWST-GO-8410. KD acknowledges support from Los Alamos National Laboratory. This work was released under LA-UR-26-26360. LYAY acknowledges support from the Taiwanese Ministry of Education (MoE) Yushan Fellowship (MOE-115-YSFMS-0002-005-P1).

\end{acknowledgments}

\begin{contribution}
%%This section gives authors the space to recognize author contributions. The text inside this environment is NOT counted towards the total word quanta. At a minimum, manuscripts are expected to include this text:
R. Larson and K. Davis are the PIs of SPAM, planned the observations, wrote the manuscript, and contributed equally to this work. 
H.B.A. reduced all the NIRCam data.
S.L.F. produced the photometry catalog and ran the photometric redshift fitting.
K.C. ran Dense Basis on the catalog and provided the SFHs for the 3rd science case.
All other authors contributed to the SPAM collaboration and provided comments and scientific expertise.

%% But authors are expected to provide more specific details, e.g. 
%%
%%SC was responsible for writing and submitting the manuscript.
%%WWM came up with the initial research concept and edited the manuscript.
%%OTS obtained the funding and edited the manuscript.
%%EBF provided the formal analysis and validation. He also edited the manuscript.
%%GEH Supervised the undergraduates, wrote the software and administers the project github and Zenodo repositories.
%%
%% Authors can use the Contributor Role Taxonomy (CRediT) at
%% for ideas on how write a good statement tailored to their needs.

\end{contribution}

\facilities{JWST(NIRCam)}

\software{ \texttt{astropy} \citep{2013A&A...558A..33A,2018AJ....156..123A,2022ApJ...935..167A},  
          \texttt{Cloudy} \citep{2013RMxAA..49..137F}, 
          \texttt{Source Extractor} \citep{1996A&AS..117..393B}
          , \texttt{Matplotlib} \citep{Hunter2007}, \texttt{NumPy} \citep{numpy}, \texttt{Pandas} \citep{pandas1, pandas2}, \texttt{Scipy} \citep{scipy}, \texttt{SAOImage DS9} \citep{ds9,ds9_2026}, \texttt{IDL} \citep{idl_software}
          }


\appendix

\section{Improved Depths over Legacy CEERS}\label{sec:depths}

\begin{figure*}[h!]
    \centering
    \includegraphics[width=0.49\linewidth]{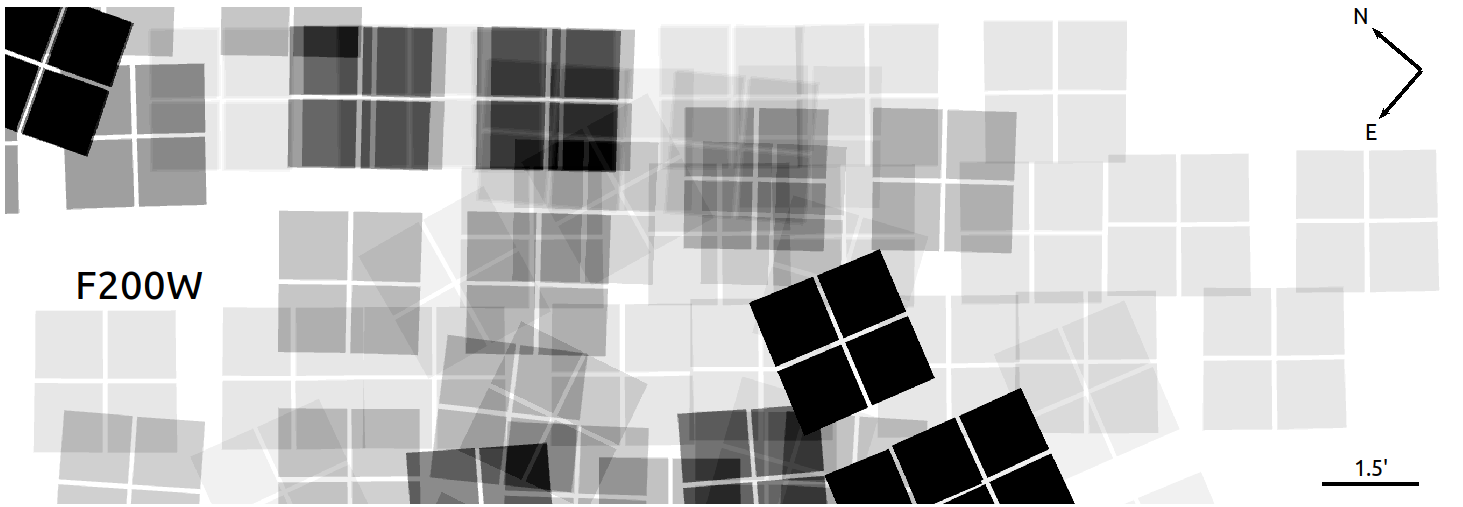}
    \includegraphics[width=0.49\linewidth]{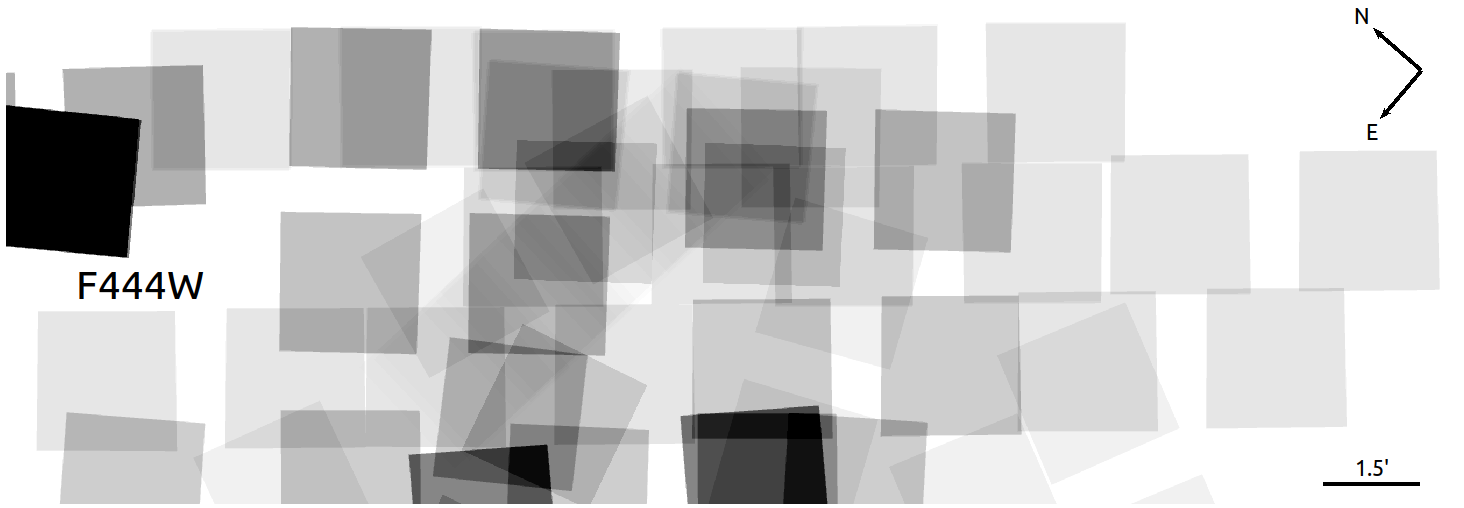}
    \caption{\textbf{Left: }F200W exposure map for all the public \jwst/NIRCam data included in the SPAM catalog. \textbf{Right: }Same, but for the F444W Filter. These highlight the extended footprint of the SPAM catalog, including coverage over the legacy CEERS field plus additional public data. }
    \label{fig:exposuremaps}
\end{figure*}

Our SPAM catalog includes data from all public \jwst/NIRCam imaging near the legacy CEERS footprint. We reduced all data using the same pipeline and process, then mosaicked them to create a CEERS tile that encompasses a $0.370 \times 0.123$ degree rectangle around the original CEERS NIRCam footprint (which is the same as the SPAM footprint). Figure \ref{fig:exposuremaps} shows this extended coverage, with the complete F200W coverage on the left and the F444W coverage on the right.

\begin{figure*}[h!]
    \centering
    \includegraphics[width=0.99\linewidth]{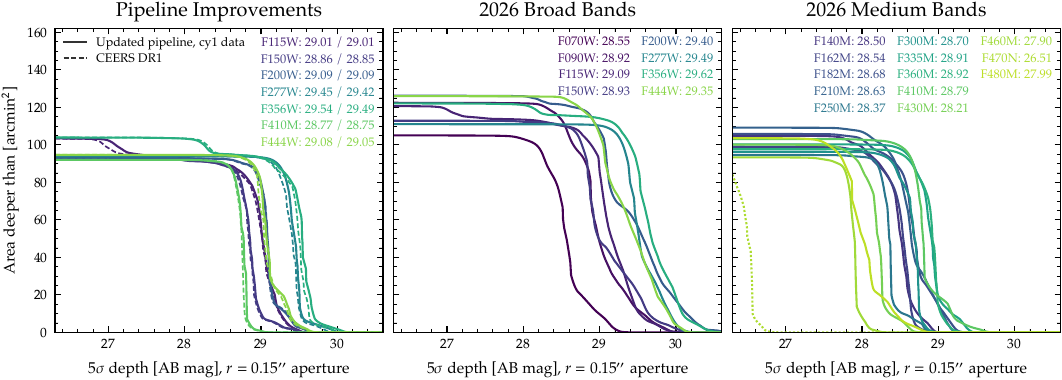}
    \caption{\textbf{Left:} Depth increase vs. area due to pipeline improvements since the CEERS DR1 (dotted; \citealt{Bagley2023}) and with the newest pipeline (solid). \textbf{Middle:} Depth vs. area for all of the broad-band data in the CEERS footprint (See Figure \ref{fig:exposuremaps}) that was reduced, mosaicked, and cataloged for this paper. Compared with the left plot, this shows the improved depth from additional imaging in each filter from ancillary data. \textbf{Right:} Depth vs. area for the new medium-band imaging in the CEERS footprint from SPAM, MINERVA, and any other ancillary data in the field as of publication. Annotations give the 5$\sigma$ depth at 50\% area. }
    \label{fig:improvements}
\end{figure*}

We calculate the 5$\sigma$ point-source depths reported in Figure \ref{fig:depths} and Table \ref{tbl:filters} using the measured noise function in the field. We use an aperture diameter of 0.2 arcsec and correct to total flux based on the flux enclosed for a given band's PSF. The improved depths over the legacy CEERS data we report are largely due to the increased coverage from additional imaging over the field, but pipeline improvements do play a small part (See Figure \ref{fig:improvements}, Left). These pipeline improvements include new flats, which now use all Cycle 1-3 data and most Cycle 4 data. New darks have been added since the original CEERS data release, and the readnoise and superbiases have also been improved. We have also tuned the 1/f correction using a Gaussian process model.  

Most of the depth improvements we report come from additional data over the field in each filter from other public NIRCam programs. Coverage is not uniform across the field in any filter, as shown in Figure \ref{fig:exposuremaps}. This means we calculate a median depth over the full footprint, but the minimum and maximum depths span a broad range. We show this by plotting the area vs. depth in Figure \ref{fig:improvements}. The center panel shows depth vs. area for all public BB data in the CEERS footprint (See Figure \ref{fig:exposuremaps}) available at the time of publication. Comparing the center panel to the left panel shows improved depths relative to the original CEERS DR1 \citep{Bagley2023} data due to the addition of ancillary data in the field. The right panel of Figure \ref{fig:improvements} shows depth vs. area for all MB and one narrow-band (NB) filter in the mosaics from both SPAM and MINERVA, as well as a few ancillary pointings in the CEERS footprint. Annotations indicated the 5$\sigma$ depths at 50\% area.

%% For this sample we use BibTeX plus aasjournalv7.bst to generate the

\bibliography{SPAM}{}

@INPROCEEDINGS{Perez-Gonzales2013,
       author = {{P{\'e}rez-Gonz{\'a}lez}, P.~G. and {Cava}, A.},
        title = "{SHARDS: Survey for High-z Absorption Red \& Dead Sources}",
    booktitle = {Revista Mexicana de Astronomia y Astrofisica Conference Series},
         year = 2013,
       series = {Revista Mexicana de Astronomia y Astrofisica Conference Series},
       volume = {42},
        month = may,
        pages = {55-57},
          doi = {10.48550/arXiv.1201.1800},
archivePrefix = {arXiv},
       eprint = {1201.1800},
 primaryClass = {astro-ph.CO},
       adsurl = {https://ui.adsabs.harvard.edu/abs/2013RMxAC..42...55P}
}

@ARTICLE{Whitler2026,
       author = {{Whitler}, Lily and {Stark}, Daniel P. and {Mason}, Charlotte A. and {Tang}, Mengtao and {Chen}, Zuyi and {Lu}, Ting-Yi and {Prieto-Lyon}, Gonzalo and {Hutter}, Anne},
        title = "{Deep JWST spectroscopy of galaxies in a candidate ionized bubble at z = 8.7: probing reionization at pMpc scales with Ly {\ensuremath{\alpha}} emission}",
      journal = {\mnras},
         year = 2026,
        month = may,
       volume = {548},
       number = {2},
          eid = {stag639},
        pages = {stag639},
          doi = {10.1093/mnras/stag639},
archivePrefix = {arXiv},
       eprint = {2510.12019},
 primaryClass = {astro-ph.GA},
       adsurl = {https://ui.adsabs.harvard.edu/abs/2026MNRAS.548ag639W}
}

@ARTICLE{bezanson24,
       author = {{Bezanson}, Rachel and {Labbe}, Ivo and {Whitaker}, Katherine E. and {Leja}, Joel and {Price}, Sedona H. and {Franx}, Marijn and {Brammer}, Gabriel and {Marchesini}, Danilo and {Zitrin}, Adi and {Wang}, Bingjie and {Weaver}, John R. and {Furtak}, Lukas J. and {Atek}, Hakim and {Coe}, Dan and {Cutler}, Sam E. and {Dayal}, Pratika and {van Dokkum}, Pieter and {Feldmann}, Robert and {F{\"o}rster Schreiber}, Natascha M. and {Fujimoto}, Seiji and {Geha}, Marla and {Glazebrook}, Karl and {de Graaff}, Anna and {Greene}, Jenny E. and {Juneau}, St{\'e}phanie and {Kassin}, Susan and {Kriek}, Mariska and {Khullar}, Gourav and {Maseda}, Michael and {Mowla}, Lamiya A. and {Muzzin}, Adam and {Nanayakkara}, Themiya and {Nelson}, Erica J. and {Oesch}, Pascal A. and {Pacifici}, Camilla and {Pan}, Richard and {Papovich}, Casey and {Setton}, David J. and {Shapley}, Alice E. and {Smit}, Renske and {Stefanon}, Mauro and {Taylor}, Edward N. and {Williams}, Christina C.},
        title = "{The JWST UNCOVER Treasury Survey: Ultradeep NIRSpec and NIRCam Observations before the Epoch of Reionization}",
      journal = {\apj},
         year = 2024,
        month = oct,
       volume = {974},
       number = {1},
          eid = {92},
        pages = {92},
          doi = {10.3847/1538-4357/ad66cf},
archivePrefix = {arXiv},
       eprint = {2212.04026},
 primaryClass = {astro-ph.GA},
       adsurl = {https://ui.adsabs.harvard.edu/abs/2024ApJ...974...92B}
}

@software{idl_software,
  author = {{NV5 Global, Inc.}},
  title = {{IDL: Interactive Data Language}},
  version = {8.9},
  year = 2026,
  publisher = {NV5 Geospatial},
  address = {Boulder, CO, USA},
  url = {https://www.nv5geospatialsoftware.com/},
  note = {IDL is a registered trademark of NV5 Global, Inc.}
}

@ARTICLE{ds9_2026,
       author = {{Fruscione}, Antonella and {Glotfelty}, Kenny and {Joye}, William and {McDowell}, Jonathan},
        title = "{SAOImageDS9: An Essential Tool for Astronomical Exploration}",
      journal = {\aj},
         year = 2026,
        month = aug,
       volume = {172},
       number = {2},
          eid = {116},
        pages = {116},
          doi = {10.3847/1538-3881/ae84c1},
archivePrefix = {arXiv},
       eprint = {2606.30897},
 primaryClass = {astro-ph.IM},
       adsurl = {https://ui.adsabs.harvard.edu/abs/2026AJ....172..116F}
}

@INPROCEEDINGS{ds9,
       author = {{Joye}, W.~A. and {Mandel}, E.},
        title = "{New Features of SAOImage DS9}",
    booktitle = {Astronomical Data Analysis Software and Systems XII},
         year = 2003,
       editor = {{Payne}, H.~E. and {Jedrzejewski}, R.~I. and {Hook}, R.~N.},
       series = {Astronomical Society of the Pacific Conference Series},
       volume = {295},
        month = jan,
        pages = {489},
       adsurl = {https://ui.adsabs.harvard.edu/abs/2003ASPC..295..489J}
}

@ARTICLE{Sersic63,
       author = {{S{\'e}rsic}, J.~L.},
        title = "{Influence of the atmospheric and instrumental dispersion on the brightness distribution in a galaxy}",
      journal = {Boletin de la Asociacion Argentina de Astronomia La Plata Argentina},
         year = 1963,
        month = feb,
       volume = {6},
        pages = {41-43},
       adsurl = {https://ui.adsabs.harvard.edu/abs/1963BAAA....6...41S}
}

@ARTICLE{AntwiDanso25,
       author = {{Antwi-Danso}, Jacqueline and {Muzzin}, Adam and {Robbins}, Luke and {Asada}, Yoshihisa and {Marchesini}, Danilo and {Sawicki}, Marcin and {Iyer}, Kartheik and {Whitaker}, Kate and {Speagle}, Joshua and {Papovich}, Casey and {Willott}, Chris and {Brada{\v{c}}}, Maru{\v{s}}a and {Desprez}, Guillaume and {Markov}, Vladan and {Martis}, Nicholas and {Noirot}, Ga{\"e}l and {Sarrouh}, Ghassan and {Kannan}, Rahul and {Abraham}, Roberto and {Fujimoto}, Seiji and {Myers}, Katherine},
        title = "{An Ancient Descendant of the First Galaxies}",
      journal = {arXiv e-prints},
         year = 2025,
        month = dec,
          eid = {arXiv:2512.03154},
        pages = {arXiv:2512.03154},
          doi = {10.48550/arXiv.2512.03154},
archivePrefix = {arXiv},
       eprint = {2512.03154},
 primaryClass = {astro-ph.GA},
       adsurl = {https://ui.adsabs.harvard.edu/abs/2025arXiv251203154A}
}

@ARTICLE{Hamadouche26,
       author = {{Hamadouche}, Massissilia L. and {Whitaker}, Katherine E. and {Valentino}, Francesco and {Antwi-Danso}, Jacqueline and {Ito}, Kei and {Beverage}, Aliza and {Zhu}, Pengpei and {Brammer}, Gabriel and {Kokorev}, Vasily and {de Lucia}, Gabriella and {Baker}, William M. and {Farcy}, Marion and {Gallazzi}, Anna and {Gillman}, Steven and {Gottumukkala}, Rashmi and {Hirschmann}, Michaela and {Jespersen}, Christian Kragh and {Kakimoto}, Takumi and {Lee}, Minju M. and {Onodera}, Masato and {Shimakawa}, Rhythm and {Tanaka}, Masayuki and {Weaver}, John R. and {Wu}, Po-Feng},
        title = "{DeepDive: Tracing the early quenching pathways of massive quiescent galaxies at $z>3$ from their star-formation histories and chemical abundances}",
      journal = {arXiv e-prints},
         year = 2026,
        month = feb,
          eid = {arXiv:2602.02485},
        pages = {arXiv:2602.02485},
          doi = {10.48550/arXiv.2602.02485},
archivePrefix = {arXiv},
       eprint = {2602.02485},
 primaryClass = {astro-ph.GA},
       adsurl = {https://ui.adsabs.harvard.edu/abs/2026arXiv260202485H}
}

@ARTICLE{Edward26,
       author = {{Edward}, Adit H. and {Antwi-Danso}, Jacqueline and {Muzzin}, Adam and {Forrest}, Ben and {McConachie}, Ian and {Beverage}, Aliza and {Chang}, Wenjun and {Cooper}, M.~C. and {Gomez}, Percy and {Hamadouche}, Massissilia L. and {Henry}, Aur{\'e}lien and {Lei}, Han and {Marchesini}, Danilo and {Noble}, Allison and {Urbano Stawinski}, Stephanie M. and {Wilson}, Gillian and {Wisz}, M.~E.},
        title = "{MAGAZ3NE: Spatially Resolved Ages and Chemical Abundances of Ultra-Massive Quiescent Galaxies at z $\sim$ 3.5 using JWST/NIRSpec IFU}",
      journal = {arXiv e-prints},
         year = 2026,
        month = may,
          eid = {arXiv:2605.27555},
        pages = {arXiv:2605.27555},
          doi = {10.48550/arXiv.2605.27555},
archivePrefix = {arXiv},
       eprint = {2605.27555},
 primaryClass = {astro-ph.GA},
       adsurl = {https://ui.adsabs.harvard.edu/abs/2026arXiv260527555E}
}

@ARTICLE{Baker25,
       author = {{Baker}, William M. and {Lim}, Seunghwan and {D'Eugenio}, Francesco and {Maiolino}, Roberto and {Ji}, Zhiyuan and {Arribas}, Santiago and {Bunker}, Andrew J. and {Carniani}, Stefano and {Charlot}, Stephane and {de Graaff}, Anna and {Hainline}, Kevin and {Looser}, Tobias J. and {Lyu}, Jianwei and {Rinaldi}, Pierluigi and {Robertson}, Brant and {Schaller}, Matthieu and {Schaye}, Joop and {Scholtz}, Jan and {{\"U}bler}, Hannah and {Williams}, Christina C. and {Willmer}, Christopher N.~A. and {Willott}, Chris and {Zhu}, Yongda},
        title = "{The abundance and nature of high-redshift quiescent galaxies from JADES spectroscopy and the FLAMINGO simulations}",
      journal = {\mnras},
         year = 2025,
        month = may,
       volume = {539},
       number = {1},
        pages = {557-589},
          doi = {10.1093/mnras/staf475},
archivePrefix = {arXiv},
       eprint = {2410.14773},
 primaryClass = {astro-ph.GA},
       adsurl = {https://ui.adsabs.harvard.edu/abs/2025MNRAS.539..557B}
}

@ARTICLE{Lagos24,
       author = {{Lagos}, Claudia del P. and {Bravo}, Mat{\'\i}as and {Tobar}, Rodrigo and {Obreschkow}, Danail and {Power}, Chris and {Robotham}, Aaron S.~G. and {Proctor}, Katy L. and {Hansen}, Samuel and {Chandro-G{\'o}mez}, {\'A}ngel and {Carrivick}, Julian},
        title = "{Quenching massive galaxies across cosmic time with the semi-analytic model SHARK V2.0}",
      journal = {\mnras},
         year = 2024,
        month = jul,
       volume = {531},
       number = {3},
        pages = {3551-3578},
          doi = {10.1093/mnras/stae1024},
archivePrefix = {arXiv},
       eprint = {2309.02310},
 primaryClass = {astro-ph.GA},
       adsurl = {https://ui.adsabs.harvard.edu/abs/2024MNRAS.531.3551L}
}

@ARTICLE{Valentino23,
       author = {{Valentino}, Francesco and {Brammer}, Gabriel and {Gould}, Katriona M.~L. and {Kokorev}, Vasily and {Fujimoto}, Seiji and {Jespersen}, Christian Kragh and {Vijayan}, Aswin P. and {Weaver}, John R. and {Ito}, Kei and {Tanaka}, Masayuki and {Ilbert}, Olivier and {Magdis}, Georgios E. and {Whitaker}, Katherine E. and {Faisst}, Andreas L. and {Gallazzi}, Anna and {Gillman}, Steven and {Gim{\'e}nez-Arteaga}, Clara and {G{\'o}mez-Guijarro}, Carlos and {Kubo}, Mariko and {Heintz}, Kasper E. and {Hirschmann}, Michaela and {Oesch}, Pascal and {Onodera}, Masato and {Rizzo}, Francesca and {Lee}, Minju and {Strait}, Victoria and {Toft}, Sune},
        title = "{An Atlas of Color-selected Quiescent Galaxies at z > 3 in Public JWST Fields}",
      journal = {\apj},
         year = 2023,
        month = apr,
       volume = {947},
       number = {1},
          eid = {20},
        pages = {20},
          doi = {10.3847/1538-4357/acbefa},
archivePrefix = {arXiv},
       eprint = {2302.10936},
 primaryClass = {astro-ph.GA},
       adsurl = {https://ui.adsabs.harvard.edu/abs/2023ApJ...947...20V}
}

@ARTICLE{williams23,
       author = {{Williams}, Christina C. and {Tacchella}, Sandro and {Maseda}, Michael V. and {Robertson}, Brant E. and {Johnson}, Benjamin D. and {Willott}, Chris J. and {Eisenstein}, Daniel J. and {Willmer}, Christopher N.~A. and {Ji}, Zhiyuan and {Hainline}, Kevin N. and {Helton}, Jakob M. and {Alberts}, Stacey and {Baum}, Stefi and {Bhatawdekar}, Rachana and {Boyett}, Kristan and {Bunker}, Andrew J. and {Carniani}, Stefano and {Charlot}, Stephane and {Chevallard}, Jacopo and {Curtis-Lake}, Emma and {de Graaff}, Anna and {Egami}, Eiichi and {Franx}, Marijn and {Kumari}, Nimisha and {Maiolino}, Roberto and {Nelson}, Erica J. and {Rieke}, Marcia J. and {Sandles}, Lester and {Shivaei}, Irene and {Simmonds}, Charlotte and {Smit}, Renske and {Suess}, Katherine A. and {Sun}, Fengwu and {{\"U}bler}, Hannah and {Witstok}, Joris},
        title = "{JEMS: A Deep Medium-band Imaging Survey in the Hubble Ultra Deep Field with JWST NIRCam and NIRISS}",
      journal = {\apjs},
         year = 2023,
        month = oct,
       volume = {268},
       number = {2},
          eid = {64},
        pages = {64},
          doi = {10.3847/1538-4365/acf130},
archivePrefix = {arXiv},
       eprint = {2301.09780},
 primaryClass = {astro-ph.GA},
       adsurl = {https://ui.adsabs.harvard.edu/abs/2023ApJS..268...64W}
}

@ARTICLE{larson22b,
       author = {{Larson}, Rebecca L. and {Hutchison}, Taylor A. and {Bagley}, Micaela and {Finkelstein}, Steven L. and {Yung}, L.~Y. Aaron and {Somerville}, Rachel S. and {Hirschmann}, Michaela and {Brammer}, Gabriel and {Holwerda}, Benne W. and {Papovich}, Casey and {Morales}, Alexa M. and {Wilkins}, Stephen M.},
        title = "{Spectral Templates Optimal for Selecting Galaxies at z > 8 with the JWST}",
      journal = {\apj},
         year = 2023,
        month = dec,
       volume = {958},
       number = {2},
          eid = {141},
        pages = {141},
          doi = {10.3847/1538-4357/acfed4},
archivePrefix = {arXiv},
       eprint = {2211.10035},
 primaryClass = {astro-ph.GA},
       adsurl = {https://ui.adsabs.harvard.edu/abs/2023ApJ...958..141L}
}

@ARTICLE{finkelstein23,
       author = {{Finkelstein}, Steven L. and {Bagley}, Micaela B. and {Ferguson}, Henry C. and {Wilkins}, Stephen M. and {Kartaltepe}, Jeyhan S. and {Papovich}, Casey and {Yung}, L.~Y. Aaron and {Arrabal Haro}, Pablo and {Behroozi}, Peter and {Dickinson}, Mark and {Kocevski}, Dale D. and {Koekemoer}, Anton M. and {Larson}, Rebecca L. and {Le Bail}, Aur{\'e}lien and {Morales}, Alexa M. and {P{\'e}rez-Gonz{\'a}lez}, Pablo G. and {Burgarella}, Denis and {Dav{\'e}}, Romeel and {Hirschmann}, Michaela and {Somerville}, Rachel S. and {Wuyts}, Stijn and {Bromm}, Volker and {Casey}, Caitlin M. and {Fontana}, Adriano and {Fujimoto}, Seiji and {Gardner}, Jonathan P. and {Giavalisco}, Mauro and {Grazian}, Andrea and {Grogin}, Norman A. and {Hathi}, Nimish P. and {Hutchison}, Taylor A. and {Jha}, Saurabh W. and {Jogee}, Shardha and {Kewley}, Lisa J. and {Kirkpatrick}, Allison and {Long}, Arianna S. and {Lotz}, Jennifer M. and {Pentericci}, Laura and {Pierel}, Justin D.~R. and {Pirzkal}, Nor and {Ravindranath}, Swara and {Ryan}, Russell E. and {Trump}, Jonathan R. and {Yang}, Guang and {Bhatawdekar}, Rachana and {Bisigello}, Laura and {Buat}, V{\'e}ronique and {Calabr{\`o}}, Antonello and {Castellano}, Marco and {Cleri}, Nikko J. and {Cooper}, M.~C. and {Croton}, Darren and {Daddi}, Emanuele and {Dekel}, Avishai and {Elbaz}, David and {Franco}, Maximilien and {Gawiser}, Eric and {Holwerda}, Benne W. and {Huertas-Company}, Marc and {Jaskot}, Anne E. and {Leung}, Gene C.~K. and {Lucas}, Ray A. and {Mobasher}, Bahram and {Pandya}, Viraj and {Tacchella}, Sandro and {Weiner}, Benjamin J. and {Zavala}, Jorge A.},
        title = "{CEERS Key Paper. I. An Early Look into the First 500 Myr of Galaxy Formation with JWST}",
      journal = {\apjl},
         year = 2023,
        month = mar,
       volume = {946},
       number = {1},
          eid = {L13},
        pages = {L13},
          doi = {10.3847/2041-8213/acade4},
archivePrefix = {arXiv},
       eprint = {2211.05792},
 primaryClass = {astro-ph.GA},
       adsurl = {https://ui.adsabs.harvard.edu/abs/2023ApJ...946L..13F}
}

@software{rest2023jwsthst,
       author = {{Rest}, Armin and {Pierel}, Justin and {Correnti}, Matteo and {Canipe}, Alicia and {Hilbert}, Bryan and {Engesser}, Mike and {Sunnquist}, Ben and {Fox}, Ori},
        title = "{arminrest/jhat: The JWST HST Alignment Tool (JHAT)}",
         year = 2023,
        month = may,
          eid = {10.5281/zenodo.7892935},
          doi = {10.5281/zenodo.7892935},
      version = {v2},
    publisher = {Zenodo},
       adsurl = {https://ui.adsabs.harvard.edu/abs/2023zndo...7892935R}
}

@ARTICLE{franco2026cosmos,
       author = {{Franco}, Maximilien and {Casey}, Caitlin M. and {Koekemoer}, Anton M. and {Liu}, Daizhong and {Bagley}, Micaela B. and {McCracken}, Henry Joy and {Kartaltepe}, Jeyhan S. and {Akins}, Hollis B. and {Ilbert}, Olivier and {Shuntov}, Marko and {Harish}, Santosh and {Robertson}, Brant E. and {Arango-Toro}, Rafael C. and {Battisti}, Andrew J. and {Chartab}, Nima and {Drakos}, Nicole E. and {Faisst}, Andreas L. and {Flayhart}, Carter and {Gozaliasl}, Ghassem and {Hirschmann}, Michaela and {Massey}, Richard and {Rhodes}, Jason and {Sattari}, Zahra and {Scognamiglio}, Diana and {Weaver}, John R. and {Yang}, Lilan and {Zavala}, Jorge A. and {Berman}, Edward M. and {Gentile}, Fabrizio and {Gillman}, Steven and {Long}, Arianna S. and {Magdis}, Georgios and {McCleary}, Jacqueline E. and {McKinney}, Jed and {Mobasher}, Bahram and {Paquereau}, Louise and {Rest}, Armin and {Sanders}, David B. and {Toft}, Sune and {Yu}, Si-Yue},
        title = "{COSMOS-Web: Comprehensive Data Reduction for Wide-area JWST NIRCam Imaging}",
      journal = {\apj},
         year = 2026,
        month = mar,
       volume = {999},
       number = {2},
          eid = {200},
        pages = {200},
          doi = {10.3847/1538-4357/ae3aa1},
archivePrefix = {arXiv},
       eprint = {2506.03256},
 primaryClass = {astro-ph.IM},
       adsurl = {https://ui.adsabs.harvard.edu/abs/2026ApJ...999..200F}
}

@ARTICLE{urbanostawinski24,
       author = {{Urbano Stawinski}, Stephanie M. and {Cooper}, M.~C. and {Finkelstein}, Steven L. and {Jung}, Intae and {P{\'e}rez-Gonz{\'a}lez}, Pablo G. and {Bagley}, Micaela B. and {Casey}, Caitlin M. and {Cooper}, Olivia R. and {Hathi}, Nimish P. and {Holwerda}, Benne W. and {Koekemoer}, Anton M. and {Kartaltepe}, Jeyhan S. and {Fern{\'a}ndez}, Vital and {Larson}, Rebecca L. and {Lucas}, Ray A. and {Yung}, L.~Y. Aaron},
        title = "{Deeper than DEEP: a spectroscopic survey of z > 3 Ly {\ensuremath{\alpha}} emitters in the Extended Groth Strip}",
      journal = {\mnras},
         year = 2024,
        month = mar,
       volume = {528},
       number = {4},
        pages = {5624-5632},
          doi = {10.1093/mnras/stae361},
archivePrefix = {arXiv},
       eprint = {2307.04782},
 primaryClass = {astro-ph.GA},
       adsurl = {https://ui.adsabs.harvard.edu/abs/2024MNRAS.528.5624U}
}

@ARTICLE{jung18,
   author = {{Jung}, I. and {Finkelstein}, S.~L. and {Livermore}, R.~C. and 
	{Dickinson}, M. and {Larson}, R.~L. and {Papovich}, C. and {Song}, M. and 
	{Tilvi}, V. and {Wold}, I.},
    title = "{Texas Spectroscopic Search for Ly{$\alpha$} Emission at the End of Reionization I. Constraining the Ly{$\alpha$} Equivalent-width Distribution at 6.0 $<$ 7.0}",
  journal = {\apj},
archivePrefix = "arXiv",
   eprint = {1803.01870},
     year = 2018,
    month = sep,
   volume = 864,
      eid = {103},
    pages = {103}}

@ARTICLE{larson22,
       author = {{Larson}, Rebecca L. and {Finkelstein}, Steven L. and {Hutchison}, Taylor A. and {Papovich}, Casey and {Bagley}, Micaela and {Dickinson}, Mark and {Rojas-Ruiz}, Sof{\'\i}a and {Ferguson}, Harry C. and {Jung}, Intae and {Giavalisco}, Mauro and {Grazian}, Andrea and {Pentericci}, Laura and {Tacchella}, Sandro},
        title = "{Searching for Islands of Reionization: A Potential Ionized Bubble Powered by a Spectroscopic Overdensity at z = 8.7}",
      journal = {\apj},
         year = 2022,
        month = may,
       volume = {930},
       number = {2},
          eid = {104},
        pages = {104},
          doi = {10.3847/1538-4357/ac5dbd},
archivePrefix = {arXiv},
       eprint = {2203.08461},
 primaryClass = {astro-ph.GA},
       adsurl = {https://ui.adsabs.harvard.edu/abs/2022ApJ...930..104L}
}

@ARTICLE{mosdef,
       author = {{Sanders}, Ryan L. and {Shapley}, Alice E. and {Jones}, Tucker and {Reddy}, Naveen A. and {Kriek}, Mariska and {Siana}, Brian and {Coil}, Alison L. and {Mobasher}, Bahram and {Shivaei}, Irene and {Dav{\'e}}, Romeel and {Azadi}, Mojegan and {Price}, Sedona H. and {Leung}, Gene and {Freeman}, William R. and {Fetherolf}, Tara and {de Groot}, Laura and {Zick}, Tom and {Barro}, Guillermo},
        title = "{The MOSDEF Survey: The Evolution of the Mass-Metallicity Relation from z = 0 to z 3.3}",
      journal = {\apj},
         year = 2021,
        month = jun,
       volume = {914},
       number = {1},
          eid = {19},
        pages = {19},
          doi = {10.3847/1538-4357/abf4c1},
archivePrefix = {arXiv},
       eprint = {2009.07292},
 primaryClass = {astro-ph.GA},
       adsurl = {https://ui.adsabs.harvard.edu/abs/2021ApJ...914...19S}
}

@ARTICLE{stefanon17,
       author = {{Stefanon}, Mauro and {Yan}, Haojing and {Mobasher}, Bahram and
         {Barro}, Guillermo and {Donley}, Jennifer L. and {Fontana}, Adriano and
         {Hemmati}, Shoubaneh and {Koekemoer}, Anton M. and {Lee}, BoMee and
         {Lee}, Seong-Kook and {Nayyeri}, Hooshang and {Peth}, Michael and
         {Pforr}, Janine and {Salvato}, Mara and {Wiklind}, Tommy and
         {Wuyts}, Stijn and {Ashby}, Matthew L.~N. and {Castellano}, Marco and
         {Conselice}, Christopher J. and {Cooper}, Michael C. and
         {Cooray}, Asantha R. and {Dolch}, Timothy and {Ferguson}, Henry and
         {Galametz}, Audrey and {Giavalisco}, Mauro and {Guo}, Yicheng and
         {Willner}, Steven P. and {Dickinson}, Mark E. and {Faber}, Sandra M. and
         {Fazio}, Giovanni G. and {Gardner}, Jonathan P. and {Gawiser}, Eric and
         {Grazian}, Andrea and {Grogin}, Norman A. and {Kocevski}, Dale and
         {Koo}, David C. and {Lee}, Kyoung-Soo and {Lucas}, Ray A. and
         {McGrath}, Elizabeth J. and {Nandra}, Kirpal and {Newman}, Jeffrey A. and
         {van der Wel}, Arjen},
        title = "{CANDELS Multi-wavelength Catalogs: Source Identification and Photometry in the CANDELS Extended Groth Strip}",
      journal = {\apjs},
         year = "2017",
        month = "Apr",
       volume = {229},
       number = {2},
          eid = {32},
        pages = {32},
          doi = {10.3847/1538-4365/aa66cb},
archivePrefix = {arXiv},
       eprint = {1703.05768},
 primaryClass = {astro-ph.GA},
       adsurl = {https://ui.adsabs.harvard.edu/abs/2017ApJS..229...32S}}

@ARTICLE{heintz24,
       author = {{Heintz}, Kasper E. and {Watson}, Darach and {Brammer}, Gabriel and {Vejlgaard}, Simone and {Hutter}, Anne and {Strait}, Victoria B. and {Matthee}, Jorryt and {Oesch}, Pascal A. and {Jakobsson}, P{\'a}ll and {Tanvir}, Nial R. and {Laursen}, Peter and {Naidu}, Rohan P. and {Mason}, Charlotte A. and {Killi}, Meghana and {Jung}, Intae and {Hsiao}, Tiger Yu-Yang and {Abdurro'uf} and {Coe}, Dan and {Arrabal Haro}, Pablo and {Finkelstein}, Steven L. and {Toft}, Sune},
        title = "{Strong damped Lyman-{\ensuremath{\alpha}} absorption in young star-forming galaxies at redshifts 9 to 11}",
      journal = {Science},
         year = 2024,
        month = may,
       volume = {384},
       number = {6698},
        pages = {890-894},
          doi = {10.1126/science.adj0343},
archivePrefix = {arXiv},
       eprint = {2306.00647},
 primaryClass = {astro-ph.GA},
       adsurl = {https://ui.adsabs.harvard.edu/abs/2024Sci...384..890H}
}

@ARTICLE{degraaff24,
       author = {{de Graaff}, Anna and {Rix}, Hans-Walter and {Carniani}, Stefano and {Suess}, Katherine A. and {Charlot}, St{\'e}phane and {Curtis-Lake}, Emma and {Arribas}, Santiago and {Baker}, William M. and {Boyett}, Kristan and {Bunker}, Andrew J. and {Cameron}, Alex J. and {Chevallard}, Jacopo and {Curti}, Mirko and {Eisenstein}, Daniel J. and {Franx}, Marijn and {Hainline}, Kevin and {Hausen}, Ryan and {Ji}, Zhiyuan and {Johnson}, Benjamin D. and {Jones}, Gareth C. and {Maiolino}, Roberto and {Maseda}, Michael V. and {Nelson}, Erica and {Parlanti}, Eleonora and {Rawle}, Tim and {Robertson}, Brant and {Tacchella}, Sandro and {{\"U}bler}, Hannah and {Williams}, Christina C. and {Willmer}, Christopher N.~A. and {Willott}, Chris},
        title = "{Ionised gas kinematics and dynamical masses of z {\ensuremath{\gtrsim}} 6 galaxies from JADES/NIRSpec high-resolution spectroscopy}",
      journal = {\aap},
         year = 2024,
        month = apr,
       volume = {684},
          eid = {A87},
        pages = {A87},
          doi = {10.1051/0004-6361/202347755},
archivePrefix = {arXiv},
       eprint = {2308.09742},
 primaryClass = {astro-ph.GA},
       adsurl = {https://ui.adsabs.harvard.edu/abs/2024A&A...684A..87D}
}

@ARTICLE{davis07,
   author = {{Davis}, M. and {Guhathakurta}, P. and {Konidaris}, N.~P. and 
	{Newman}, J.~A. and {Ashby}, M.~L.~N. and {Biggs}, A.~D. and 
	{Barmby}, P. and {Bundy}, K. and {Chapman}, S.~C. and {Coil}, A.~L. and 
	{Conselice}, C.~J. and {Cooper}, M.~C. and {Croton}, D.~J. and 
	{Eisenhardt}, P.~R.~M. and {Ellis}, R.~S. and {Faber}, S.~M. and 
	{Fang}, T. and {Fazio}, G.~G. and {Georgakakis}, A. and {Gerke}, B.~F. and 
	{Goss}, W.~M. and {Gwyn}, S. and {Harker}, J. and {Hopkins}, A.~M. and 
	{Huang}, J.-S. and {Ivison}, R.~J. and {Kassin}, S.~A. and {Kirby}, E.~N. and 
	{Koekemoer}, A.~M. and {Koo}, D.~C. and {Laird}, E.~S. and {Le Floc'h}, E. and 
	{Lin}, L. and {Lotz}, J.~M. and {Marshall}, P.~J. and {Martin}, D.~C. and 
	{Metevier}, A.~J. and {Moustakas}, L.~A. and {Nandra}, K. and 
	{Noeske}, K.~G. and {Papovich}, C. and {Phillips}, A.~C. and 
	{Rich}, R.~M. and {Rieke}, G.~H. and {Rigopoulou}, D. and {Salim}, S. and 
	{Schiminovich}, D. and {Simard}, L. and {Smail}, I. and {Small}, T.~A. and 
	{Weiner}, B.~J. and {Willmer}, C.~N.~A. and {Willner}, S.~P. and 
	{Wilson}, G. and {Wright}, E.~L. and {Yan}, R.},
    title = "{The All-Wavelength Extended Groth Strip International Survey (AEGIS) Data Sets}",
  journal = {\apjl},
     year = 2007,
    month = may,
   volume = 660,
    pages = {L1-L6},}

@ARTICLE{grogin11,
   author = {{Grogin}, N.~A. and {Kocevski}, D.~D. and {Faber}, S.~M. and 
	{Ferguson}, H.~C. and {Koekemoer}, A.~M. and {Riess}, A.~G. and 
	{Acquaviva}, V. and {Alexander}, D.~M. and {Almaini}, O. and 
	{Ashby}, M.~L.~N. and {Barden}, M. and {Bell}, E.~F. and {Bournaud}, F. and 
	{Brown}, T.~M. and {Caputi}, K.~I. and {Casertano}, S. and {Cassata}, P. and 
	{Castellano}, M. and {Challis}, P. and {Chary}, R.-R. and {Cheung}, E. and 
	{Cirasuolo}, M. and {Conselice}, C.~J. and {Roshan Cooray}, A. and 
	{Croton}, D.~J. and {Daddi}, E. and {Dahlen}, T. and {Dav{\'e}}, R. and 
	{de Mello}, D.~F. and {Dekel}, A. and {Dickinson}, M. and {Dolch}, T. and 
	{Donley}, J.~L. and {Dunlop}, J.~S. and {Dutton}, A.~A. and 
	{Elbaz}, D. and {Fazio}, G.~G. and {Filippenko}, A.~V. and {Finkelstein}, S.~L. and 
	{Fontana}, A. and {Gardner}, J.~P. and {Garnavich}, P.~M. and 
	{Gawiser}, E. and {Giavalisco}, M. and {Grazian}, A. and {Guo}, Y. and 
	{Hathi}, N.~P. and {H{\"a}ussler}, B. and {Hopkins}, P.~F. and 
	{Huang}, J.-S. and {Huang}, K.-H. and {Jha}, S.~W. and {Kartaltepe}, J.~S. and 
	{Kirshner}, R.~P. and {Koo}, D.~C. and {Lai}, K. and {Lee}, K.-S. and 
	{Li}, W. and {Lotz}, J.~M. and {Lucas}, R.~A. and {Madau}, P. and 
	{McCarthy}, P.~J. and {McGrath}, E.~J. and {McIntosh}, D.~H. and 
	{McLure}, R.~J. and {Mobasher}, B. and {Moustakas}, L.~A. and 
	{Mozena}, M. and {Nandra}, K. and {Newman}, J.~A. and {Niemi}, S.-M. and 
	{Noeske}, K.~G. and {Papovich}, C.~J. and {Pentericci}, L. and 
	{Pope}, A. and {Primack}, J.~R. and {Rajan}, A. and {Ravindranath}, S. and 
	{Reddy}, N.~A. and {Renzini}, A. and {Rix}, H.-W. and {Robaina}, A.~R. and 
	{Rodney}, S.~A. and {Rosario}, D.~J. and {Rosati}, P. and {Salimbeni}, S. and 
	{Scarlata}, C. and {Siana}, B. and {Simard}, L. and {Smidt}, J. and 
	{Somerville}, R.~S. and {Spinrad}, H. and {Straughn}, A.~N. and 
	{Strolger}, L.-G. and {Telford}, O. and {Teplitz}, H.~I. and 
	{Trump}, J.~R. and {van der Wel}, A. and {Villforth}, C. and 
	{Wechsler}, R.~H. and {Weiner}, B.~J. and {Wiklind}, T. and 
	{Wild}, V. and {Wilson}, G. and {Wuyts}, S. and {Yan}, H.-J. and 
	{Yun}, M.~S.},
    title = "{CANDELS: The Cosmic Assembly Near-infrared Deep Extragalactic Legacy Survey}",
  journal = {\apjs},
archivePrefix = "arXiv",
   eprint = {1105.3753},
 primaryClass = "astro-ph.CO",
     year = 2011,
    month = dec,
   volume = 197,
      eid = {35},
    pages = {35}}

@ARTICLE{rieke23,
       author = {{Rieke}, Marcia J. and {Kelly}, Douglas M. and {Misselt}, Karl and {Stansberry}, John and {Boyer}, Martha and {Beatty}, Thomas and {Egami}, Eiichi and {Florian}, Michael and {Greene}, Thomas P. and {Hainline}, Kevin and {Leisenring}, Jarron and {Roellig}, Thomas and {Schlawin}, Everett and {Sun}, Fengwu and {Tinnin}, Lee and {Williams}, Christina C. and {Willmer}, Christopher N.~A. and {Wilson}, Debra and {Clark}, Charles R. and {Rohrbach}, Scott and {Brooks}, Brian and {Canipe}, Alicia and {Correnti}, Matteo and {DiFelice}, Audrey and {Gennaro}, Mario and {Girard}, Julien H. and {Hartig}, George and {Hilbert}, Bryan and {Koekemoer}, Anton M. and {Nikolov}, Nikolay K. and {Pirzkal}, Norbert and {Rest}, Armin and {Robberto}, Massimo and {Sunnquist}, Ben and {Telfer}, Randal and {Wu}, Chi Rai and {Ferry}, Malcolm and {Lewis}, Dan and {Baum}, Stefi and {Beichman}, Charles and {Doyon}, Ren{\'e} and {Dressler}, Alan and {Eisenstein}, Daniel J. and {Ferrarese}, Laura and {Hodapp}, Klaus and {Horner}, Scott and {Jaffe}, Daniel T. and {Johnstone}, Doug and {Krist}, John and {Martin}, Peter and {McCarthy}, Donald W. and {Meyer}, Michael and {Rieke}, George H. and {Trauger}, John and {Young}, Erick T.},
        title = "{Performance of NIRCam on JWST in Flight}",
      journal = {\pasp},
         year = 2023,
        month = feb,
       volume = {135},
       number = {1044},
          eid = {028001},
        pages = {028001},
          doi = {10.1088/1538-3873/acac53},
archivePrefix = {arXiv},
       eprint = {2212.12069},
 primaryClass = {astro-ph.IM},
       adsurl = {https://ui.adsabs.harvard.edu/abs/2023PASP..135b8001R}
}

@ARTICLE{finkelstein25,
       author = {{Finkelstein}, Steven L. and {Bagley}, Micaela B. and {Arrabal Haro}, Pablo and {Dickinson}, Mark and {Ferguson}, Henry C. and {Kartaltepe}, Jeyhan S. and {Kocevski}, Dale D. and {Koekemoer}, Anton M. and {Lotz}, Jennifer M. and {Papovich}, Casey and {P{\'e}rez-Gonz{\'a}lez}, Pablo G. and {Pirzkal}, Nor and {Somerville}, Rachel S. and {Trump}, Jonathan R. and {Yang}, Guang and {Yung}, L.~Y. Aaron and {Fontana}, Adriano and {Grazian}, Andrea and {Grogin}, Norman A. and {Kewley}, Lisa J. and {Kirkpatrick}, Allison and {Larson}, Rebecca L. and {Pentericci}, Laura and {Ravindranath}, Swara and {Wilkins}, Stephen M. and {Almaini}, Omar and {Amor{\'\i}n}, Ricardo O. and {Barro}, Guillermo and {Bhatawdekar}, Rachana and {Bisigello}, Laura and {Brooks}, Madisyn and {Buat}, V{\'e}ronique and {Buitrago}, Fernando and {Burgarella}, Denis and {Calabr{\`o}}, Antonello and {Castellano}, Marco and {Cheng}, Yingjie and {Cleri}, Nikko J. and {Cole}, Justin W. and {Cooper}, M.~C. and {Cooper}, Olivia R. and {Costantin}, Luca and {Cox}, Isa G. and {Croton}, Darren and {Daddi}, Emanuele and {Davis}, Kelcey and {Dekel}, Avishai and {Elbaz}, David and {Fern{\'a}ndez}, Vital and {Fujimoto}, Seiji and {Gandolfi}, Giovanni and {Gardner}, Jonathan P. and {Gawiser}, Eric and {Giavalisco}, Mauro and {G{\'o}mez-Guijarro}, Carlos and {Guo}, Yuchen and {Gupta}, Ansh R. and {Hathi}, Nimish P. and {Harish}, Santosh and {Henry}, Aur{\'e}lien and {Hirschmann}, Michaela and {Hu}, Weida and {Hutchison}, Taylor A. and {Iyer}, Kartheik G. and {Jaskot}, Anne E. and {Jha}, Saurabh W. and {Jung}, Intae and {Kassin}, Susan A. and {Kokorev}, Vasily and {Kurczynski}, Peter and {Leung}, Gene C.~K. and {Llerena}, Mario and {Long}, Arianna S. and {Lucas}, Ray A. and {Lu}, Shiying and {McGrath}, Elizabeth J. and {McIntosh}, Daniel H. and {Merlin}, Emiliano and {Mobasher}, Bahram and {Morales}, Alexa M. and {Napolitano}, Lorenzo and {Pacucci}, Fabio and {Pandya}, Viraj and {Rafelski}, Marc and {Rodighiero}, Giulia and {Rose}, Caitlin and {Santini}, Paola and {Seill{\'e}}, Lise-Marie and {Simons}, Raymond C. and {Shen}, Lu and {Straughn}, Amber N. and {Tacchella}, Sandro and {Taylor}, Anthony J. and {Vanderhoof}, Brittany N. and {Vega-Ferrero}, Jes{\'u}s and {Weiner}, Benjamin J. and {Willmer}, Christopher N.~A. and {Zhu}, Peixin and {Bell}, Eric F. and {Wuyts}, Stijn and {Holwerda}, Benne W. and {Wang}, Xin and {Wang}, Weichen and {Zavala}, Jorge A. and {CEERS Collaboration}},
        title = "{The Cosmic Evolution Early Release Science Survey (CEERS)}",
      journal = {\apjl},
         year = 2025,
        month = apr,
       volume = {983},
       number = {1},
          eid = {L4},
        pages = {L4},
          doi = {10.3847/2041-8213/adbbd3},
archivePrefix = {arXiv},
       eprint = {2501.04085},
 primaryClass = {astro-ph.GA},
       adsurl = {https://ui.adsabs.harvard.edu/abs/2025ApJ...983L...4F}
}

@ARTICLE{Arrabal2023a,
       author = {{Arrabal Haro}, Pablo and {Dickinson}, Mark and {Finkelstein}, Steven L. and {Fujimoto}, Seiji and {Fern{\'a}ndez}, Vital and {Kartaltepe}, Jeyhan S. and {Jung}, Intae and {Cole}, Justin W. and {Burgarella}, Denis and {Chworowsky}, Katherine and {Hutchison}, Taylor A. and {Morales}, Alexa M. and {Papovich}, Casey and {Simons}, Raymond C. and {Amor{\'\i}n}, Ricardo O. and {Backhaus}, Bren E. and {Bagley}, Micaela B. and {Bisigello}, Laura and {Calabr{\`o}}, Antonello and {Castellano}, Marco and {Cleri}, Nikko J. and {Dav{\'e}}, Romeel and {Dekel}, Avishai and {Ferguson}, Henry C. and {Fontana}, Adriano and {Gawiser}, Eric and {Giavalisco}, Mauro and {Harish}, Santosh and {Hathi}, Nimish P. and {Hirschmann}, Michaela and {Holwerda}, Benne W. and {Huertas-Company}, Marc and {Koekemoer}, Anton M. and {Larson}, Rebecca L. and {Lucas}, Ray A. and {Mobasher}, Bahram and {P{\'e}rez-Gonz{\'a}lez}, Pablo G. and {Pirzkal}, Nor and {Rose}, Caitlin and {Santini}, Paola and {Trump}, Jonathan R. and {de la Vega}, Alexander and {Wang}, Xin and {Weiner}, Benjamin J. and {Wilkins}, Stephen M. and {Yang}, Guang and {Yung}, L.~Y. Aaron and {Zavala}, Jorge A.},
        title = "{Spectroscopic Confirmation of CEERS NIRCam-selected Galaxies at z ≃ 8-10}",
      journal = {\apjl},
         year = 2023,
        month = jul,
       volume = {951},
       number = {1},
          eid = {L22},
        pages = {L22},
          doi = {10.3847/2041-8213/acdd54},
archivePrefix = {arXiv},
       eprint = {2304.05378},
 primaryClass = {astro-ph.GA},
       adsurl = {https://ui.adsabs.harvard.edu/abs/2023ApJ...951L..22A}
}

@ARTICLE{Arrabal2023b,
       author = {{Arrabal Haro}, Pablo and {Dickinson}, Mark and {Finkelstein}, Steven L. and {Kartaltepe}, Jeyhan S. and {Donnan}, Callum T. and {Burgarella}, Denis and {Carnall}, Adam C. and {Cullen}, Fergus and {Dunlop}, James S. and {Fern{\'a}ndez}, Vital and {Fujimoto}, Seiji and {Jung}, Intae and {Krips}, Melanie and {Larson}, Rebecca L. and {Papovich}, Casey and {P{\'e}rez-Gonz{\'a}lez}, Pablo G. and {Amor{\'\i}n}, Ricardo O. and {Bagley}, Micaela B. and {Buat}, V{\'e}ronique and {Casey}, Caitlin M. and {Chworowsky}, Katherine and {Cohen}, Seth H. and {Ferguson}, Henry C. and {Giavalisco}, Mauro and {Huertas-Company}, Marc and {Hutchison}, Taylor A. and {Kocevski}, Dale D. and {Koekemoer}, Anton M. and {Lucas}, Ray A. and {McLeod}, Derek J. and {McLure}, Ross J. and {Pirzkal}, Norbert and {Seill{\'e}}, Lise-Marie and {Trump}, Jonathan R. and {Weiner}, Benjamin J. and {Wilkins}, Stephen M. and {Zavala}, Jorge A.},
        title = "{Confirmation and refutation of very luminous galaxies in the early Universe}",
      journal = {\nat},
         year = 2023,
        month = oct,
       volume = {622},
       number = {7984},
        pages = {707-711},
          doi = {10.1038/s41586-023-06521-7},
archivePrefix = {arXiv},
       eprint = {2303.15431},
 primaryClass = {astro-ph.GA},
       adsurl = {https://ui.adsabs.harvard.edu/abs/2023Natur.622..707A}
}

@ARTICLE{Bagley2023,
       author = {{Bagley}, Micaela B. and {Finkelstein}, Steven L. and {Koekemoer}, Anton M. and {Ferguson}, Henry C. and {Arrabal Haro}, Pablo and {Dickinson}, Mark and {Kartaltepe}, Jeyhan S. and {Papovich}, Casey and {P{\'e}rez-Gonz{\'a}lez}, Pablo G. and {Pirzkal}, Nor and {Somerville}, Rachel S. and {Willmer}, Christopher N.~A. and {Yang}, Guang and {Yung}, L.~Y. Aaron and {Fontana}, Adriano and {Grazian}, Andrea and {Grogin}, Norman A. and {Hirschmann}, Michaela and {Kewley}, Lisa J. and {Kirkpatrick}, Allison and {Kocevski}, Dale D. and {Lotz}, Jennifer M. and {Medrano}, Aubrey and {Morales}, Alexa M. and {Pentericci}, Laura and {Ravindranath}, Swara and {Trump}, Jonathan R. and {Wilkins}, Stephen M. and {Calabr{\`o}}, Antonello and {Cooper}, M.~C. and {Costantin}, Luca and {de la Vega}, Alexander and {Hilbert}, Bryan and {Hutchison}, Taylor A. and {Larson}, Rebecca L. and {Lucas}, Ray A. and {McGrath}, Elizabeth J. and {Ryan}, Russell and {Wang}, Xin and {Wuyts}, Stijn},
        title = "{CEERS Epoch 1 NIRCam Imaging: Reduction Methods and Simulations Enabling Early JWST Science Results}",
      journal = {\apjl},
         year = 2023,
        month = mar,
       volume = {946},
       number = {1},
          eid = {L12},
        pages = {L12},
          doi = {10.3847/2041-8213/acbb08},
archivePrefix = {arXiv},
       eprint = {2211.02495},
 primaryClass = {astro-ph.IM},
       adsurl = {https://ui.adsabs.harvard.edu/abs/2023ApJ...946L..12B}
}

@ARTICLE{Carnall2022,
       author = {{Carnall}, Adam C. and {McLure}, Ross J. and {Dunlop}, James S. and {Hamadouche}, Massissilia and {Cullen}, Fergus and {McLeod}, Derek J. and {Begley}, Ryan and {Amorin}, Ricardo and {Bolzonella}, Micol and {Castellano}, Marco and {Cimatti}, Andrea and {Fontanot}, Fabio and {Gargiulo}, Adriana and {Garilli}, Bianca and {Mannucci}, Filippo and {Pentericci}, Laura and {Talia}, Margherita and {Zamorani}, Giovani and {Calabro}, Antonello and {Cresci}, Giovanni and {Hathi}, Nimish P.},
        title = "{The Stellar Metallicities of Massive Quiescent Galaxies at 1.0 < z < 1.3 from KMOS + VANDELS}",
      journal = {\apj},
         year = 2022,
        month = apr,
       volume = {929},
       number = {2},
          eid = {131},
        pages = {131},
          doi = {10.3847/1538-4357/ac5b62},
archivePrefix = {arXiv},
       eprint = {2108.13430},
 primaryClass = {astro-ph.GA},
       adsurl = {https://ui.adsabs.harvard.edu/abs/2022ApJ...929..131C}
}

@ARTICLE{Davis2024,
       author = {{Davis}, Kelcey and {Trump}, Jonathan R. and {Simons}, Raymond C. and {McGrath}, Elizabeth J. and {Wilkins}, Stephen M. and {Arrabal Haro}, Pablo and {Bagley}, Micaela B. and {Dickinson}, Mark and {Fern{\'a}ndez}, Vital and {Amor{\'\i}n}, Ricardo O. and {Backhaus}, Bren E. and {Cleri}, Nikko J. and {Llerena}, Mario and {Brunker}, Samantha W. and {Barro}, Guillermo and {Bisigello}, Laura and {Brooks}, Madisyn and {Costantin}, Luca and {de La Vega}, Alexander and {Dekel}, Avishai and {Finkelstein}, Steven L. and {Hathi}, Nimish P. and {Hirschmann}, Michaela and {Kartaltepe}, Jeyhan S. and {Koekemoer}, Anton M. and {Lucas}, Ray A. and {Papovich}, Casey and {P{\'e}rez-Gonz{\'a}lez}, Pablo G. and {Pirzkal}, Nor and {Rodighiero}, Giulia and {Rose}, Caitlin and {Yung}, L.~Y. Aaron and {Ceers Collaborators}},
        title = "{A Census from JWST of Extreme Emission-line Galaxies Spanning the Epoch of Reionization in CEERS}",
      journal = {\apj},
         year = 2024,
        month = oct,
       volume = {974},
       number = {1},
          eid = {42},
        pages = {42},
          doi = {10.3847/1538-4357/ad6865},
archivePrefix = {arXiv},
       eprint = {2312.07799},
 primaryClass = {astro-ph.GA},
       adsurl = {https://ui.adsabs.harvard.edu/abs/2024ApJ...974...42D}
}

@ARTICLE{Donnan2023,
       author = {{Donnan}, C.~T. and {McLeod}, D.~J. and {Dunlop}, J.~S. and {McLure}, R.~J. and {Carnall}, A.~C. and {Begley}, R. and {Cullen}, F. and {Hamadouche}, M.~L. and {Bowler}, R.~A.~A. and {Magee}, D. and {McCracken}, H.~J. and {Milvang-Jensen}, B. and {Moneti}, A. and {Targett}, T.},
        title = "{The evolution of the galaxy UV luminosity function at redshifts z ≃ 8 - 15 from deep JWST and ground-based near-infrared imaging}",
      journal = {\mnras},
         year = 2023,
        month = feb,
       volume = {518},
       number = {4},
        pages = {6011-6040},
          doi = {10.1093/mnras/stac3472},
archivePrefix = {arXiv},
       eprint = {2207.12356},
 primaryClass = {astro-ph.GA},
       adsurl = {https://ui.adsabs.harvard.edu/abs/2023MNRAS.518.6011D}
}

@ARTICLE{Endsley2024,
       author = {{Endsley}, Ryan and {Stark}, Daniel P. and {Whitler}, Lily and {Topping}, Michael W. and {Johnson}, Benjamin D. and {Robertson}, Brant and {Tacchella}, Sandro and {Alberts}, Stacey and {Baker}, William M. and {Bhatawdekar}, Rachana and {Boyett}, Kristan and {Bunker}, Andrew J. and {Cameron}, Alex J. and {Carniani}, Stefano and {Charlot}, Stephane and {Chen}, Zuyi and {Chevallard}, Jacopo and {Curtis-Lake}, Emma and {Danhaive}, A. Lola and {Egami}, Eiichi and {Eisenstein}, Daniel J. and {Hainline}, Kevin and {Helton}, Jakob M. and {Ji}, Zhiyuan and {Looser}, Tobias J. and {Maiolino}, Roberto and {Nelson}, Erica and {Pusk{\'a}s}, D{\'a}vid and {Rieke}, George and {Rieke}, Marcia and {Rix}, Hans-Walter and {Sandles}, Lester and {Saxena}, Aayush and {Simmonds}, Charlotte and {Smit}, Renske and {Sun}, Fengwu and {Williams}, Christina C. and {Willmer}, Christopher N.~A. and {Willott}, Chris and {Witstok}, Joris},
        title = "{The star-forming and ionizing properties of dwarf z 6-9 galaxies in JADES: insights on bursty star formation and ionized bubble growth}",
      journal = {\mnras},
         year = 2024,
        month = sep,
       volume = {533},
       number = {1},
        pages = {1111-1142},
          doi = {10.1093/mnras/stae1857},
archivePrefix = {arXiv},
       eprint = {2306.05295},
 primaryClass = {astro-ph.GA},
       adsurl = {https://ui.adsabs.harvard.edu/abs/2024MNRAS.533.1111E}
}

@ARTICLE{Fujimoto2023,
       author = {{Fujimoto}, Seiji and {Arrabal Haro}, Pablo and {Dickinson}, Mark and {Finkelstein}, Steven L. and {Kartaltepe}, Jeyhan S. and {Larson}, Rebecca L. and {Burgarella}, Denis and {Bagley}, Micaela B. and {Behroozi}, Peter and {Chworowsky}, Katherine and {Hirschmann}, Michaela and {Trump}, Jonathan R. and {Wilkins}, Stephen M. and {Yung}, L.~Y. Aaron and {Koekemoer}, Anton M. and {Papovich}, Casey and {Pirzkal}, Nor and {Ferguson}, Henry C. and {Fontana}, Adriano and {Grogin}, Norman A. and {Grazian}, Andrea and {Kewley}, Lisa J. and {Kocevski}, Dale D. and {Lotz}, Jennifer M. and {Pentericci}, Laura and {Ravindranath}, Swara and {Somerville}, Rachel S. and {Wilkins}, Stephen M. and {Amor{\'\i}n}, Ricardo O. and {Backhaus}, Bren E. and {Calabr{\`o}}, Antonello and {Casey}, Caitlin M. and {Cooper}, M.~C. and {Fern{\'a}ndez}, Vital and {Franco}, Maximilien and {Giavalisco}, Mauro and {Hathi}, Nimish P. and {Harish}, Santosh and {Hutchison}, Taylor A. and {Iyer}, Kartheik G. and {Jung}, Intae and {Lucas}, Ray A. and {Zavala}, Jorge A.},
        title = "{CEERS Spectroscopic Confirmation of NIRCam-selected z {\ensuremath{\gtrsim}} 8 Galaxy Candidates with JWST/NIRSpec: Initial Characterization of Their Properties}",
      journal = {\apjl},
         year = 2023,
        month = jun,
       volume = {949},
       number = {2},
          eid = {L25},
        pages = {L25},
          doi = {10.3847/2041-8213/acd2d9},
archivePrefix = {arXiv},
       eprint = {2301.09482},
 primaryClass = {astro-ph.GA},
       adsurl = {https://ui.adsabs.harvard.edu/abs/2023ApJ...949L..25F}
}

@ARTICLE{Finkelstein2019,
       author = {{Finkelstein}, Steven L. and {D'Aloisio}, Anson and {Paardekooper}, Jan-Pieter and {Ryan}, Russell, Jr. and {Behroozi}, Peter and {Finlator}, Kristian and {Livermore}, Rachael and {Upton Sanderbeck}, Phoebe R. and {Dalla Vecchia}, Claudio and {Khochfar}, Sadegh},
        title = "{Conditions for Reionizing the Universe with a Low Galaxy Ionizing Photon Escape Fraction}",
      journal = {\apj},
         year = 2019,
        month = jul,
       volume = {879},
       number = {1},
          eid = {36},
        pages = {36},
          doi = {10.3847/1538-4357/ab1ea8},
archivePrefix = {arXiv},
       eprint = {1902.02792},
 primaryClass = {astro-ph.CO},
       adsurl = {https://ui.adsabs.harvard.edu/abs/2019ApJ...879...36F}
}

@ARTICLE{Finkelstein2022a,
       author = {{Finkelstein}, Steven L. and {Bagley}, Micaela and {Song}, Mimi and {Larson}, Rebecca and {Papovich}, Casey and {Dickinson}, Mark and {Finkelstein}, Keely D. and {Koekemoer}, Anton M. and {Pirzkal}, Norbert and {Somerville}, Rachel S. and {Yung}, L.~Y. Aaron and {Behroozi}, Peter and {Ferguson}, Harry and {Giavalisco}, Mauro and {Grogin}, Norman and {Hathi}, Nimish and {Hutchison}, Taylor A. and {Jung}, Intae and {Kocevski}, Dale and {Kawinwanichakij}, Lalitwadee and {Rojas-Ruiz}, Sof{\'\i}a and {Ryan}, Russell and {Snyder}, Gregory F. and {Tacchella}, Sandro},
        title = "{A Census of the Bright z = 8.5-11 Universe with the Hubble and Spitzer Space Telescopes in the CANDELS Fields}",
      journal = {\apj},
         year = 2022,
        month = mar,
       volume = {928},
       number = {1},
          eid = {52},
        pages = {52},
          doi = {10.3847/1538-4357/ac3aed},
archivePrefix = {arXiv},
       eprint = {2106.13813},
 primaryClass = {astro-ph.GA},
       adsurl = {https://ui.adsabs.harvard.edu/abs/2022ApJ...928...52F}
}

@ARTICLE{Finkelstein2022c,
       author = {{Finkelstein}, Steven L. and {Bagley}, Micaela B. and {Arrabal Haro}, Pablo and {Dickinson}, Mark and {Ferguson}, Henry C. and {Kartaltepe}, Jeyhan S. and {Papovich}, Casey and {Burgarella}, Denis and {Kocevski}, Dale D. and {Huertas-Company}, Marc and {Iyer}, Kartheik G. and {Koekemoer}, Anton M. and {Larson}, Rebecca L. and {P{\'e}rez-Gonz{\'a}lez}, Pablo G. and {Rose}, Caitlin and {Tacchella}, Sandro and {Wilkins}, Stephen M. and {Chworowsky}, Katherine and {Medrano}, Aubrey and {Morales}, Alexa M. and {Somerville}, Rachel S. and {Yung}, L.~Y. Aaron and {Fontana}, Adriano and {Giavalisco}, Mauro and {Grazian}, Andrea and {Grogin}, Norman A. and {Kewley}, Lisa J. and {Kirkpatrick}, Allison and {Kurczynski}, Peter and {Lotz}, Jennifer M. and {Pentericci}, Laura and {Pirzkal}, Nor and {Ravindranath}, Swara and {Ryan}, Russell E. and {Trump}, Jonathan R. and {Yang}, Guang and {Almaini}, Omar and {Amor{\'\i}n}, Ricardo O. and {Annunziatella}, Marianna and {Backhaus}, Bren E. and {Barro}, Guillermo and {Behroozi}, Peter and {Bell}, Eric F. and {Bhatawdekar}, Rachana and {Bisigello}, Laura and {Bromm}, Volker and {Buat}, V{\'e}ronique and {Buitrago}, Fernando and {Calabr{\`o}}, Antonello and {Casey}, Caitlin M. and {Castellano}, Marco and {Ch{\'a}vez Ortiz}, {\'O}scar A. and {Ciesla}, Laure and {Cleri}, Nikko J. and {Cohen}, Seth H. and {Cole}, Justin W. and {Cooke}, Kevin C. and {Cooper}, M.~C. and {Cooray}, Asantha R. and {Costantin}, Luca and {Cox}, Isabella G. and {Croton}, Darren and {Daddi}, Emanuele and {Dav{\'e}}, Romeel and {de La Vega}, Alexander and {Dekel}, Avishai and {Elbaz}, David and {Estrada-Carpenter}, Vicente and {Faber}, Sandra M. and {Fern{\'a}ndez}, Vital and {Finkelstein}, Keely D. and {Freundlich}, Jonathan and {Fujimoto}, Seiji and {Garc{\'\i}a-Argum{\'a}nez}, {\'A}ngela and {Gardner}, Jonathan P. and {Gawiser}, Eric and {G{\'o}mez-Guijarro}, Carlos and {Guo}, Yuchen and {Hamblin}, Kurt and {Hamilton}, Timothy S. and {Hathi}, Nimish P. and {Holwerda}, Benne W. and {Hirschmann}, Michaela and {Hutchison}, Taylor A. and {Jaskot}, Anne E. and {Jha}, Saurabh W. and {Jogee}, Shardha and {Juneau}, St{\'e}phanie and {Jung}, Intae and {Kassin}, Susan A. and {Le Bail}, Aur{\'e}lien and {Leung}, Gene C.~K. and {Lucas}, Ray A. and {Magnelli}, Benjamin and {Mantha}, Kameswara Bharadwaj and {Matharu}, Jasleen and {McGrath}, Elizabeth J. and {McIntosh}, Daniel H. and {Merlin}, Emiliano and {Mobasher}, Bahram and {Newman}, Jeffrey A. and {Nicholls}, David C. and {Pandya}, Viraj and {Rafelski}, Marc and {Ronayne}, Kaila and {Santini}, Paola and {Seill{\'e}}, Lise-Marie and {Shah}, Ekta A. and {Shen}, Lu and {Simons}, Raymond C. and {Snyder}, Gregory F. and {Stanway}, Elizabeth R. and {Straughn}, Amber N. and {Teplitz}, Harry I. and {Vanderhoof}, Brittany N. and {Vega-Ferrero}, Jes{\'u}s and {Wang}, Weichen and {Weiner}, Benjamin J. and {Willmer}, Christopher N.~A. and {Wuyts}, Stijn and {Zavala}, Jorge A. and {Ceers Team}},
        title = "{A Long Time Ago in a Galaxy Far, Far Away: A Candidate z {\ensuremath{\sim}} 12 Galaxy in Early JWST CEERS Imaging}",
      journal = {\apjl},
         year = 2022,
        month = dec,
       volume = {940},
       number = {2},
          eid = {L55},
        pages = {L55},
          doi = {10.3847/2041-8213/ac966e},
archivePrefix = {arXiv},
       eprint = {2207.12474},
 primaryClass = {astro-ph.GA},
       adsurl = {https://ui.adsabs.harvard.edu/abs/2022ApJ...940L..55F}
}

@ARTICLE{Glazebrook2023,
       author = {{Glazebrook}, K. and {Nanayakkara}, T. and {Jacobs}, C. and {Leethochawalit}, N. and {Calabr{\`o}}, A. and {Bonchi}, A. and {Castellano}, M. and {Fontana}, A. and {Mason}, C. and {Merlin}, E. and {Morishita}, T. and {Paris}, D. and {Trenti}, M. and {Treu}, T. and {Santini}, P. and {Wang}, X. and {Boyett}, K. and {Bradac}, Marusa and {Brammer}, G. and {Jones}, T. and {Marchesini}, D. and {Nonino}, M. and {Vulcani}, B.},
        title = "{Early Results from GLASS-JWST. XV. Properties of the Faintest Red Sources in the NIRCAM Deep Fields}",
      journal = {\apjl},
         year = 2023,
        month = apr,
       volume = {947},
       number = {2},
          eid = {L25},
        pages = {L25},
          doi = {10.3847/2041-8213/acba8b},
archivePrefix = {arXiv},
       eprint = {2208.03468},
 primaryClass = {astro-ph.GA},
       adsurl = {https://ui.adsabs.harvard.edu/abs/2023ApJ...947L..25G}
}

@ARTICLE{Wold2024,
       author = {{Wold}, Isak G.~B. and {Malhotra}, Sangeeta and {Rhoads}, James E. and {Weaver}, John R. and {Wang}, Bingjie},
        title = "{UNCOVERing the Faint-End of the z=7 [OIII] Luminosity Function with JWST's F410M Medium Bandpass Filter}",
      journal = {arXiv e-prints},
         year = 2024,
        month = jul,
          eid = {arXiv:2407.19023},
        pages = {arXiv:2407.19023},
          doi = {10.48550/arXiv.2407.19023},
archivePrefix = {arXiv},
       eprint = {2407.19023},
 primaryClass = {astro-ph.GA},
       adsurl = {https://ui.adsabs.harvard.edu/abs/2024arXiv240719023W}
}

@ARTICLE{Guo2024,
       author = {{Guo}, Jingsong and {Onoue}, Masafusa and {Inayoshi}, Kohei and {Kocevski}, Dale D. and {Finkelstein}, Steven L. and {Bagley}, Micaela B. and {McGrath}, Elizabeth J.},
        title = "{A Search for $z=5$ H$\alpha$ and H$\beta+$[O III] Dual-Line Emitting Galaxies in the JWST CEERS Field: Implications for the AGN Abundance}",
      journal = {arXiv e-prints},
         year = 2024,
        month = sep,
          eid = {arXiv:2409.19205},
        pages = {arXiv:2409.19205},
          doi = {10.48550/arXiv.2409.19205},
archivePrefix = {arXiv},
       eprint = {2409.19205},
 primaryClass = {astro-ph.GA},
       adsurl = {https://ui.adsabs.harvard.edu/abs/2024arXiv240919205G}
}

@ARTICLE{Harikane2024,
       author = {{Harikane}, Yuichi and {Nakajima}, Kimihiko and {Ouchi}, Masami and {Umeda}, Hiroya and {Isobe}, Yuki and {Ono}, Yoshiaki and {Xu}, Yi and {Zhang}, Yechi},
        title = "{Pure Spectroscopic Constraints on UV Luminosity Functions and Cosmic Star Formation History from 25 Galaxies at z $_{spec}$ = 8.61-13.20 Confirmed with JWST/NIRSpec}",
      journal = {\apj},
         year = 2024,
        month = jan,
       volume = {960},
       number = {1},
          eid = {56},
        pages = {56},
          doi = {10.3847/1538-4357/ad0b7e},
archivePrefix = {arXiv},
       eprint = {2304.06658},
 primaryClass = {astro-ph.GA},
       adsurl = {https://ui.adsabs.harvard.edu/abs/2024ApJ...960...56H}
}

@ARTICLE{Jung2024,
       author = {{Jung}, Intae and {Finkelstein}, Steven L. and {Arrabal Haro}, Pablo and {Dickinson}, Mark and {Ferguson}, Henry C. and {Hutchison}, Taylor A. and {Kartaltepe}, Jeyhan S. and {Larson}, Rebecca L. and {Simons}, Raymond C. and {Papovich}, Casey and {Park}, Hyunbae and {Pentericci}, Laura and {Trump}, Jonathan R. and {Amor{\'\i}n}, Ricardo O. and {Backhaus}, Bren E. and {Bagley}, Micaela B. and {Casey}, Caitlin M. and {Cheng}, Yingjie and {Cleri}, Nikko J. and {Cooper}, M.~C. and {Cooper}, Olivia R. and {Gardner}, Jonathan P. and {Gawiser}, Eric and {Grazian}, Andrea and {Hathi}, Nimish P. and {Hirschmann}, Michaela and {Koekemoer}, Anton M. and {Lucas}, Ray A. and {Mobasher}, Bahram and {Pirzkal}, Nor and {Ravindranath}, Swara and {Straughn}, Amber N. and {Yung}, L.~Y. Aaron and {de la Vega}, Alexander},
        title = "{CEERS: Diversity of Ly{\ensuremath{\alpha}} Emitters during the Epoch of Reionization}",
      journal = {\apj},
         year = 2024,
        month = may,
       volume = {967},
       number = {1},
          eid = {73},
        pages = {73},
          doi = {10.3847/1538-4357/ad3913},
archivePrefix = {arXiv},
       eprint = {2304.05385},
 primaryClass = {astro-ph.GA},
       adsurl = {https://ui.adsabs.harvard.edu/abs/2024ApJ...967...73J}
}

@ARTICLE{Kennicutt2012,
       author = {{Kennicutt}, Robert C. and {Evans}, Neal J.},
        title = "{Star Formation in the Milky Way and Nearby Galaxies}",
      journal = {\araa},
         year = 2012,
        month = sep,
       volume = {50},
        pages = {531-608},
          doi = {10.1146/annurev-astro-081811-125610},
archivePrefix = {arXiv},
       eprint = {1204.3552},
 primaryClass = {astro-ph.GA},
       adsurl = {https://ui.adsabs.harvard.edu/abs/2012ARA&A..50..531K}
}

@ARTICLE{Mason2015,
       author = {{Mason}, Charlotte A. and {Trenti}, Michele and {Treu}, Tommaso},
        title = "{The Galaxy UV Luminosity Function before the Epoch of Reionization}",
      journal = {\apj},
         year = 2015,
        month = nov,
       volume = {813},
       number = {1},
          eid = {21},
        pages = {21},
          doi = {10.1088/0004-637X/813/1/21},
archivePrefix = {arXiv},
       eprint = {1508.01204},
 primaryClass = {astro-ph.GA},
       adsurl = {https://ui.adsabs.harvard.edu/abs/2015ApJ...813...21M}
}

@ARTICLE{Rinaldi2023,
       author = {{Rinaldi}, P. and {Caputi}, K.~I. and {Costantin}, L. and {Gillman}, S. and {Iani}, E. and {P{\'e}rez-Gonz{\'a}lez}, P.~G. and {{\"O}stlin}, G. and {Colina}, L. and {Greve}, T.~R. and {Noorgard-Nielsen}, H.~U. and {Wright}, G.~S. and {Alonso-Herrero}, A. and {{\'A}lvarez-M{\'a}rquez}, J. and {Eckart}, A. and {Garc{\'\i}a-Mar{\'\i}n}, M. and {Hjorth}, J. and {Ilbert}, O. and {Kendrew}, S. and {Labiano}, A. and {Le F{\`e}vre}, O. and {Pye}, J. and {Tikkanen}, T. and {Walter}, F. and {van der Werf}, P. and {Ward}, M. and {Annunziatella}, M. and {Azzollini}, R. and {Bik}, A. and {Boogaard}, L. and {Bosman}, S.~E.~I. and {Crespo G{\'o}mez}, A. and {Jermann}, I. and {Langeroodi}, D. and {Melinder}, J. and {Meyer}, R.~A. and {Moutard}, T. and {Peissker}, F. and {Topinka}, M. and {van Dishoeck}, E. and {G{\"u}del}, M. and {Henning}, Th. and {Lagage}, P. -O. and {Ray}, T. and {Vandenbussche}, B. and {Waelkens}, C. and {Navarro-Carrera}, R. and {Kokorev}, V.},
        title = "{MIDIS: Strong (H{\ensuremath{\beta}}+[O III]) and H{\ensuremath{\alpha}} Emitters at Redshift z ≃ 7-8 Unveiled with JWST NIRCam and MIRI Imaging in the Hubble eXtreme Deep Field}",
      journal = {\apj},
         year = 2023,
        month = aug,
       volume = {952},
       number = {2},
          eid = {143},
        pages = {143},
          doi = {10.3847/1538-4357/acdc27},
archivePrefix = {arXiv},
       eprint = {2301.10717},
 primaryClass = {astro-ph.GA},
       adsurl = {https://ui.adsabs.harvard.edu/abs/2023ApJ...952..143R}
}

@ARTICLE{Robertson2015,
       author = {{Robertson}, Brant E. and {Ellis}, Richard S. and {Furlanetto}, Steven R. and {Dunlop}, James S.},
        title = "{Cosmic Reionization and Early Star-forming Galaxies: A Joint Analysis of New Constraints from Planck and the Hubble Space Telescope}",
      journal = {\apjl},
         year = 2015,
        month = apr,
       volume = {802},
       number = {2},
          eid = {L19},
        pages = {L19},
          doi = {10.1088/2041-8205/802/2/L19},
archivePrefix = {arXiv},
       eprint = {1502.02024},
 primaryClass = {astro-ph.CO},
       adsurl = {https://ui.adsabs.harvard.edu/abs/2015ApJ...802L..19R}
}

@ARTICLE{Rosdahl2022,
       author = {{Rosdahl}, Joakim and {Blaizot}, J{\'e}r{\'e}my and {Katz}, Harley and {Kimm}, Taysun and {Garel}, Thibault and {Haehnelt}, Martin and {Keating}, Laura C. and {Martin-Alvarez}, Sergio and {Michel-Dansac}, L{\'e}o and {Ocvirk}, Pierre},
        title = "{LyC escape from SPHINX galaxies in the Epoch of Reionization}",
      journal = {\mnras},
         year = 2022,
        month = sep,
       volume = {515},
       number = {2},
        pages = {2386-2414},
          doi = {10.1093/mnras/stac1942},
archivePrefix = {arXiv},
       eprint = {2207.03232},
 primaryClass = {astro-ph.GA},
       adsurl = {https://ui.adsabs.harvard.edu/abs/2022MNRAS.515.2386R}
}

@ARTICLE{Tang2023,
       author = {{Tang}, Mengtao and {Stark}, Daniel P. and {Chen}, Zuyi and {Mason}, Charlotte and {Topping}, Michael and {Endsley}, Ryan and {Senchyna}, Peter and {Plat}, Ad{\`e}le and {Lu}, Ting-Yi and {Whitler}, Lily and {Robertson}, Brant and {Charlot}, St{\'e}phane},
        title = "{JWST/NIRSpec spectroscopy of z = 7-9 star-forming galaxies with CEERS: new insight into bright Ly{\ensuremath{\alpha}} emitters in ionized bubbles}",
      journal = {\mnras},
         year = 2023,
        month = dec,
       volume = {526},
       number = {2},
        pages = {1657-1686},
          doi = {10.1093/mnras/stad2763},
archivePrefix = {arXiv},
       eprint = {2301.07072},
 primaryClass = {astro-ph.GA},
       adsurl = {https://ui.adsabs.harvard.edu/abs/2023MNRAS.526.1657T}
}

@ARTICLE{Whitler2024,
       author = {{Whitler}, Lily and {Stark}, Daniel P. and {Endsley}, Ryan and {Chen}, Zuyi and {Mason}, Charlotte and {Topping}, Michael W. and {Charlot}, St{\'e}phane},
        title = "{Insight from JWST/Near Infrared Camera into galaxy overdensities around bright Lyman-alpha emitters during reionization: implications for ionized bubbles at z   9}",
      journal = {\mnras},
         year = 2024,
        month = apr,
       volume = {529},
       number = {2},
        pages = {855-872},
          doi = {10.1093/mnras/stae516},
archivePrefix = {arXiv},
       eprint = {2305.16670},
 primaryClass = {astro-ph.GA},
       adsurl = {https://ui.adsabs.harvard.edu/abs/2024MNRAS.529..855W}
}

@ARTICLE{Xiao2023,
       author = {{Xiao}, Mengyuan and {Oesch}, Pascal and {Elbaz}, David and {Bing}, Longji and {Nelson}, Erica and {Weibel}, Andrea and {Illingworth}, Garth and {van Dokkum}, Pieter and {Naidu}, Rohan and {Daddi}, Emanuele and {Bouwens}, Rychard and {Matthee}, Jorryt and {Wuyts}, Stijn and {Chisholm}, John and {Brammer}, Gabriel and {Dickinson}, Mark and {Magnelli}, Benjamin and {Leroy}, Lucas and {Schaerer}, Daniel and {Herard-Demanche}, Thomas and {Lim}, Seunghwan and {Barrufet}, Laia and {Endsley}, Ryan and {Fudamoto}, Yoshinobu and {G{\'o}mez-Guijarro}, Carlos and {Gottumukkala}, Rashmi and {Labbe}, Ivo and {Magee}, Daniel and {Marchesini}, Danilo and {Maseda}, Michael and {Qin}, Yuxiang and {Reddy}, Naveen and {Shapley}, Alice and {Shivaei}, Irene and {Shuntov}, Marko and {Stefanon}, Mauro and {Whitaker}, Katherine and {Wyithe}, J. Stuart},
        title = "{Accelerated Formation of Ultra-Massive Galaxies in the First Billion Years}",
      journal = {arXiv e-prints},
         year = 2023,
        month = sep,
          eid = {arXiv:2309.02492},
        pages = {arXiv:2309.02492},
          doi = {10.48550/arXiv.2309.02492},
archivePrefix = {arXiv},
       eprint = {2309.02492},
 primaryClass = {astro-ph.GA},
       adsurl = {https://ui.adsabs.harvard.edu/abs/2023arXiv230902492X}
}

@ARTICLE{Sun2023,
       author = {{Sun}, Guochao and {Faucher-Gigu{\`e}re}, Claude-Andr{\'e} and {Hayward}, Christopher C. and {Shen}, Xuejian},
        title = "{Seen and unseen: bursty star formation and its implications for observations of high-redshift galaxies with JWST}",
      journal = {\mnras},
         year = 2023,
        month = dec,
       volume = {526},
       number = {2},
        pages = {2665-2672},
          doi = {10.1093/mnras/stad2902},
archivePrefix = {arXiv},
       eprint = {2305.02713},
 primaryClass = {astro-ph.GA},
       adsurl = {https://ui.adsabs.harvard.edu/abs/2023MNRAS.526.2665S}
}

@ARTICLE{Perez-Gonzlez2023,
       author = {{P{\'e}rez-Gonz{\'a}lez}, Pablo G. and {Barro}, Guillermo and {Annunziatella}, Marianna and {Costantin}, Luca and {Garc{\'\i}a-Argum{\'a}nez}, {\'A}ngela and {McGrath}, Elizabeth J. and {M{\'e}rida}, Rosa M. and {Zavala}, Jorge A. and {Arrabal Haro}, Pablo and {Bagley}, Micaela B. and {Backhaus}, Bren E. and {Behroozi}, Peter and {Bell}, Eric F. and {Bisigello}, Laura and {Buat}, V{\'e}ronique and {Calabr{\`o}}, Antonello and {Casey}, Caitlin M. and {Cleri}, Nikko J. and {Coogan}, Rosemary T. and {Cooper}, M.~C. and {Cooray}, Asantha R. and {Dekel}, Avishai and {Dickinson}, Mark and {Elbaz}, David and {Ferguson}, Henry C. and {Finkelstein}, Steven L. and {Fontana}, Adriano and {Franco}, Maximilien and {Gardner}, Jonathan P. and {Giavalisco}, Mauro and {G{\'o}mez-Guijarro}, Carlos and {Grazian}, Andrea and {Grogin}, Norman A. and {Guo}, Yuchen and {Huertas-Company}, Marc and {Jogee}, Shardha and {Kartaltepe}, Jeyhan S. and {Kewley}, Lisa J. and {Kirkpatrick}, Allison and {Kocevski}, Dale D. and {Koekemoer}, Anton M. and {Long}, Arianna S. and {Lotz}, Jennifer M. and {Lucas}, Ray A. and {Papovich}, Casey and {Pirzkal}, Nor and {Ravindranath}, Swara and {Somerville}, Rachel S. and {Tacchella}, Sandro and {Trump}, Jonathan R. and {Wang}, Weichen and {Wilkins}, Stephen M. and {Wuyts}, Stijn and {Yang}, Guang and {Yung}, L.~Y. Aaron},
        title = "{CEERS Key Paper. IV. A Triality in the Nature of HST-dark Galaxies}",
      journal = {\apjl},
         year = 2023,
        month = mar,
       volume = {946},
       number = {1},
          eid = {L16},
        pages = {L16},
          doi = {10.3847/2041-8213/acb3a5},
archivePrefix = {arXiv},
       eprint = {2211.00045},
 primaryClass = {astro-ph.GA},
       adsurl = {https://ui.adsabs.harvard.edu/abs/2023ApJ...946L..16P}
}

@ARTICLE{Chworowsky2024,
       author = {{Chworowsky}, Katherine and {Finkelstein}, Steven L. and {Boylan-Kolchin}, Michael and {McGrath}, Elizabeth J. and {Iyer}, Kartheik G. and {Papovich}, Casey and {Dickinson}, Mark and {Taylor}, Anthony J. and {Yung}, L.~Y. Aaron and {Arrabal Haro}, Pablo and {Bagley}, Micaela B. and {Backhaus}, Bren E. and {Bhatawdekar}, Rachana and {Cheng}, Yingjie and {Cleri}, Nikko J. and {Cole}, Justin W. and {Cooper}, M.~C. and {Costantin}, Luca and {Dekel}, Avishai and {Franco}, Maximilien and {Fujimoto}, Seiji and {Hayward}, Christopher C. and {Holwerda}, Benne W. and {Huertas-Company}, Marc and {Hirschmann}, Michaela and {Hutchison}, Taylor A. and {Koekemoer}, Anton M. and {Larson}, Rebecca L. and {Li}, Zhaozhou and {Long}, Arianna S. and {Lucas}, Ray A. and {Pirzkal}, Nor and {Rodighiero}, Giulia and {Somerville}, Rachel S. and {Vanderhoof}, Brittany N. and {de la Vega}, Alexander and {Wilkins}, Stephen M. and {Yang}, Guang and {Zavala}, Jorge A.},
        title = "{Evidence for a Shallow Evolution in the Volume Densities of Massive Galaxies at z = 4{\textendash}8 from CEERS}",
      journal = {\aj},
         year = 2024,
        month = sep,
       volume = {168},
       number = {3},
          eid = {113},
        pages = {113},
          doi = {10.3847/1538-3881/ad57c1},
archivePrefix = {arXiv},
       eprint = {2311.14804},
 primaryClass = {astro-ph.GA},
       adsurl = {https://ui.adsabs.harvard.edu/abs/2024AJ....168..113C}
}

@ARTICLE{2022ApJ...935..167A,
       author = {{Astropy Collaboration} and {Price-Whelan}, Adrian M. and {Lim}, Pey Lian and {Earl}, Nicholas and {Starkman}, Nathaniel and {Bradley}, Larry and {Shupe}, David L. and {Patil}, Aarya A. and {Corrales}, Lia and {Brasseur}, C.~E. and {N{\"o}the}, Maximilian and {Donath}, Axel and {Tollerud}, Erik and {Morris}, Brett M. and {Ginsburg}, Adam and {Vaher}, Eero and {Weaver}, Benjamin A. and {Tocknell}, James and {Jamieson}, William and {van Kerkwijk}, Marten H. and {Robitaille}, Thomas P. and {Merry}, Bruce and {Bachetti}, Matteo and {G{\"u}nther}, H. Moritz and {Aldcroft}, Thomas L. and {Alvarado-Montes}, Jaime A. and {Archibald}, Anne M. and {B{\'o}di}, Attila and {Bapat}, Shreyas and {Barentsen}, Geert and {Baz{\'a}n}, Juanjo and {Biswas}, Manish and {Boquien}, M{\'e}d{\'e}ric and {Burke}, D.~J. and {Cara}, Daria and {Cara}, Mihai and {Conroy}, Kyle E. and {Conseil}, Simon and {Craig}, Matthew W. and {Cross}, Robert M. and {Cruz}, Kelle L. and {D'Eugenio}, Francesco and {Dencheva}, Nadia and {Devillepoix}, Hadrien A.~R. and {Dietrich}, J{\"o}rg P. and {Eigenbrot}, Arthur Davis and {Erben}, Thomas and {Ferreira}, Leonardo and {Foreman-Mackey}, Daniel and {Fox}, Ryan and {Freij}, Nabil and {Garg}, Suyog and {Geda}, Robel and {Glattly}, Lauren and {Gondhalekar}, Yash and {Gordon}, Karl D. and {Grant}, David and {Greenfield}, Perry and {Groener}, Austen M. and {Guest}, Steve and {Gurovich}, Sebastian and {Handberg}, Rasmus and {Hart}, Akeem and {Hatfield-Dodds}, Zac and {Homeier}, Derek and {Hosseinzadeh}, Griffin and {Jenness}, Tim and {Jones}, Craig K. and {Joseph}, Prajwel and {Kalmbach}, J. Bryce and {Karamehmetoglu}, Emir and {Ka{\l}uszy{\'n}ski}, Miko{\l}aj and {Kelley}, Michael S.~P. and {Kern}, Nicholas and {Kerzendorf}, Wolfgang E. and {Koch}, Eric W. and {Kulumani}, Shankar and {Lee}, Antony and {Ly}, Chun and {Ma}, Zhiyuan and {MacBride}, Conor and {Maljaars}, Jakob M. and {Muna}, Demitri and {Murphy}, N.~A. and {Norman}, Henrik and {O'Steen}, Richard and {Oman}, Kyle A. and {Pacifici}, Camilla and {Pascual}, Sergio and {Pascual-Granado}, J. and {Patil}, Rohit R. and {Perren}, Gabriel I. and {Pickering}, Timothy E. and {Rastogi}, Tanuj and {Roulston}, Benjamin R. and {Ryan}, Daniel F. and {Rykoff}, Eli S. and {Sabater}, Jose and {Sakurikar}, Parikshit and {Salgado}, Jes{\'u}s and {Sanghi}, Aniket and {Saunders}, Nicholas and {Savchenko}, Volodymyr and {Schwardt}, Ludwig and {Seifert-Eckert}, Michael and {Shih}, Albert Y. and {Jain}, Anany Shrey and {Shukla}, Gyanendra and {Sick}, Jonathan and {Simpson}, Chris and {Singanamalla}, Sudheesh and {Singer}, Leo P. and {Singhal}, Jaladh and {Sinha}, Manodeep and {Sip{\H{o}}cz}, Brigitta M. and {Spitler}, Lee R. and {Stansby}, David and {Streicher}, Ole and {{\v{S}}umak}, Jani and {Swinbank}, John D. and {Taranu}, Dan S. and {Tewary}, Nikita and {Tremblay}, Grant R. and {de Val-Borro}, Miguel and {Van Kooten}, Samuel J. and {Vasovi{\'c}}, Zlatan and {Verma}, Shresth and {de Miranda Cardoso}, Jos{\'e} Vin{\'\i}cius and {Williams}, Peter K.~G. and {Wilson}, Tom J. and {Winkel}, Benjamin and {Wood-Vasey}, W.~M. and {Xue}, Rui and {Yoachim}, Peter and {Zhang}, Chen and {Zonca}, Andrea and {Astropy Project Contributors}},
        title = "{The Astropy Project: Sustaining and Growing a Community-oriented Open-source Project and the Latest Major Release (v5.0) of the Core Package}",
      journal = {\apj},
         year = 2022,
        month = aug,
       volume = {935},
       number = {2},
          eid = {167},
        pages = {167},
          doi = {10.3847/1538-4357/ac7c74},
archivePrefix = {arXiv},
       eprint = {2206.14220},
 primaryClass = {astro-ph.IM},
       adsurl = {https://ui.adsabs.harvard.edu/abs/2022ApJ...935..167A}
}

@ARTICLE{2018AJ....156..123A,
       author = {{Astropy Collaboration} and {Price-Whelan}, A.~M. and {Sip{\H{o}}cz}, B.~M. and {G{\"u}nther}, H.~M. and {Lim}, P.~L. and {Crawford}, S.~M. and {Conseil}, S. and {Shupe}, D.~L. and {Craig}, M.~W. and {Dencheva}, N. and {Ginsburg}, A. and {VanderPlas}, J.~T. and {Bradley}, L.~D. and {P{\'e}rez-Su{\'a}rez}, D. and {de Val-Borro}, M. and {Aldcroft}, T.~L. and {Cruz}, K.~L. and {Robitaille}, T.~P. and {Tollerud}, E.~J. and {Ardelean}, C. and {Babej}, T. and {Bach}, Y.~P. and {Bachetti}, M. and {Bakanov}, A.~V. and {Bamford}, S.~P. and {Barentsen}, G. and {Barmby}, P. and {Baumbach}, A. and {Berry}, K.~L. and {Biscani}, F. and {Boquien}, M. and {Bostroem}, K.~A. and {Bouma}, L.~G. and {Brammer}, G.~B. and {Bray}, E.~M. and {Breytenbach}, H. and {Buddelmeijer}, H. and {Burke}, D.~J. and {Calderone}, G. and {Cano Rodr{\'\i}guez}, J.~L. and {Cara}, M. and {Cardoso}, J.~V.~M. and {Cheedella}, S. and {Copin}, Y. and {Corrales}, L. and {Crichton}, D. and {D'Avella}, D. and {Deil}, C. and {Depagne}, {\'E}. and {Dietrich}, J.~P. and {Donath}, A. and {Droettboom}, M. and {Earl}, N. and {Erben}, T. and {Fabbro}, S. and {Ferreira}, L.~A. and {Finethy}, T. and {Fox}, R.~T. and {Garrison}, L.~H. and {Gibbons}, S.~L.~J. and {Goldstein}, D.~A. and {Gommers}, R. and {Greco}, J.~P. and {Greenfield}, P. and {Groener}, A.~M. and {Grollier}, F. and {Hagen}, A. and {Hirst}, P. and {Homeier}, D. and {Horton}, A.~J. and {Hosseinzadeh}, G. and {Hu}, L. and {Hunkeler}, J.~S. and {Ivezi{\'c}}, {\v{Z}}. and {Jain}, A. and {Jenness}, T. and {Kanarek}, G. and {Kendrew}, S. and {Kern}, N.~S. and {Kerzendorf}, W.~E. and {Khvalko}, A. and {King}, J. and {Kirkby}, D. and {Kulkarni}, A.~M. and {Kumar}, A. and {Lee}, A. and {Lenz}, D. and {Littlefair}, S.~P. and {Ma}, Z. and {Macleod}, D.~M. and {Mastropietro}, M. and {McCully}, C. and {Montagnac}, S. and {Morris}, B.~M. and {Mueller}, M. and {Mumford}, S.~J. and {Muna}, D. and {Murphy}, N.~A. and {Nelson}, S. and {Nguyen}, G.~H. and {Ninan}, J.~P. and {N{\"o}the}, M. and {Ogaz}, S. and {Oh}, S. and {Parejko}, J.~K. and {Parley}, N. and {Pascual}, S. and {Patil}, R. and {Patil}, A.~A. and {Plunkett}, A.~L. and {Prochaska}, J.~X. and {Rastogi}, T. and {Reddy Janga}, V. and {Sabater}, J. and {Sakurikar}, P. and {Seifert}, M. and {Sherbert}, L.~E. and {Sherwood-Taylor}, H. and {Shih}, A.~Y. and {Sick}, J. and {Silbiger}, M.~T. and {Singanamalla}, S. and {Singer}, L.~P. and {Sladen}, P.~H. and {Sooley}, K.~A. and {Sornarajah}, S. and {Streicher}, O. and {Teuben}, P. and {Thomas}, S.~W. and {Tremblay}, G.~R. and {Turner}, J.~E.~H. and {Terr{\'o}n}, V. and {van Kerkwijk}, M.~H. and {de la Vega}, A. and {Watkins}, L.~L. and {Weaver}, B.~A. and {Whitmore}, J.~B. and {Woillez}, J. and {Zabalza}, V. and {Astropy Contributors}},
        title = "{The Astropy Project: Building an Open-science Project and Status of the v2.0 Core Package}",
      journal = {\aj},
         year = 2018,
        month = sep,
       volume = {156},
       number = {3},
          eid = {123},
        pages = {123},
          doi = {10.3847/1538-3881/aabc4f},
archivePrefix = {arXiv},
       eprint = {1801.02634},
 primaryClass = {astro-ph.IM},
       adsurl = {https://ui.adsabs.harvard.edu/abs/2018AJ....156..123A}
}

@ARTICLE{2013A&A...558A..33A,
       author = {{Astropy Collaboration} and {Robitaille}, Thomas P. and
         {Tollerud}, Erik J. and {Greenfield}, Perry and {Droettboom}, Michael and
         {Bray}, Erik and {Aldcroft}, Tom and {Davis}, Matt and
         {Ginsburg}, Adam and {Price-Whelan}, Adrian M. and
         {Kerzendorf}, Wolfgang E. and {Conley}, Alexander and {Crighton}, Neil and
         {Barbary}, Kyle and {Muna}, Demitri and {Ferguson}, Henry and
         {Grollier}, Fr{\'e}d{\'e}ric and {Parikh}, Madhura M. and
         {Nair}, Prasanth H. and {Unther}, Hans M. and {Deil}, Christoph and
         {Woillez}, Julien and {Conseil}, Simon and {Kramer}, Roban and
         {Turner}, James E.~H. and {Singer}, Leo and {Fox}, Ryan and
         {Weaver}, Benjamin A. and {Zabalza}, Victor and {Edwards}, Zachary I. and
         {Azalee Bostroem}, K. and {Burke}, D.~J. and {Casey}, Andrew R. and
         {Crawford}, Steven M. and {Dencheva}, Nadia and {Ely}, Justin and
         {Jenness}, Tim and {Labrie}, Kathleen and {Lim}, Pey Lian and
         {Pierfederici}, Francesco and {Pontzen}, Andrew and {Ptak}, Andy and
         {Refsdal}, Brian and {Servillat}, Mathieu and {Streicher}, Ole},
        title = "{Astropy: A community Python package for astronomy}",
      journal = {\aap},
         year = "2013",
        month = "Oct",
       volume = {558},
          eid = {A33},
        pages = {A33},
          doi = {10.1051/0004-6361/201322068},
archivePrefix = {arXiv},
       eprint = {1307.6212},
 primaryClass = {astro-ph.IM},
       adsurl = {https://ui.adsabs.harvard.edu/abs/2013A&A...558A..33A}
}

@ARTICLE{1996A&AS..117..393B,
       author = {{Bertin}, E. and {Arnouts}, S.},
        title = "{SExtractor: Software for source extraction.}",
      journal = {\aaps},
         year = "1996",
        month = "Jun",
       volume = {117},
        pages = {393-404},
          doi = {10.1051/aas:1996164},
       adsurl = {https://ui.adsabs.harvard.edu/abs/1996A&AS..117..393B}
}

@ARTICLE{2013RMxAA..49..137F,
       author = {{Ferland}, G.~J. and {Porter}, R.~L. and {van Hoof}, P.~A.~M. and
         {Williams}, R.~J.~R. and {Abel}, N.~P. and {Lykins}, M.~L. and
         {Shaw}, G. and {Henney}, W.~J. and {Stancil}, P.~C.},
        title = "{The 2013 Release of Cloudy}",
      journal = {\rmxaa},
         year = "2013",
        month = "Apr",
       volume = {49},
        pages = {137-163},
archivePrefix = {arXiv},
       eprint = {1302.4485},
 primaryClass = {astro-ph.GA},
       adsurl = {https://ui.adsabs.harvard.edu/abs/2013RMxAA..49..137F}
}

@ARTICLE{Muzzin2025,
       author = {{Muzzin}, Adam and {Suess}, Katherine A. and {Marchesini}, Danilo and {Robbins}, Luke and {Willott}, Chris J. and {Alberts}, Stacey and {Antwi-Danso}, Jacqueline and {Asada}, Yoshihisa and {Brammer}, Gabriel and {Cutler}, Sam E. and {Iyer}, Kartheik G. and {Labbe}, Ivo and {Martis}, Nicholas S. and {Miller}, Tim B. and {Mitsuhashi}, Ikki and {Pope}, Alexandra and {Sajina}, Anna and {Sarrouh}, Ghassan T.~E. and {Sharma}, Monu and {Stefanon}, Mauro and {Whitaker}, Katherine E. and {Abraham}, Roberto and {Atek}, Hakim and {Bradac}, Marusa and {Berek}, Samantha and {Bezanson}, Rachel and {Brown}, Westley and {Burgasser}, Adam J. and {Chicoine}, Nathalie and {Cloonan}, Aidan P. and {Cooper}, Olivia R. and {Dayal}, Pratika and {de Graaff}, Anna and {Desprez}, Guillaume and {Feldmann}, Robert and {Forrest}, Ben and {Franx}, Marijn and {Fudamoto}, Yoshinobu and {Fujimoto}, Seiji and {Furtak}, Lukas J. and {Glazebrook}, Karl and {Goovaerts}, Ilias and {Greene}, Jenny E. and {Jagga}, Naadiyah and {Jarvis}, William W.~H. and {Kriek}, Mariska and {Khullar}, Gourav and {La Torre}, Valentina and {Leja}, Joel and {Lin}, Jamie and {Lorenz}, Brian and {Lyon}, Daniel and {Markov}, Vladan and {Maseda}, Michael V. and {McConachie}, Ian and {Merchant}, Maya and {Merida}, Rosa M. and {Mowla}, Lamiya and {Myers}, Katherine and {Naidu}, Rohan P. and {Nanayakkara}, Themiya and {Nelson}, Erica J. and {Noirot}, Gael and {Oesch}, Pascal A. and {Omori}, Kiyoaki C. and {Pan}, Richard and {Porraz Barrera}, Natalia and {Price}, Sedona H. and {Ravindranath}, Swara and {Sawicki}, Marcin and {Setton}, David J. and {Smit}, Renske and {Sok}, Visal and {Speagle}, Joshua S. and {Taylor}, Edward N. and {Tan}, Vivian Yun Yan and {Tripodi}, Roberta and {van der Wel}, Arjen and {Perez Vidal}, Edgar and {Wang}, Bingjie and {Weaver}, John R. and {Williams}, Christina C. and {Withers}, Sunna and {Zaidi}, Kumail},
        title = "{MINERVA: A NIRCam Medium Band and MIRI Imaging Survey to Unlock the Hidden Gems of the Distant Universe}",
      journal = {arXiv e-prints},
         year = 2025,
        month = jul,
          eid = {arXiv:2507.19706},
        pages = {arXiv:2507.19706},
          doi = {10.48550/arXiv.2507.19706},
archivePrefix = {arXiv},
       eprint = {2507.19706},
 primaryClass = {astro-ph.GA},
       adsurl = {https://ui.adsabs.harvard.edu/abs/2025arXiv250719706M}
}

@ARTICLE{Davis2026,
       author = {{Davis}, Kelcey and {Brooks}, Madisyn and {Trump}, Jonathan R. and {Fern{\'a}ndez}, Vital and {Hutchison}, Taylor A. and {Larson}, Rebecca L. and {Taylor}, Anthony J. and {McGrath}, Elizabeth J. and {Barro}, Guillermo and {Koekemoer}, Anton M. and {Arrabal Haro}, Pablo and {Dickinson}, Mark and {Backhaus}, Bren E. and {Cleri}, Nikko J. and {Finkelstein}, Steven L. and {Ganapathy}, Ananya and {Simons}, Raymond C. and {Amor{\'\i}n}, Ricardo O. and {de la Vega}, Alexander and {Grogin}, Norman A. and {Hirschmann}, Michaela and {Hu}, Weida and {Johnson}, Jarrett L. and {Kartaltepe}, Jeyhan S. and {Kocevski}, Dale and {Llerena}, Mario and {Lucas}, Ray A. and {Marshall}, Madeline A. and {Pacucci}, Fabio and {Pentericci}, Laura and {Upton Sanderbeck}, Phoebe R.},
        title = "{Extreme Emission Line Galaxies in CEERS Are Powered by Star Formation, not AGN}",
      journal = {arXiv e-prints},
         year = 2026,
        month = feb,
          eid = {arXiv:2602.23310},
        pages = {arXiv:2602.23310},
          doi = {10.48550/arXiv.2602.23310},
archivePrefix = {arXiv},
       eprint = {2602.23310},
 primaryClass = {astro-ph.GA},
       adsurl = {https://ui.adsabs.harvard.edu/abs/2026arXiv260223310D}
}

@ARTICLE{Davis2023,
       author = {{Davis}, Kelcey and {Trump}, Jonathan R. and {Simons}, Raymond C. and {McGrath}, Elizabeth J. and {Wilkins}, Stephen M. and {Arrabal Haro}, Pablo and {Bagley}, Micaela B. and {Dickinson}, Mark and {Fern{\'a}ndez}, Vital and {Amor{\'\i}n}, Ricardo O. and {Backhaus}, Bren E. and {Cleri}, Nikko J. and {Llerena}, Mario and {Brunker}, Samantha W. and {Barro}, Guillermo and {Bisigello}, Laura and {Brooks}, Madisyn and {Costantin}, Luca and {de La Vega}, Alexander and {Dekel}, Avishai and {Finkelstein}, Steven L. and {Hathi}, Nimish P. and {Hirschmann}, Michaela and {Kartaltepe}, Jeyhan S. and {Koekemoer}, Anton M. and {Lucas}, Ray A. and {Papovich}, Casey and {P{\'e}rez-Gonz{\'a}lez}, Pablo G. and {Pirzkal}, Nor and {Rodighiero}, Giulia and {Rose}, Caitlin and {Yung}, L.~Y. Aaron and {Ceers Collaborators}},
        title = "{A Census from JWST of Extreme Emission-line Galaxies Spanning the Epoch of Reionization in CEERS}",
      journal = {\apj},
         year = 2024,
        month = oct,
       volume = {974},
       number = {1},
          eid = {42},
        pages = {42},
          doi = {10.3847/1538-4357/ad6865},
archivePrefix = {arXiv},
       eprint = {2312.07799},
 primaryClass = {astro-ph.GA},
       adsurl = {https://ui.adsabs.harvard.edu/abs/2024ApJ...974...42D}
}

@ARTICLE{Oke1983, 
       author = {{Oke}, J.~B. and {Gunn}, J.~E.},
        title = "{Secondary standard stars for absolute spectrophotometry.}",
      journal = {\apj},
         year = 1983,
        month = mar,
       volume = {266},
        pages = {713-717},
          doi = {10.1086/160817},
       adsurl = {https://ui.adsabs.harvard.edu/abs/1983ApJ...266..713O}
}

@ARTICLE{Koekemoer11,
       author = {{Koekemoer}, Anton M. and {Faber}, S.~M. and {Ferguson}, Henry C. and {Grogin}, Norman A. and {Kocevski}, Dale D. and {Koo}, David C. and {Lai}, Kamson and {Lotz}, Jennifer M. and {Lucas}, Ray A. and {McGrath}, Elizabeth J. and {Ogaz}, Sara and {Rajan}, Abhijith and {Riess}, Adam G. and {Rodney}, Steve A. and {Strolger}, Louis and {Casertano}, Stefano and {Castellano}, Marco and {Dahlen}, Tomas and {Dickinson}, Mark and {Dolch}, Timothy and {Fontana}, Adriano and {Giavalisco}, Mauro and {Grazian}, Andrea and {Guo}, Yicheng and {Hathi}, Nimish P. and {Huang}, Kuang-Han and {van der Wel}, Arjen and {Yan}, Hao-Jing and {Acquaviva}, Viviana and {Alexander}, David M. and {Almaini}, Omar and {Ashby}, Matthew L.~N. and {Barden}, Marco and {Bell}, Eric F. and {Bournaud}, Fr{\'e}d{\'e}ric and {Brown}, Thomas M. and {Caputi}, Karina I. and {Cassata}, Paolo and {Challis}, Peter J. and {Chary}, Ranga-Ram and {Cheung}, Edmond and {Cirasuolo}, Michele and {Conselice}, Christopher J. and {Roshan Cooray}, Asantha and {Croton}, Darren J. and {Daddi}, Emanuele and {Dav{\'e}}, Romeel and {de Mello}, Duilia F. and {de Ravel}, Loic and {Dekel}, Avishai and {Donley}, Jennifer L. and {Dunlop}, James S. and {Dutton}, Aaron A. and {Elbaz}, David and {Fazio}, Giovanni G. and {Filippenko}, Alexei V. and {Finkelstein}, Steven L. and {Frazer}, Chris and {Gardner}, Jonathan P. and {Garnavich}, Peter M. and {Gawiser}, Eric and {Gruetzbauch}, Ruth and {Hartley}, Will G. and {H{\"a}ussler}, Boris and {Herrington}, Jessica and {Hopkins}, Philip F. and {Huang}, Jia-Sheng and {Jha}, Saurabh W. and {Johnson}, Andrew and {Kartaltepe}, Jeyhan S. and {Khostovan}, Ali A. and {Kirshner}, Robert P. and {Lani}, Caterina and {Lee}, Kyoung-Soo and {Li}, Weidong and {Madau}, Piero and {McCarthy}, Patrick J. and {McIntosh}, Daniel H. and {McLure}, Ross J. and {McPartland}, Conor and {Mobasher}, Bahram and {Moreira}, Heidi and {Mortlock}, Alice and {Moustakas}, Leonidas A. and {Mozena}, Mark and {Nandra}, Kirpal and {Newman}, Jeffrey A. and {Nielsen}, Jennifer L. and {Niemi}, Sami and {Noeske}, Kai G. and {Papovich}, Casey J. and {Pentericci}, Laura and {Pope}, Alexandra and {Primack}, Joel R. and {Ravindranath}, Swara and {Reddy}, Naveen A. and {Renzini}, Alvio and {Rix}, Hans-Walter and {Robaina}, Aday R. and {Rosario}, David J. and {Rosati}, Piero and {Salimbeni}, Sara and {Scarlata}, Claudia and {Siana}, Brian and {Simard}, Luc and {Smidt}, Joseph and {Snyder}, Diana and {Somerville}, Rachel S. and {Spinrad}, Hyron and {Straughn}, Amber N. and {Telford}, Olivia and {Teplitz}, Harry I. and {Trump}, Jonathan R. and {Vargas}, Carlos and {Villforth}, Carolin and {Wagner}, Cory R. and {Wandro}, Pat and {Wechsler}, Risa H. and {Weiner}, Benjamin J. and {Wiklind}, Tommy and {Wild}, Vivienne and {Wilson}, Grant and {Wuyts}, Stijn and {Yun}, Min S.},
        title = "{CANDELS: The Cosmic Assembly Near-infrared Deep Extragalactic Legacy Survey{\textemdash}The Hubble Space Telescope Observations, Imaging Data Products, and Mosaics}",
      journal = {\apjs},
         year = 2011,
        month = dec,
       volume = {197},
       number = {2},
          eid = {36},
        pages = {36},
          doi = {10.1088/0067-0049/197/2/36},
archivePrefix = {arXiv},
       eprint = {1105.3754},
 primaryClass = {astro-ph.CO},
       adsurl = {https://ui.adsabs.harvard.edu/abs/2011ApJS..197...36K}
}

@software{bushouse_2025_16280965,
  author       = {Bushouse, Howard and Eisenhamer, Jonathan and
                  Dencheva, Nadia and Davies, James and
                  Greenfield, Perry and Morrison, Jane and
                  Hodge, Phil and Simon, Bernie and
                  Grumm, David and Droettboom, Michael and
                  Slavich, Edward and Sosey, Megan and
                  Pauly, Tyler and Miller, Todd and
                  Jedrzejewski, Robert and Hack, Warren and
                  Davis, David and Crawford, Steven and
                  Law, David and Gordon, Karl},
  title        = {{JWST Calibration Pipeline}},
  year         = 2025,
  publisher    = {Zenodo},
  doi          = {10.5281/zenodo.16280965},
  url          = {https://doi.org/10.5281/zenodo.16280965}
}

@ARTICLE{Kocevski2024,
       author = {{Kocevski}, Dale D. and {Finkelstein}, Steven L. and {Barro}, Guillermo and {Taylor}, Anthony J. and {Calabr{\`o}}, Antonello and {Laloux}, Brivael and {Buchner}, Johannes and {Trump}, Jonathan R. and {Leung}, Gene C.~K. and {Yang}, Guang and {Dickinson}, Mark and {P{\'e}rez-Gonz{\'a}lez}, Pablo G. and {Pacucci}, Fabio and {Inayoshi}, Kohei and {Somerville}, Rachel S. and {McGrath}, Elizabeth J. and {Akins}, Hollis B. and {Bagley}, Micaela B. and {Bowler}, Rebecca A.~A. and {Bisigello}, Laura and {Carnall}, Adam and {Casey}, Caitlin M. and {Cheng}, Yingjie and {Cleri}, Nikko J. and {Costantin}, Luca and {Cullen}, Fergus and {Davis}, Kelcey and {Donnan}, Callum T. and {Dunlop}, James S. and {Ellis}, Richard S. and {Ferguson}, Henry C. and {Fujimoto}, Seiji and {Fontana}, Adriano and {Giavalisco}, Mauro and {Grazian}, Andrea and {Grogin}, Norman A. and {Hathi}, Nimish P. and {Hirschmann}, Michaela and {Huertas-Company}, Marc and {Holwerda}, Benne W. and {Illingworth}, Garth and {Juneau}, St{\'e}phanie and {Kartaltepe}, Jeyhan S. and {Koekemoer}, Anton M. and {Li}, Wenxiu and {Lucas}, Ray A. and {Magee}, Dan and {Mason}, Charlotte and {McLeod}, Derek J. and {McLure}, Ross J. and {Napolitano}, Lorenzo and {Papovich}, Casey and {Pirzkal}, Nor and {Rodighiero}, Giulia and {Santini}, Paola and {Wilkins}, Stephen M. and {Yung}, L.~Y. Aaron},
        title = "{The Rise of Faint, Red Active Galactic Nuclei at z > 4: A Sample of Little Red Dots in the JWST Extragalactic Legacy Fields}",
      journal = {\apj},
         year = 2025,
        month = jun,
       volume = {986},
       number = {2},
          eid = {126},
        pages = {126},
          doi = {10.3847/1538-4357/adbc7d},
archivePrefix = {arXiv},
       eprint = {2404.03576},
 primaryClass = {astro-ph.GA},
       adsurl = {https://ui.adsabs.harvard.edu/abs/2025ApJ...986..126K}
}

@Article{Hunter2007,
  Author    = {Hunter, J. D.},
  Title     = {Matplotlib: A 2D graphics environment},
  Journal   = {Computing in Science \& Engineering},
  Volume    = {9},
  Number    = {3},
  Pages     = {90--95},
  publisher = {IEEE COMPUTER SOC},
  doi       = {10.1109/MCSE.2007.55},
  year      = 2007
}

@Article{         numpy,
 title         = {Array programming with {NumPy}},
 author        = {Charles R. Harris and K. Jarrod Millman and St{\'{e}}fan J.
                 van der Walt and Ralf Gommers and Pauli Virtanen and David
                 Cournapeau and Eric Wieser and Julian Taylor and Sebastian
                 Berg and Nathaniel J. Smith and Robert Kern and Matti Picus
                 and Stephan Hoyer and Marten H. van Kerkwijk and Matthew
                 Brett and Allan Haldane and Jaime Fern{\'{a}}ndez del
                 R{\'{i}}o and Mark Wiebe and Pearu Peterson and Pierre
                 G{\'{e}}rard-Marchant and Kevin Sheppard and Tyler Reddy and
                 Warren Weckesser and Hameer Abbasi and Christoph Gohlke and
                 Travis E. Oliphant},
 year          = {2020},
 month         = sep,
 journal       = {Nature},
 volume        = {585},
 number        = {7825},
 pages         = {357--362},
 doi           = {10.1038/s41586-020-2649-2},
 publisher     = {Springer Science and Business Media {LLC}},
 url           = {https://doi.org/10.1038/s41586-020-2649-2}
}

@software{pandas1,
    author       = {The pandas development team},
    title        = {pandas-dev/pandas: Pandas},
    month        = feb,
    year         = 2020,
    publisher    = {Zenodo},
    version      = {1.3.1},
    doi          = {10.5281/zenodo.3509134},
    url          = {https://doi.org/10.5281/zenodo.3509134}
}

@InProceedings{ pandas2,
  author    = { {W}es {M}c{K}inney },
  title     = { {D}ata {S}tructures for {S}tatistical {C}omputing in {P}ython },
  booktitle = { {P}roceedings of the 9th {P}ython in {S}cience {C}onference },
  pages     = { 56 - 61 },
  year      = { 2010 },
  editor    = { {S}t\'efan van der {W}alt and {J}arrod {M}illman },
  doi       = { 10.25080/Majora-92bf1922-00a }
}

@ARTICLE{scipy,
  author  = {Virtanen, Pauli and Gommers, Ralf and Oliphant, Travis E. and
            Haberland, Matt and Reddy, Tyler and Cournapeau, David and
            Burovski, Evgeni and Peterson, Pearu and Weckesser, Warren and
            Bright, Jonathan and {van der Walt}, St{\'e}fan J. and
            Brett, Matthew and Wilson, Joshua and Millman, K. Jarrod and
            Mayorov, Nikolay and Nelson, Andrew R. J. and Jones, Eric and
            Kern, Robert and Larson, Eric and Carey, C J and
            Polat, {\.I}lhan and Feng, Yu and Moore, Eric W. and
            {VanderPlas}, Jake and Laxalde, Denis and Perktold, Josef and
            Cimrman, Robert and Henriksen, Ian and Quintero, E. A. and
            Harris, Charles R. and Archibald, Anne M. and
            Ribeiro, Ant{\^o}nio H. and Pedregosa, Fabian and
            {van Mulbregt}, Paul and {SciPy 1.0 Contributors}},
  title   = {{{SciPy} 1.0: Fundamental Algorithms for Scientific
            Computing in Python}},
  journal = {Nature Methods},
  year    = {2020},
  volume  = {17},
  pages   = {261--272},
  adsurl  = {https://rdcu.be/b08Wh},
  doi     = {10.1038/s41592-019-0686-2},
}

@ARTICLE{Llerena2024,
       author = {{Llerena}, M. and {Amor{\'\i}n}, R. and {Pentericci}, L. and {Arrabal Haro}, P. and {Backhaus}, B.~E. and {Bagley}, M.~B. and {Calabr{\`o}}, A. and {Cleri}, N.~J. and {Davis}, K. and {Dickinson}, M. and {Finkelstein}, S.~L. and {Gawiser}, E. and {Grogin}, N.~A. and {Hathi}, N.~P. and {Hirschmann}, M. and {Kartaltepe}, J.~S. and {Koekemoer}, A.~M. and {McGrath}, E.~J. and {Mobasher}, B. and {Napolitano}, L. and {Papovich}, C. and {Pirzkal}, N. and {Trump}, J.~R. and {Wilkins}, S.~M. and {Yung}, L.~Y.~A.},
        title = "{Physical properties of extreme emission-line galaxies at z {\ensuremath{\sim}} 4{\textendash}9 from the JWST CEERS survey}",
      journal = {\aap},
         year = 2024,
        month = nov,
       volume = {691},
          eid = {A59},
        pages = {A59},
          doi = {10.1051/0004-6361/202449904},
archivePrefix = {arXiv},
       eprint = {2403.05362},
 primaryClass = {astro-ph.GA},
       adsurl = {https://ui.adsabs.harvard.edu/abs/2024A&A...691A..59L}
}

@ARTICLE{Brammer2008,
       author = {{Brammer}, Gabriel B. and {van Dokkum}, Pieter G. and {Coppi}, Paolo},
        title = "{EAZY: A Fast, Public Photometric Redshift Code}",
      journal = {\apj},
         year = 2008,
        month = oct,
       volume = {686},
       number = {2},
        pages = {1503-1513},
          doi = {10.1086/591786},
archivePrefix = {arXiv},
       eprint = {0807.1533},
 primaryClass = {astro-ph},
       adsurl = {https://ui.adsabs.harvard.edu/abs/2008ApJ...686.1503B}
}
\bibliographystyle{aasjournalv7}



\end{document}